\documentclass[twocolumn]{aastex62}
\usepackage{CJK}
\usepackage{natbib}
\usepackage{graphicx}
\usepackage {amsmath,amssymb,eqnarray}
\usepackage{enumerate}
\usepackage{enumitem}
\usepackage{mathtools}

\newcommand{\hi}{{\rm H}{\textsc i}}

\newcommand{\degree}{\ensuremath{^\circ}}

\def\jykms{\rm{Jy~km~s^{-1} }}
\def\jybkms{\rm{Jy~b^{-1}~km~s^{-1} }}
\def\mjyb{\rm{mJy~b^{-1} }}
\def\mjybf{\rm{mJy~b_{F}^{-1} }}
\def\jybf{\rm{Jy~b_{F}^{-1} }}

\def\kms{\rm{km~s^{-1} }}
\def\cmsq{\rm{ cm^{-2} } }
\def\NHI{N_{\rm HI}}

\begin{document}

\begin{CJK*}{UTF8}{gbsn}

\title{ FEASTS: IGM cooling triggered by tidal interactions through the diffuse HI phase around NGC 4631}
\correspondingauthor{Jing Wang}
\email{jwang\_astro@pku.edu.cn}

\correspondingauthor{S-H. Oh}
\email{seheonoh@kasi.re.kr}

\author[0000-0002-6593-8820]{Jing Wang (王菁)}
\affiliation{ Kavli Institute for Astronomy and Astrophysics, Peking University, Beijing 100871, China}

\author{Dong Yang (杨冬)}
\affiliation{ Kavli Institute for Astronomy and Astrophysics, Peking University, Beijing 100871, China}

\author{S-H. Oh}
\affiliation{Department of Physics and Astronomy, Sejong University, 209 Neungdong-ro, Gwangjin-gu, Seoul, Republic of Korea}

\author{Lister Staveley-Smith}
\affiliation{International Centre for Radio Astronomy Research, University of Western Australia, 35 Stirling Highway, Crawley, WA 6009, Australia}
\affiliation{ARC Centre of Excellence for All-Sky Astrophysics in 3 Dimensions (ASTRO 3D), Australia}

\author{Jie Wang}
\affiliation{National Astronomical Observatories, Chinese Academy of Sciences, 20A Datun Road, Chaoyang District, Beijing, People's Republic of China}

\author{Q. Daniel Wang}
\affiliation{Astronomy Department, University of Massachusetts, Amherst, MA 01003, USA}

\author{Kelley M. Hess}
\affiliation{Instituto de Astrof\'{i}sica de Andaluc\'{i}a (CSIC), Glorieta de la Astronom\'{i}a s/n, 18008 Granada, Spain}
\affiliation{ASTRON, the Netherlands Institute for Radio Astronomy, Postbus 2, 7990 AA, Dwingeloo, The Netherlands}

\author{Luis C. Ho}
\affiliation{ Kavli Institute for Astronomy and Astrophysics, Peking University, Beijing 100871, China}
\affiliation{Department of Astronomy, School of Physics, Peking University, Beijing 100871, People's Republic of China}

\author{Ligang Hou}
\affiliation{National Astronomical Observatories, Chinese Academy of Sciences, 20A Datun Road, Chaoyang District, Beijing, People's Republic of China}

\author{Yingjie Jing}
\affiliation{National Astronomical Observatories, Chinese Academy of Sciences, 20A Datun Road, Chaoyang District, Beijing, People's Republic of China}

\author{Peter Kamphuis}
\affiliation{Ruhr University Bochum, Faculty of Physics and Astronomy, Astronomical Institute, 44780 Bochum, Germany }

\author{Fujia Li}
\affiliation{CAS Key Laboratory for Research in Galaxies and Cosmology, Department of Astronomy, University of Science and Technology of China, Hefei 230026, People's Republic of China}
\affiliation{ School of Astronomy and Space Science, University of Science and Technology of China, Hefei 230026, People's Republic of China}

\author{Xuchen Lin (林旭辰)}
\affiliation{ Kavli Institute for Astronomy and Astrophysics, Peking University, Beijing 100871, China}

\author{Ziming Liu}
\affiliation{National Astronomical Observatories, Chinese Academy of Sciences, 20A Datun Road, Chaoyang District, Beijing, People's Republic of China}

\author{Li Shao}
\affiliation{National Astronomical Observatories, Chinese Academy of Sciences, 20A Datun Road, Chaoyang District, Beijing, People's Republic of China}

\author{Shun Wang (王舜)}
\affiliation{ Kavli Institute for Astronomy and Astrophysics, Peking University, Beijing 100871, China}

\author{Ming Zhu}
\affiliation{National Astronomical Observatories, Chinese Academy of Sciences, 20A Datun Road, Chaoyang District, Beijing, People's Republic of China}

\begin{abstract}
We use the single-dish radio telescope FAST to map the $\hi$ in the tidally interacting NGC 4631 group with a resolution of 3.24$'$ (7 kpc), reaching a 5-$\sigma$ column density limit of $10^{17.9} ~\cmsq$ assuming a line width of 20 $\kms$. Taking the existing interferometric $\hi$ image from the HALOGAS project of WSRT as reference, we are able to identify and characterize a significant excess of large-scale, low-density, and diffuse $\hi$ in the group. This diffuse $\hi$ extends for more than 120 kpc across, and accounts for more than one fourth of the total $\hi$ detected by FAST in and around the galaxy NGC 4631. In the region of the tidal tails, the diffuse $\hi$ has a typical column density above $10^{19.5} ~\cmsq$, and is highly turbulent with a velocity dispersion around $50~\kms$. It increases in column density with the dense $\hi$, and tends to be associated with the kinematically ``hotter'' part of the dense $\hi$. Through simple modeling, we find that the majority of the diffuse $\hi$ in the tail region is likely to induce cooling out of the hot IGM instead of evaporating or being radiatively ionized. Given these relations of gas in different phases, the diffuse $\hi$ may represent a condensing phase of the IGM.  Active tidal interactions on-going and in the past may have produced the wide-spreading $\hi$ distribution, and triggered the gas accretion to NGC 4631 through the phase of the diffuse $\hi$.


\end{abstract}

\keywords{Galaxy evolution, interstellar medium }

\section{Introduction} 
\label{sec:introduction}
The loss and gain of $\hi$ are important drivers of galactic evolution, as $\hi$ is in the phase where the star-forming gas starts to cool and settle down onto a galactic disk.  Although $\hi$ disks can be several times more extended from the galactic center than the optical disks where most star formation occurs \citep{Swaters02, Wang13}, the integral $\hi$ richness is correlated with the amount of $\hi$ on optical disks  \citep{Wang20, Yu22}, and further with the specific star formation rate (SFR) \citep{Saintonge17, Guo21}. Such a link of SFR with $\hi$ far away extends further into the circum-galactic medium, indicated by the strengths of Lyman-$\alpha$ absorbers \citep{Borthakur16, Lan18}. These correlations imply a quasi-equilibrium state of baryonic flow through galaxies, and supports the role of $\hi$ as the reservoir of raw material for forming stars. 

Tidal interactions are significant channels for galaxies to both gain and lose $\hi$ \citep{Putman17, VerdesMontenegro01}, but the net effects on the whole and in each step of physical processes remain to be studied. For example, $\hi$ tails and clouds possibly of tidal origin are often found, including the Magellanic stream and at least some of the high-velocity cloud complexes around the Milky Way \citep{Putman12}. They indicate a redistribution of gas between galaxies due to tidal interactions. These extra-planar $\hi$ features should be prone to thermal evaporation \citep{Cowie77}, radiative ionisation, and dispersal due to Kelvin-Helmholtz instability and Rayleigh-Taylor instability, but long-lasting ones have been found in massive clusters \citep{Chung07}, loose groups \citep{Koopmann08, Zhu21}, and compact groups \citep{Serra13}. It indicates a complex interplay between the $\hi$ gas and the circum(inter)-galactic environment. 
 For another example, starbursts are often found in gas-rich interacting pairs \citep{Ellison13, Chown19}, possibly caused by gas inflows driven by tidal shocks, torques, and instabilities \citep{Blumenthal18}. Despite the enhanced consumption of gas, the integral $\hi$ masses of mergers and post-mergers are not found to decrease compared to control samples \citep{Ellison18, Zuo18, Shangguan19}. This unexpected consistency in $\hi$ amount may be due to a boosted CGM cooling out of thermal instabilities, or suppressed atomic-to-molecular conversion efficiency out of turbulent $\hi$, but the exact reason is unclear.  In most of the puzzles of this type, a major difficulty arises from the physical nature that various gravitational and hydrodynamic effects are involved and interact, and that gas exchanges between phases.

Sorting out the response of $\hi$ during tidal interactions is important for a refined evolutionary theory of galaxies of different types and in different environments.  Semi-analytical models of galaxy evolution have been plagued by the fact that environmental and internal effects have a strong degeneracy when reproducing the observed $\hi$ or SFR scaling relations of satellite galaxies \citep{Stevens17}. Because different environmental mechanisms co-exist, it is hard to separate and assess the role of each \citep{Boselli06, Cortese21}, even in groups and the outskirts of clusters where tidal interactions should dominate other environmental effects \citep{Boselli22}. The most promising way forward may be a more detailed analysis of existing and newly observed data. Characterizing the distribution and kinematics of $\hi$ in prominent tidally interacting galaxy samples will help us identify signatures to separate the tidal effects from other environmental effects; comparing quantified properties with physical models, and formulating empirical relations to be implemented into semi-analytical models, will help us break the degeneracy between internal and external causes. 

\begin{figure*} 
\centering
\includegraphics[width=16cm]{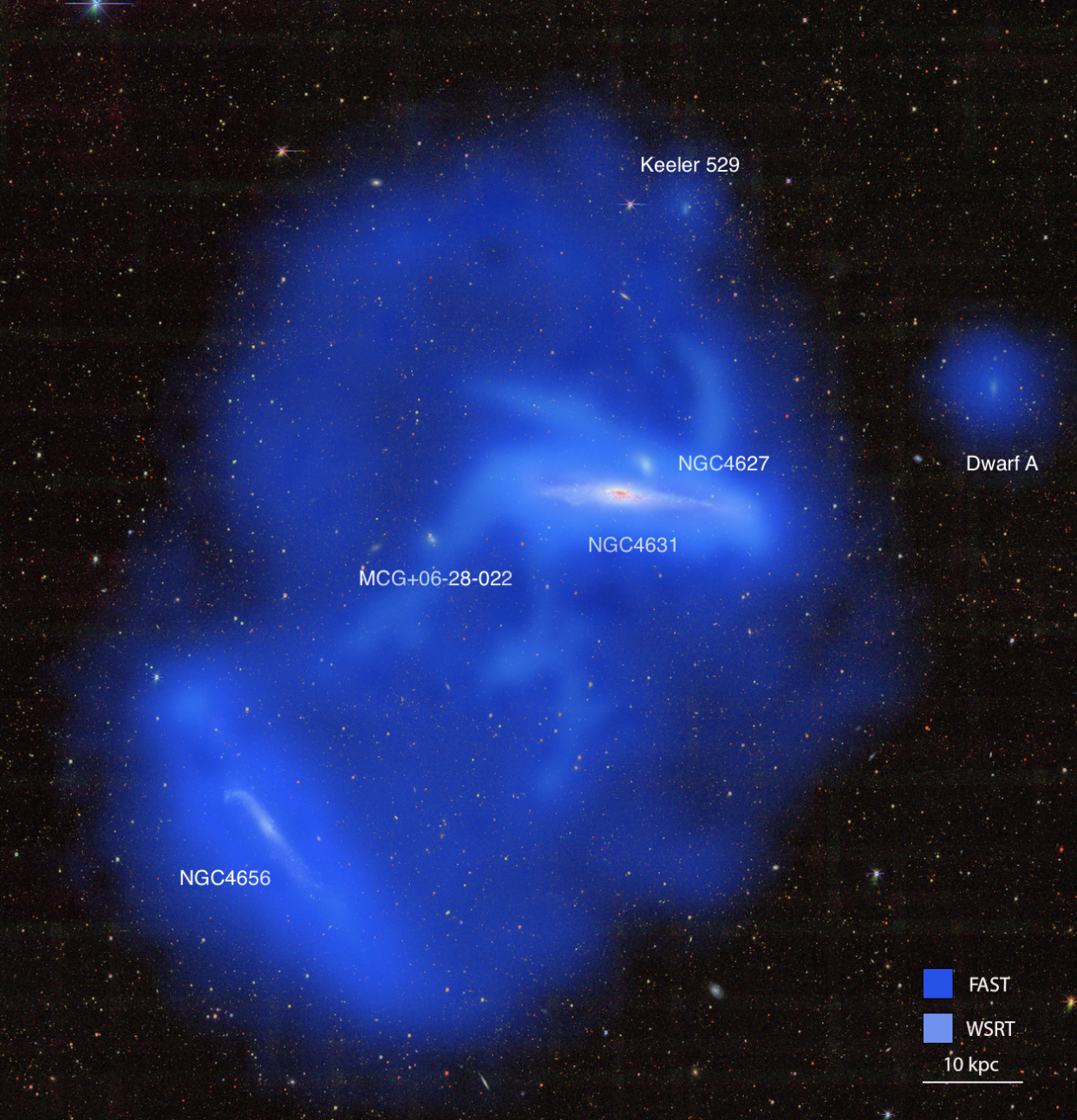}
\caption{ A false color image demonstrating the NGC 4631 group and its $\hi$ gas. On top of the optical image, the blue colored halo shows the diffuse $\hi$ flux imaged by FAST (beam FWHM=3.24$'$ or 7 kpc) in this study, while the light-blue finer structures are the denser $\hi$ previously detected in the WSRT HALOGAS \citep{Heald11} observation (beam FWHM=40$''$ or 1.46 kpc). The names of 6 relatively prominent member galaxies are denoted. }
\label{fig:atlas}
\end{figure*}

 Luckily, there have been long-lasting efforts in this direction of characterizing detailed $\hi$ properties in tidal interactions (e.g. \citealt{Rand94, Yun94, Wolfe13, Lee-Waddell19, Sorgho19, Namumba21}). A highlight among them are the systematic research on compact groups \citep{VerdesMontenegro01, VerdesMontenegro05, Borthakur10, Serra13, Hess17, Jones19}.  Built upon these benchmarking papers, in this paper we study in detail one classical interacting system, the NGC 4631 group (N4631g). We contribute the following unique inputs. We use the Five hundred meter Aperture Spherical Telescope (FAST, \citealt{Jiang19}) to obtain an $\hi$ image with a high sensitivity, and moderate resolution.  A first impression of the N4631g and its $\hi$ distribution can be obtained from Figure~\ref{fig:atlas}. 
The FAST data reveals and spatially resolves a significant excess of $\hi$ compared to a previous deep interferometric observation with the Westerbork Synthesis Radio Telescope (WSRT) by Hydrogen Accretion in LOcal GAlaxieS (HALOGAS) survey \citep{Heald11}. This paper thus addresses in particular the existence of such an extended $\hi$ envelope around NGC 4631, which the WSRT observations miss because it is too faint and too extended. The combined data show this well and allow an assessment of how much there is, how it is distributed, what its kinematics are and how it is connected to the higher density $\hi$ that the HALOGAS project found. The amount tells us about the total gas reservoir around galaxies, while the detailed properties tells us about the tidal interaction and the physics of the IGM (intra-galactic medium), CGM (circum-galactic medium), and ISM (inter-stellar medium) connection that are essential to gas accretion and depletion. 

The combination of single dish data and synthesis data, which is essential to obtain the new results in this paper, is a known but difficult problem \citep{Stanimirovic02}. This paper demonstrates the power of FAST as compared to existing attempts to add extended emission restricted to other single-dish telescopes (e.g. GBT and Parkes, \citealt{deBlok18, Das20}), which have much smaller dishes and hence have less overlap in u,v space with the synthesis data, or relatively significant side-lobes (e.g. Arecibo, \citealt{Heiles01, Hess17}). Closely relevant to this paper, \citet{Richter18} used $\hi$ image taken by the GBT in combination with the WSRT image. Limited by the resolution of the GBT image, the two types of $\hi$ data were compared mainly in a qualitative way, and the focus of that work was instead on one line-of-sight with ultraviolet spectroscopic data taken by the Hubble Space Telescope (HST). The FAST image used in this work has three times better resolution than the GBT image, has a much wider uv coverage in common with the WSRT data, and therefore enables a relatively better quantified characterization and comparison of the $\hi$ properties throughout the tidally interacting region in the group. 

This paper is organized as follows. We introduce the sample, the $\hi$ data , and the multi-wavelength data in section~\ref{sec:data}. Particularly, we describe the observation and reduction of the FAST $\hi$ data. In section~\ref{sec:comparison}, we verify that the flux calibrations are consistent between the FAST and WSRT data, and show globally the existence of excess $\hi$ detected by FAST. In section~\ref{sec:excessHI}, we conduct detailed analysis of the excess $\hi$, which is likely large-scale and low-density diffuse $\hi$. We quantify the distribution and localized kinematics of it, and its relation to the dense $\hi$ detected by WSRT.  In section~\ref{sec:environment}, we quantify the hydrodynamic and gravitational environment around the galaxy NGC 4631, and discuss the fate and motion of the (diffuse) $\hi$ in the IGM. Finally we summarize in section~\ref{sec:summary}. Throughout the paper, we assume a \citet{Chabrier03} initial mass function to estimate the stellar mass and SFR.

\section{Data  and analysis}
\label{sec:data}
\subsection{The NGC 4631 galaxy and group}
The galaxy NGC 4631, known as the Whale galaxy, is an edge-on spiral galaxy, and has remarkable $\hi$ tidal structures \citep{Weliachew78,Rand94}. It is centered at  $\alpha_{2000}=190.9905 \degree$, $\delta_{2000}=32.1682 \degree$, according to the 2MASS Extended Source Catalog \citep{Jarrett00}. 
It has a heliocentric systematic velocity of 615 $\kms$ \citep{Rand94}. We take the error weighted mean of luminosity distances derived with the TRGB method in the literature \citep{Seth05, Tully13, Radburn-Smith11, Monachesi16}, which is 7.53 Mpc.

The members of N4631g have a velocity dispersion $\sigma_c$ of 217 $\kms$ \citep{Kourkchi17}. As the brightest galaxy of the N4631g, NGC 4631 has two major companions, NGC 4656 and NGC 4627, 70.5 and 5.6 kpc away in projected distance respectively, the interaction with which should have produced most of the $\hi$ tidal structures around NGC 4631. Its interaction with NGC 4656 might start only a few hundreds of million years ago, as suggested by the age of a tidal dwarf near NGC 4656 \citep{SchechtmanRook12}, and the simulation of \citet{Combes78} in an attempt to reproduce its $\hi$ tidal tails. It also has many other fainter dwarf companions, and stellar tidal tails which do not correspond to the $\hi$ tails \citep{MartnezDelgado15}. These properties indicate a dynamic, actively interacting environment around NGC 4631. 

NGC 4631 has an active star formation possibly due to the active tidal interaction with neighbors. 
 The active star formation may have triggered powerful outflows of mass and energy. These outflows reveal themselves as super-shells and anomalous velocity features in the $\hi$ \citep{Rand94} and CO images \citep{Rand00}, filamentary structures of dust \citep{Melendez15}, ionized gas \citep{Golla96, Martin01, Strickland04, Tullmann06} extending above the disk plane, and magnetic fields perpendicular to the disk plane \citep{MoraPartiarroyo19,MoraPartiarroyo19a}. The present and past outflows may have built the prominent hot gaseous halo that is bright in the radio continuum \citep{Ekers77, Irwin12} and X-ray \citep{Wang95, Wang01}. But we point out that, the $\hi$ structure detected in this work which is large than 60 kpc in radius extends much further than the X-ray emitting hot gas halo which is roughly 10 kpc in radius.

\subsection{The FAST HI observation}
The FAST $\hi$ observations of NGC 4631 were carried out on 2022 March 25$\slash$26$\slash$27 (proposal ID: PT2021\_0071) as part of the FAST Extended Atlas of Selected Targets Survey (FEASTS)\footnote{https://github.com/FEASTS/LVgal/wiki}.  The zenith angles were $<15.7\degree$ during the observation.  A rectangle of $1.6\degree\times1.5\degree$ is targeted around $\alpha_{2000}=190.7027 \degree$, $\delta_{2000}=32.4058 \degree$, an arbitrary position (grey cross in top panel of Figure~\ref{fig:map_NHI}) between NGC 4631 and NGC 4656. The rectangle is scanned in the on-the-fly (OTF) mode with six passes, evenly divided into vertical and horizontal ones to achieve basket weaving. 
The scans are conducted with the L-band (1.05 - 1.45 GHz) 19-beam receiver rotated by 23.4$\slash$53.4\degree (horizontal$\slash$vertical), and the spacing of scanning stripes set to be 21.66$'$. We show in the left panel of Figure~\ref{fig:fast_obs} how these stripes are arranged. They cover extra regions on the four sides, in order to achieve relatively uniform sampling densities in the targeted region.

The full width half maximum (FWHM) of the raw beam is $\sim2.9'$ at a frequency of 1.42 GHz \citep{Jiang20}. The effective angular separation between scan lines is 1.15$'$, and the effective integration time per position is 235.8 s. The total integration time is 4.47 h. The observation is accompanied by a 10-K noise diode turned on for 1 s every 60 s. The data is recorded by the Spec (W+N) backend, with a sampling time of 1 s, and channel width of 7.63 kHz, or 1.61 km/s for $\hi$ 21 cm observations. 

\subsection{The FAST data reduction}
We extract a low-redshift frequency slice of 1408.7-1425.2 MHz (equivalent to 76-1609 $\kms$), and focus on this part of the data. The data reduction is carried out with a pipeline developed following the standard procedures of reducing radio single-dish image data, particularly those from Arecibo Legacy Fast ALFA Survey (ALFALFA, \citealt{Haynes18}) and HI Parkes All Sky Survey (HIPASS, \citealt{Barnes01}). It has 4 major modules, including RFI flagging, calibration, imaging, and baseline flattening. Many of these steps go backward and iterate till convergency. We briefly introduce the steps below. An early version of the pipeline is also described in Zuo et al. (2022). 

\begin{figure*} 
\centering
\includegraphics[width=9cm]{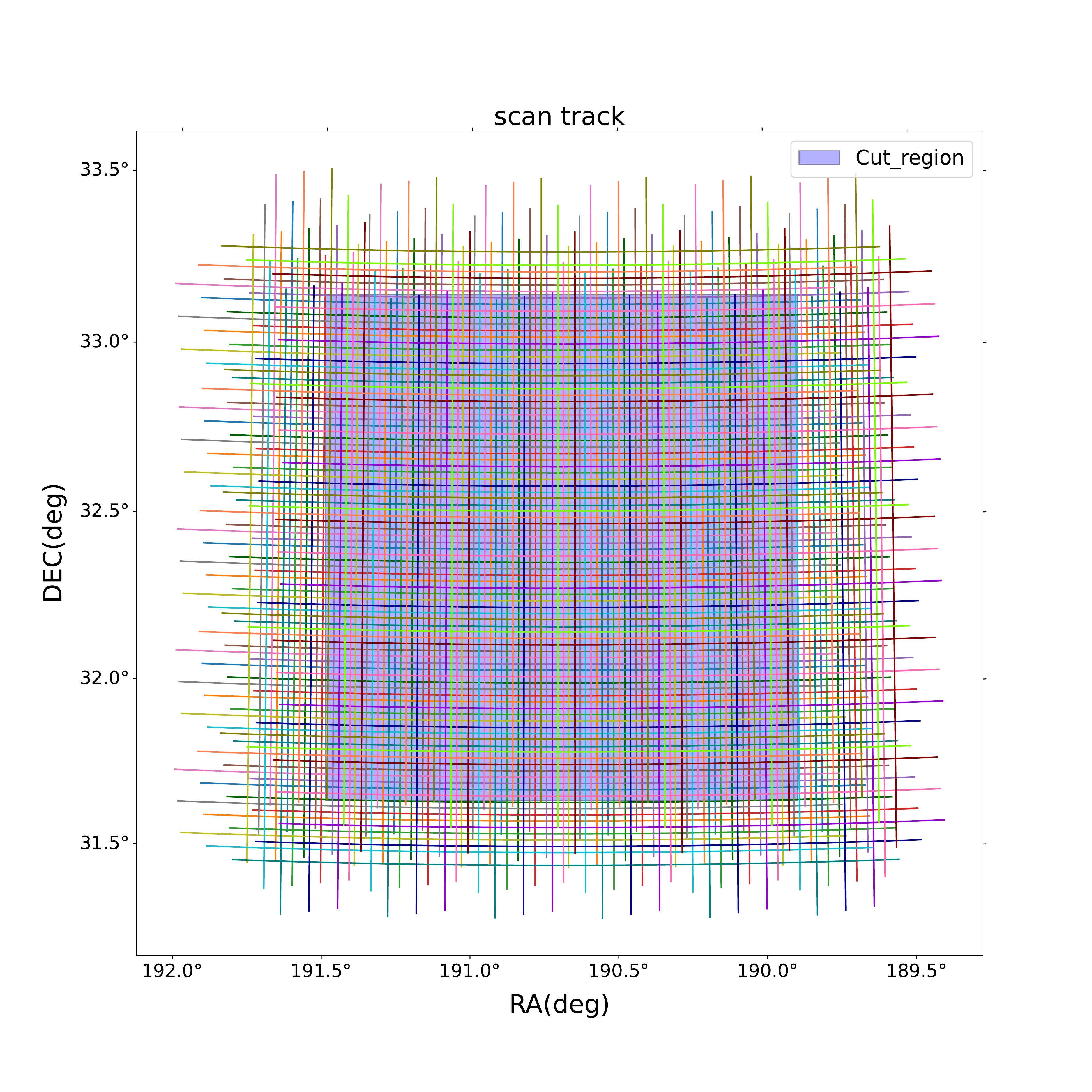}
\includegraphics[width=8.5cm]{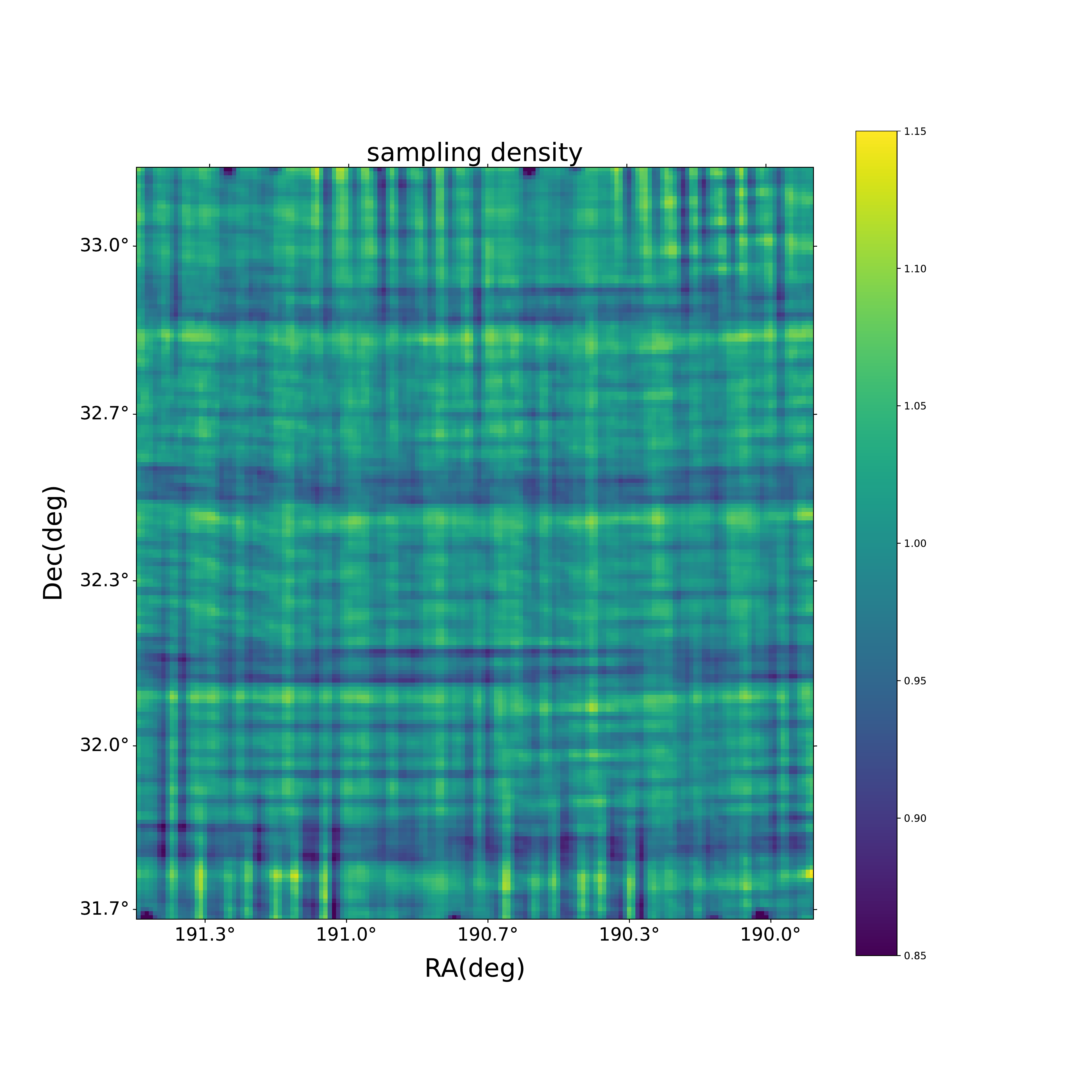}
\caption{Left: one set of vertical and horizontal scanning stripes of the observation. Lines of different colors represent different IDs of the 19 beams. The blue square represents the imaging region of the final data cube. One can see that the scanning mode is not the traditional basket weaving, but using evenly distributed horizontal and vertical scans to micmic a basket weaving. Right: the relative sampling density of the whole observed data set when gridding them into pixels. The densities are normalized to the median value.  }
\label{fig:fast_obs}
\end{figure*}

\begin{enumerate}
\item {\bf RFI flagging.}  We flag the radio frequency interferences (RFIs) in two major steps. Firstly, we use the conventional waterfall map, which is distribution of flux in the diagram of frequency versus time, to identify outstanding stripes. Secondly, the whole image region is scanned with 6 passes, so that after gridding the data by sky position for each of the 6 passes, we can use a median and 3-$\sigma$ based outlier finder to reject RFI contaminated data for the same sky position. The whole RFI flagging procedure is reviewed again after the steps of bandpass removal and flux calibration in the calibration module. The RFI contamination rate is minimal in the FAST data used in this paper. 

\item {\bf Calibration.} The bandpass is derived per beam for each stripe of the scan. 
The $\hi$ emission is masked from the waterfall map with a best effort. The mask starts with a subjective, rough region with knowledge of $\hi$ distribution from the literature \citep{Rand94}, and is adjusted later with rms level based criterion after the first round of calibration. Tests are conducted to decide an optimized smoothing width of 240 s for determining the bandpasses. The data and bandpass are calibrated against the bandpass-removed sampling of the noise diode. The mask of the $\hi$ emission is updated with the bandpass-removed and scaled data with the criterion of at least 200 connected pixels above 2-$\sigma$ threshold. The procedure goes back to the step of determining the bandpasses and is iterated for 3 times. Finally, the bandpass-removed and scaled data is corrected for a zenith angle dependent effective gain value of 13.5-16 to account for scaling differences from the perfect gain and aperture efficiency at almost zero zenith angle \citep{Jiang20}. 

\item {\bf Imaging.}  For the analysis of this paper, we produce two sets of data cubes. The first set is a conventional {\it FAST cube}, with pixel size of of 30$''$, and the channel width of 1.61 km/s.  The second set is a {\it projected FAST cube}, gridded to match the area and WCS system of the WSRT HALOGAS data (see section~\ref{sec:wsrt_data}). A Gaussian kernel with FWHM equivalent to half the FWHM of the raw beam is used to grid the data into channel maps. The FWHM of the raw beam is taken to be 2.9$'$, the median value of the 19 beams typically at the selected frequency \citep{Jiang20}. This gridding process effectively smoothes the data, increasing the FWHM of the actual beam to 3.24$'$. The right panel of Figure~\ref{fig:fast_obs} displays the the relative sampling densities of the observed data when gridding them into pixels. The density is roughly uniform with a 1-$\sigma$ scatter of 3.48\% around the median value. 

\item {\bf Baseline flattening.}  We remove the continuum in the full range 1408.7-1425.2 MHz of the selected frequency slice by modeling it with a first-order polynomial function. Before removing the continuum, the $\hi$ emissions are masked using a mask file generated by SoFiA \citep{Serra15}. We then remove the residual continuum, standing waves, and other global irregularities in the spectra, which are referred to together as the residual continuum. The residual continuum is modeled with the S-G filter with an effective polynomial order of 2, and a width of 480 $\kms$ (or 2.274 MHz), which are optimized after experiments. For reference, the major standing wave due to reflection between the dish and the receiver bin is $\sim200~\kms$ ($\sim1$ MHz) for FAST \citep{Jiang20}. This module is iterated for 3 times. 
\end{enumerate}

\subsection{The FAST data cube}
We use SoFiA \citep{Serra15} on the conventional FAST cube to generate the detection mask for $\hi$ emission. We use the threshold-based smooth$+$clipping source finding algorithm. The threshold is set to be 3-$\sigma$, and the smoothing kernels have widths of 0, 3, and 5 pixels in the sky direction, and of 0 and 3 channels in the velocity direction. The reliability module is used with a threshold of 0.99 to exclude false detections. The resulting mask is used to project the cube into moment maps and integral spectrum, and also to select emission-free regions to derive the rms level. The FAST data cube has an rms level of $0.965~ \mjybf$, where ${\rm b_F}$ denotes the beam area of FAST. It corresponds to a 5-$\sigma$ column density limit of $8.0\times10^{17} ~\cmsq$, assuming a line width of 20 $\kms$, or a 5-$\sigma$ point source mass limit of $10^{5.9}~M_{\odot}$, assuming a line width of 150 $\kms$. 

\begin{figure*} 
\centering
\includegraphics[width=9cm]{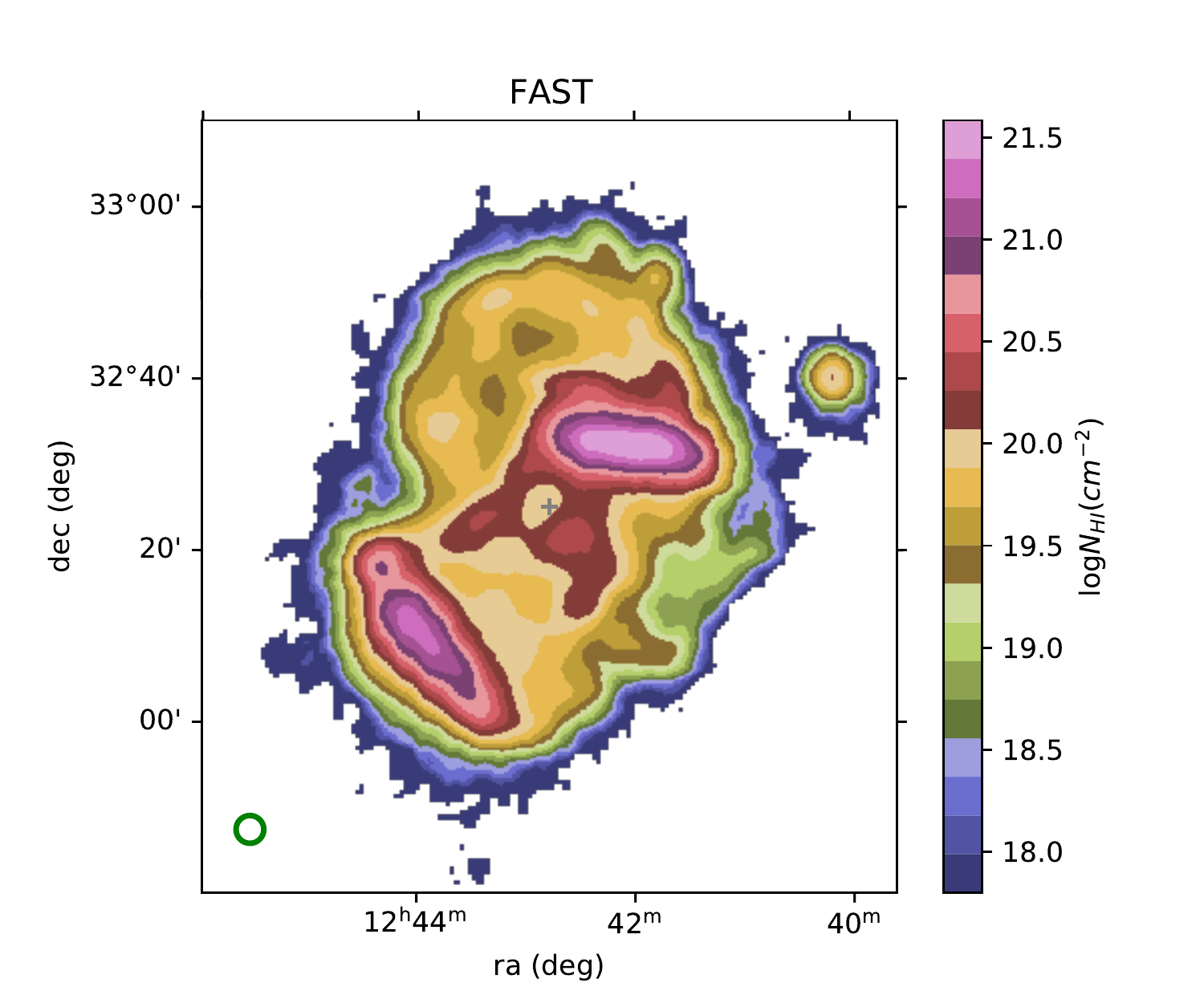}

\includegraphics[width=8.5cm]{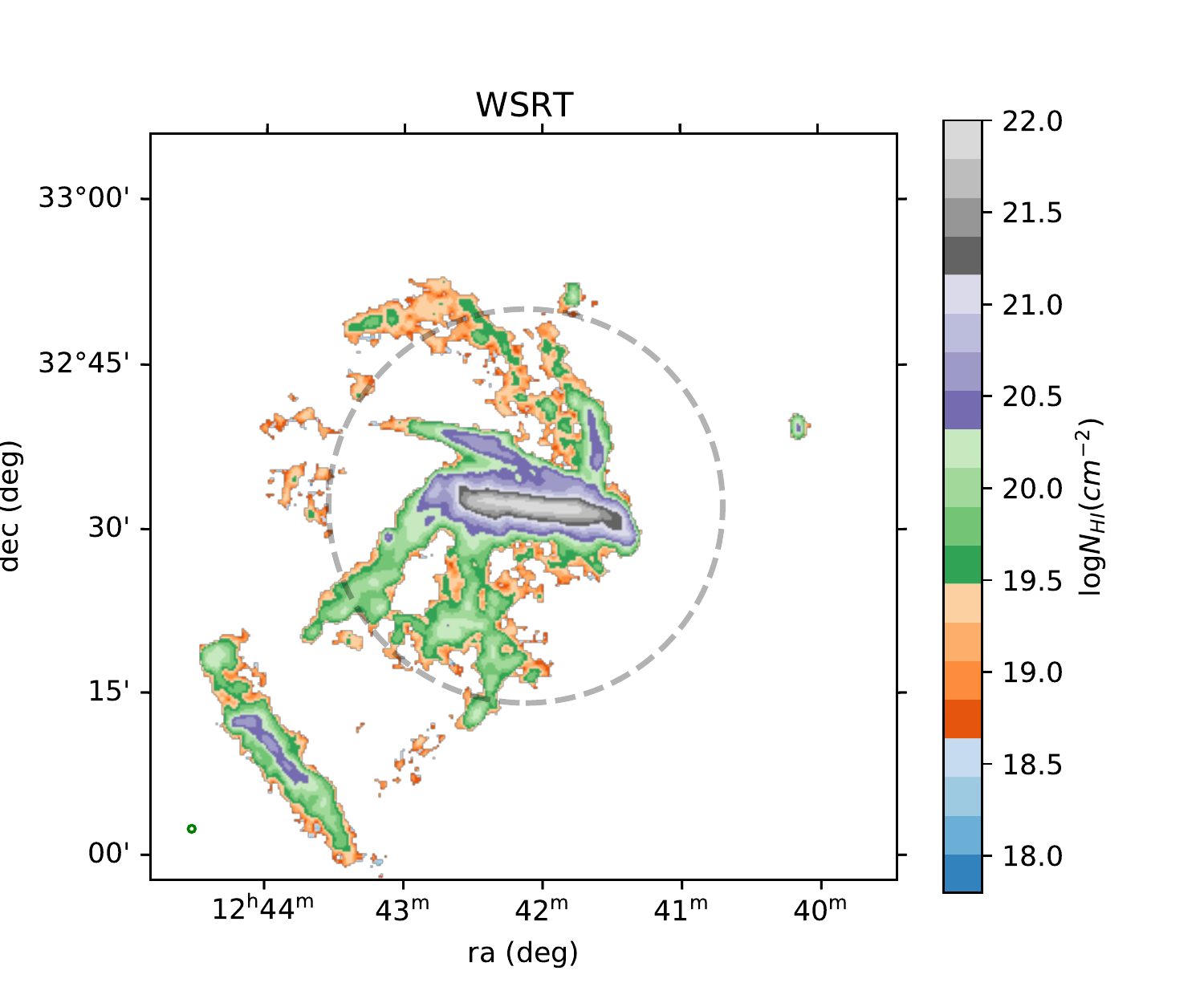}
\includegraphics[width=8.5cm]{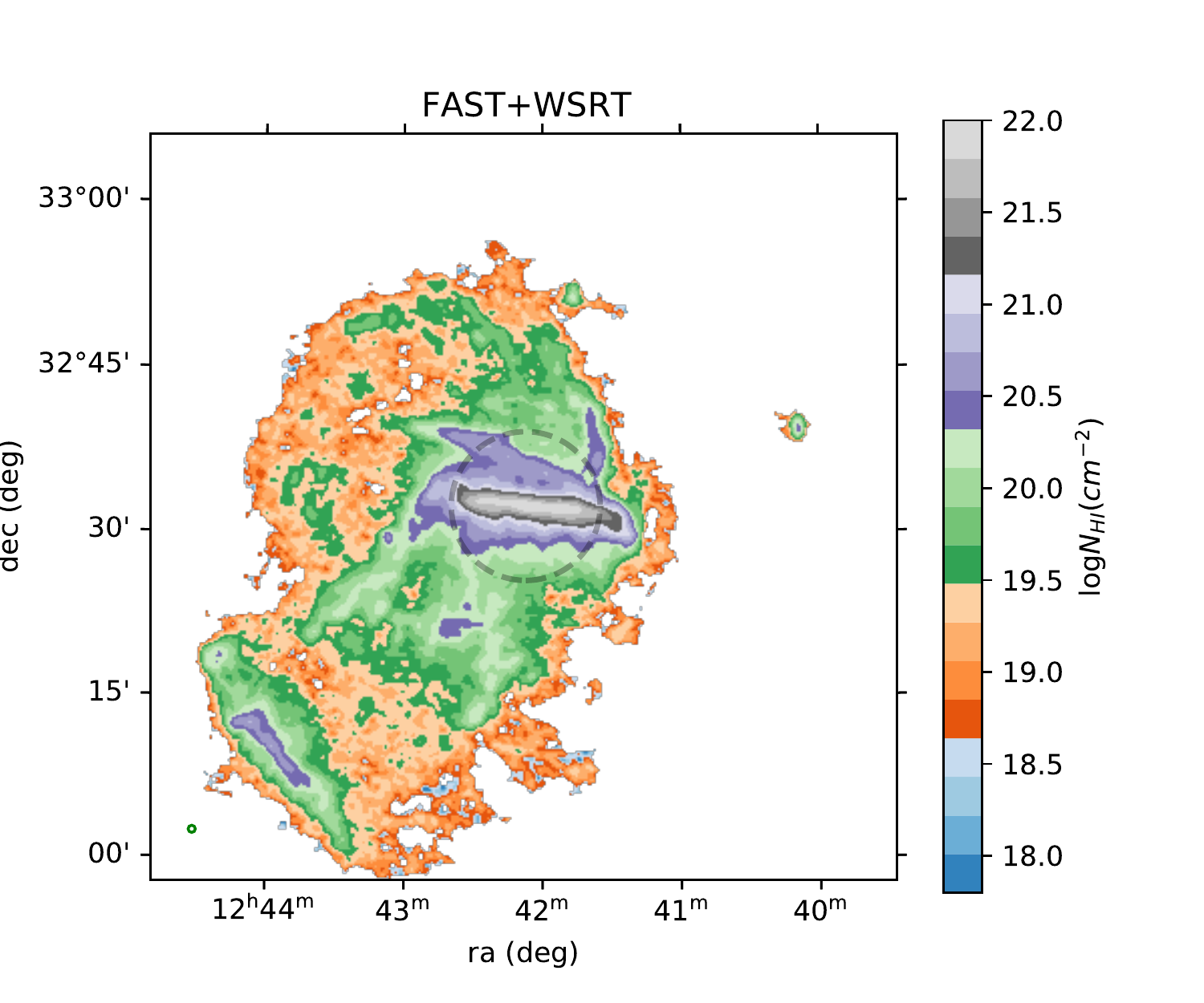}
\caption{ The $\hi$ column density maps of NGC 4631 field. The maps are derived from the FAST cube (top), WSRT cube (bottom-left), and the FAST$+$WSRT combined cube (bottom-right). In the top panel, the center of the FAST observational field is marked with a grey cross. In the bottom-left panel, the FWHM of the WSRT PB is shown as the grey dashed circle. In the bottom-right panel, the grey dashed circle has a diameter equal to the critical angular scale for WSRT to miss extended $\hi$ (see section~\ref{sec:amplitude_spectra}). The bottom-right image does not look like the sum of the other two because the combination is done in the Fourier space thus the FAST flux is conserved, and because the PB attenuation effect of the WSRT cube is applied (see section~\ref{sec:immerge}). Beam shapes are denoted as green and open ellipses at the bottom left corner of each map. 
 }
\label{fig:map_NHI}
\end{figure*}

\begin{figure} 
\centering
\includegraphics[width=9cm]{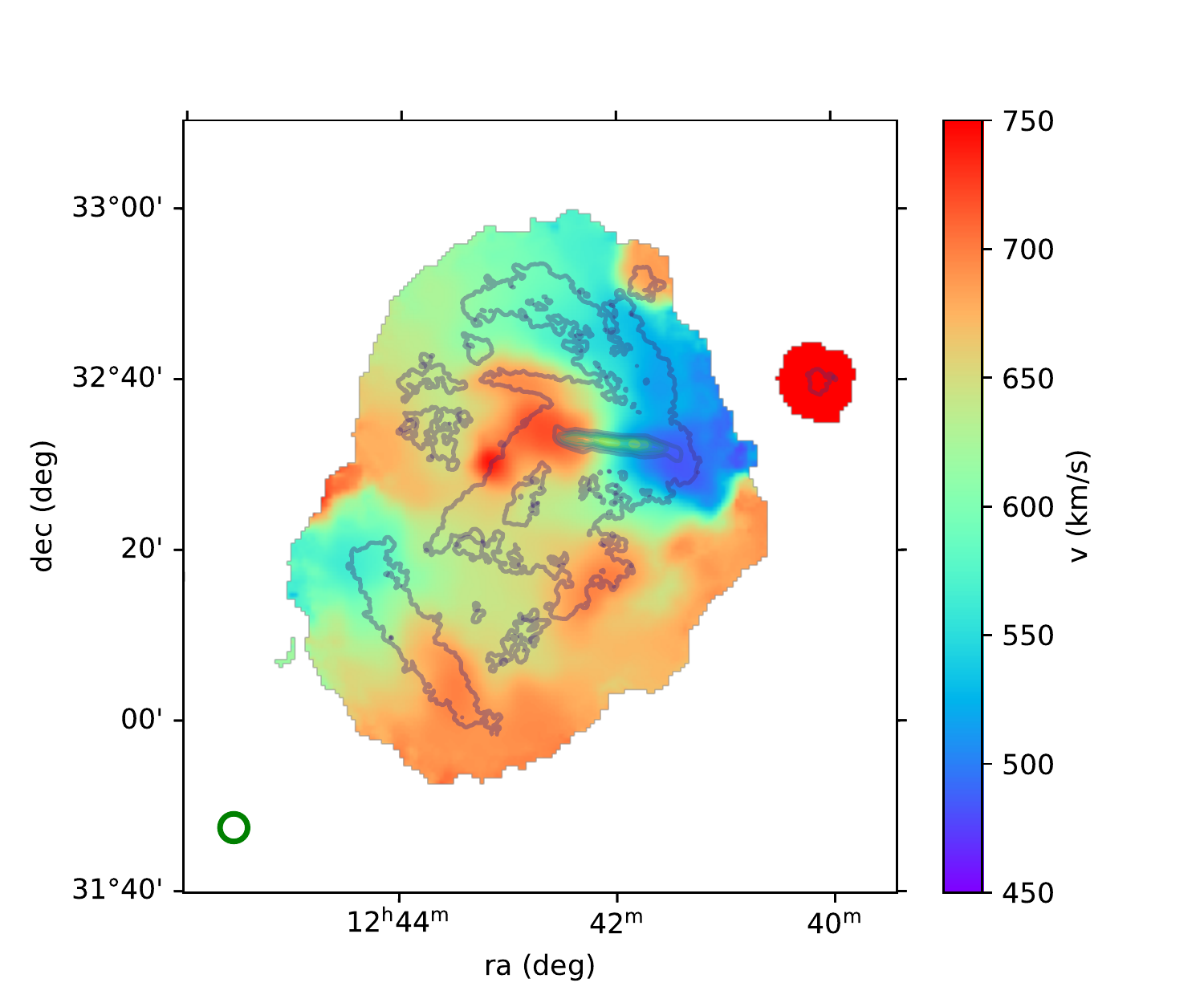}
\includegraphics[width=9cm]{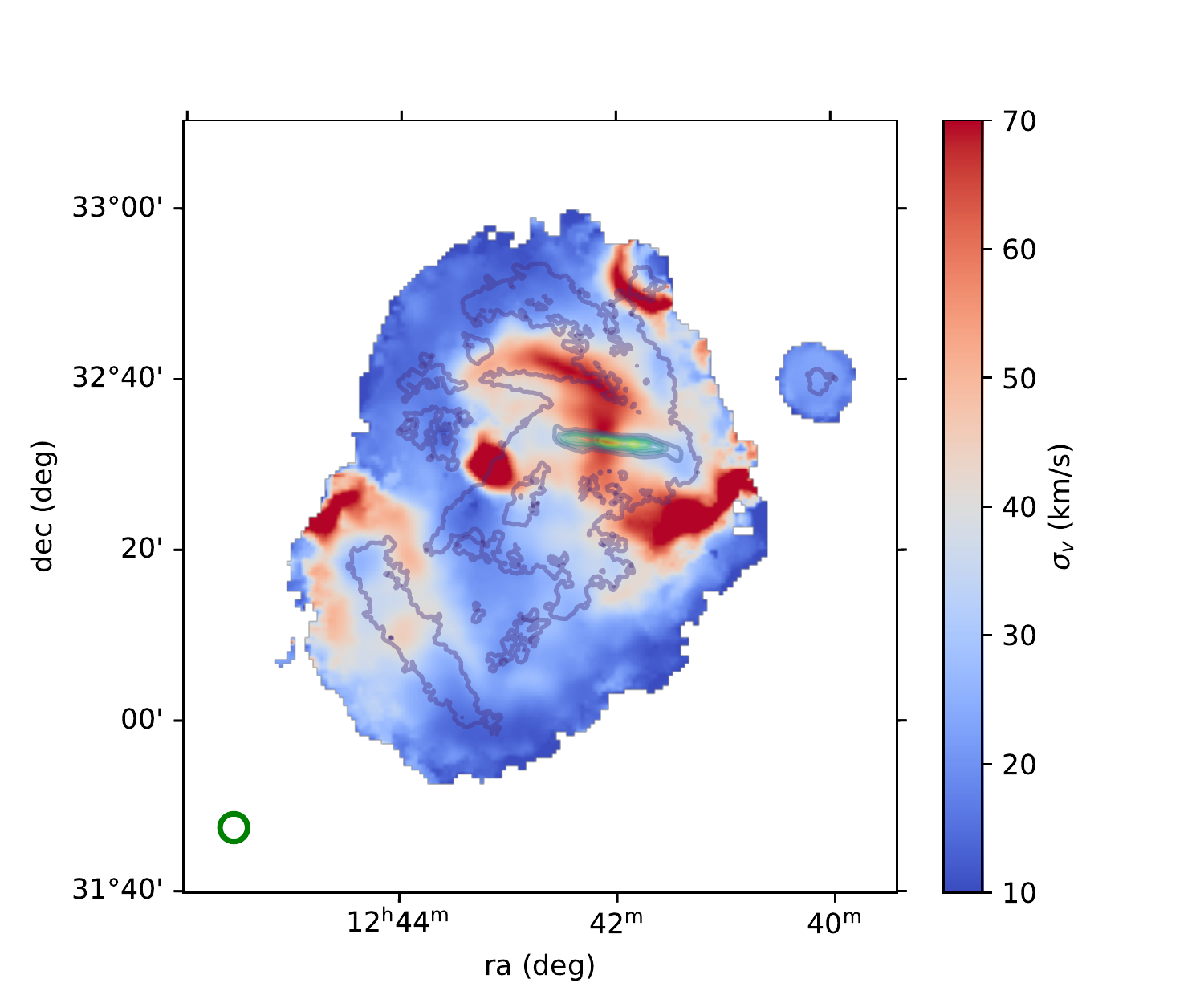}
\caption{ Moment 1 (top) and 2 (bottom) images of the NGC 4631 field. The images are derived from the FAST cube. The column density contour of the WSRT data at the level of $10^{19.5}\cmsq$ is plotted on top to guide the eye. }
\label{fig:map_mom12}
\end{figure}

We show the column density map derived from the moment-0 images in Figure~\ref{fig:map_NHI}.  Apparent from the moment images are the main target NGC 4631, its major satellite NGC 4656 to the south-east, and a known optically faint companion Dwarf A to the north-west. Another two $\hi$ bearing dwarfs previously detected in the WSRT cube of \citet{Rand94}, Keeler 529 and MCG+06-28-022, are blended into the tidal feature on the north and south-east. Due to their relatively small $\hi$ masses (each $\sim10^{7.15}~M_{\odot}$, \citealt{Rand94}), we will not distinguish them from the tidal structures in the analysis later. 
We also show the moment-1 and -2 images in Figure~\ref{fig:map_mom12}. Despite the relatively low spatial resolution, the moment-1 image shows velocity gradients in the disk regions of NGC 4631 and NGC 4656. It also shows several steep gradients in the region of tidal tails, possibly reflecting sharp turning in the direction of motions. These steep gradients are accompanied by high values in the moment-2 image, where the relative large beam of FAST tends to mix velocity structures. In the moment-2 image, the particularly high values are also caused by overlapping structures that are separated in velocity space. 

We run SoFiA similarly for the projected FAST cube, but the smoothing kernels have widths of 0, 3, 11, and 41 pixels in the sky direction instead. In unit of arcsec, the maximum extents of smoothing are actually similar for the conventional and projected FAST cubes. Expectedly, the depths and moment images from the projected FAST cube are similar to those from the conventional FAST cube.

\subsection{The WSRT data cube}
\label{sec:wsrt_data}
We use the naturally weighted data cube from the WSRT HALOGAS project \citep{Heald11}. It was observed with an integration time of 10$\times12$h. The observation has the shortest and longest baselines of WSRT around 36 m and 2.7 km, corresponding to a nominal largest and smallest angular scale of 24.5$'$ and 19.6$''$, respectively. The data cube has a synthesis beam major and minor axes of 45.0$''$ and 39.1$''$.
It has a pixel size of 4$''$, and a channel width of 4.12 $\kms$. The WSRT data cube covers an area of roughly $1\degree\times 1\degree$ around NGC 4631, so the companion NGC 4656 is near the edge of the image, and quite some of the tidal $\hi$ is near or beyond the FWHM of the WSRT primary beam (PB) which has a size of 0.3$\degree$. 

We run SoFiA on the WSRT cube to generate the detection mask. The parameter setting is similar to that for the projected FAST cube. With the SoFiA mask, we produce the moment images and integral spectra, and derive the rms level. The moment images are close to those published in \citet{Richter18}. 
The column density map derived from the moment-0 image is displayed in the bottom-left panel of Figure~\ref{fig:map_NHI}. The WSRT cube has a rms level $\sigma_W$ of $0.257~\mjyb$, where b denotes the beam area of the WSRT data. This rms level corresponds to a 5-$\sigma$ column density limit of $7.32\times10^{18}~\cmsq$ assuming a line width of 20 $\kms$ and 5-$\sigma$ point source mass limit of $10^{5.54}~M_{\odot}$ assuming a line width of 150 $\kms$. 

Through visual inspection, we find noticeable so-called ``negative bowl'' artifacts throughout the cube indicative of missing short-spacing information, particularly in the velocity range between 500 and 700 $\kms$ where tidal features are strong. They highlight the need of single-dish image to fill this missing part, but also add uncertainties and complexities when we directly compare the FAST and WSRT images to characterize the spatial distribution of the large-scale $\hi$. Luckily, the typical absolute level of those ``negative bowl'' is around 1-$\sigma$ of the WSRT data cube, and as we will show in section~\ref{sec:diffuseHI_distr} and Figure~\ref{fig:excessHI_cubedistr}, the associated cumulative absolute flux is low compared to the excess $\hi$ detected by FAST. These facts mitigate the problem, but future investigation of optimized strategy of combining the single-dish and interferometric data in the uv space may better solve this problem.

\subsection{Derived cubes}
\label{sec:ancillary}
For convenience of comparison, we produce a few derived cubes to control for the effects of the PSF (i.e. the FAST beam and the WSRT synthesis beam) and the WSRT PB attenuation.

We use the equation from \citet{Wang15} to produce a data cube of PB attenuation levels (the {\it PB cube} hereafter). The equation is a function of the distance from the image center, and was calibrated using continuum sources from NRAO VLA Sky Survey (NVSS, \citealt{Condon98}) and Faint Images of the Radio Sky at Twenty centimeters (FIRST, \citealt{Becker95}). We produce the {\it PB-corrected WSRT cube} by dividing the original WSRT cube by the PB cube.  

We produce the {\it smoothed WSRT cube} by convolving the channel maps of the WSRT cube with the FAST beam. The beam image of the FAST is derived by stacking point source images of the 19 beams with data from \citet{Jiang20}. More details and discussion regarding the beam image can be found in appendix~\ref{sec:appendix_beam}. The flux of the smoothed WSRT cube is converted to the unit of $\jybf$. 

We produce the {\it PB-attenuated FAST cube} by multiplying the projected FAST cube with the PB cube. We subtract the smoothed WSRT cube from the PB-attenuated FAST cube, and obtain the {\it PB-attenuated excess $\hi$ cube}\footnote{Strictly speaking, we should compare $(FAST~cube) * (WSRT~beam)$ with $(WSRT~cube) * (FAST~beam)$, where $*$ is the sign of operation for convolution. We thus also tried smoothing the projected FAST cube with the WSRT beam, before applying the PB attenuation. We find the two products do not differ much due to the relatively small size of the WSRT beam.}. We apply PB correction to the PB-attenuated excess $\hi$ cube, and obtain the {\it PB-free excess $\hi$ cube}. The PB-free excess $\hi$ cube is largely positive, and the very few negative regions (most apparent ones are the two small white patches near the N4631 disk in the top panel of Figure~\ref{fig:excessHI_maps}) are likely due to pointing uncertainties, deviation of real FAST beam from the adopted averaged one, and noise. 

The PB-attenuated excess $\hi$ cube has the advantage of a relatively uniform rms level, convenient for threshold based analysis, while the PB-free one has the advantage of reflecting the actual amount of excess $\hi$. We will show in section~\ref{sec:excessHI} that, the PB-free excess $\hi$ cube is practically the {\it diffuse HI cube}.

\subsection{Definition of regions} 
\begin{figure} 
\centering
\includegraphics[width=9cm]{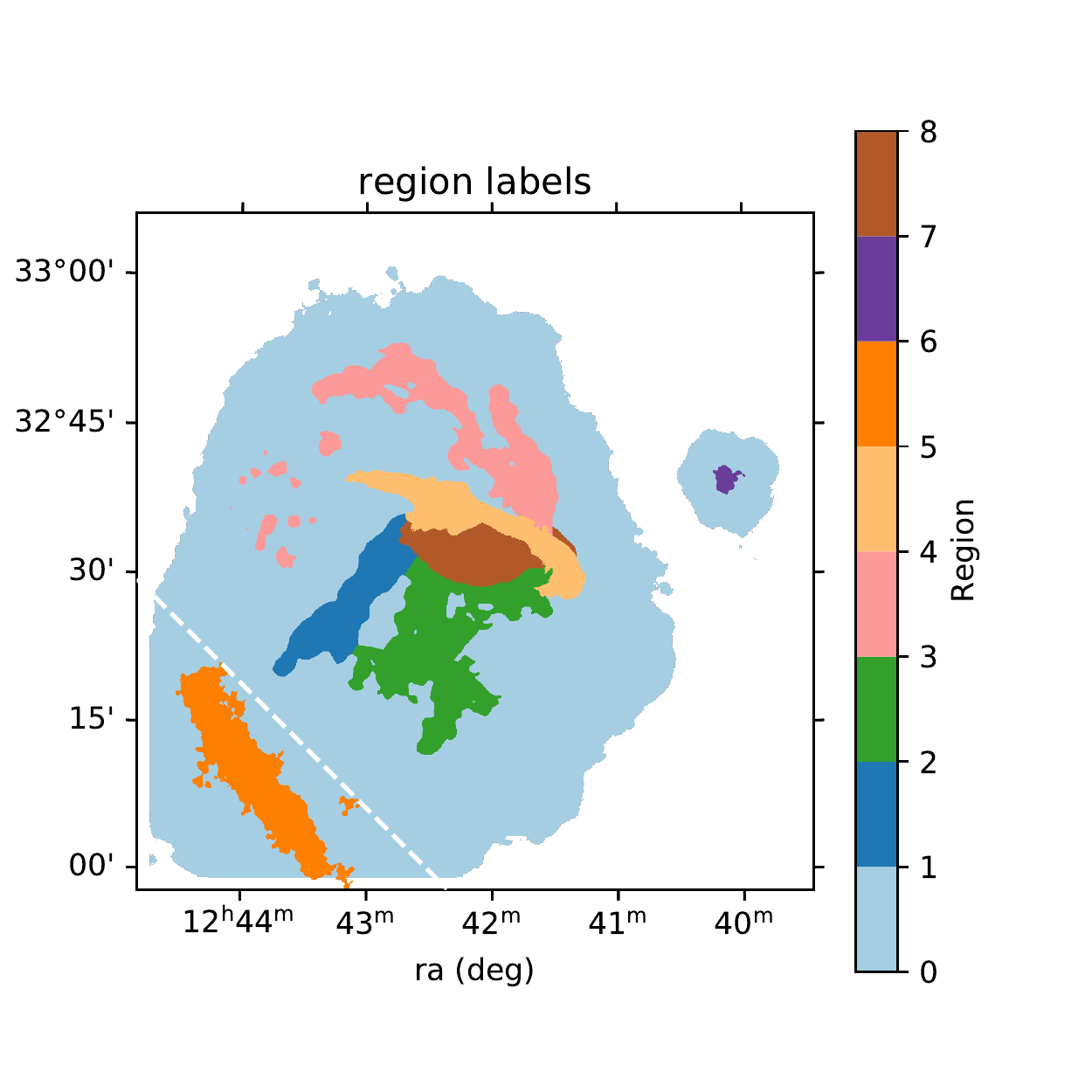}
\caption{Label of regions in the WSRT cube. The id from 1 to 7 (color from dark blue to brown) corresponds to tail 1, tail 2, tail 3, tail 4, NGC 4656,  Dwarf A, and NGC 4631 disk region respectively. The light blue color (id 0) marks the region detected in $\hi$ by the projected FAST cube. The white dashed line separates the NGC 4631 and NGC 4656 regions. }
\label{fig:map_label_region}
\end{figure}

We define  {\it the NGC 4631 region}. We take the SoFiA mask of the projected FAST cube, exclude the region of Dwarf A, and separate the region of NGC 4631 and NGC 4656 by arbitrarily drawing a line roughly along the disk direction of NGC 4656. The line and the resultant NGC 4631 region to the north-west are shown in Figure~\ref{fig:map_label_region}. This region is delineated in order to study the distribution of any excess $\hi$ detected by FAST (section~\ref{sec:diffuseHI_distr}). We exclude NGC 4656 because it is at the corner of the WSRT field of view, where the PB attenuation factor reaches 0.1 and where the rms level will thus be increased by 10 times after PB correction. 

We separate the WSRT-detected NGC 4631 region into {\it the disk region} and {\it the tail region}. The disk and tail regions are defined to compare the localized kinematics and distribution of $\hi$ fluxes detected by FAST and WSRT (section~\ref{sec:diffuseHI_vsig} and \ref{sec:diffuseHI_denseHI}). The tail region is further divided into the regions of four tails to study their bulk motions (section~\ref{sec:discuss_PSD}). These regions are defined based on the SoFiA mask of the WSRT cube and the tilted ring model of the NGC 4631 disk from \citet{Rand94}, and through the watershed algorithm.  The technique details are presented in appendix~\ref{sec:appendix_label_region}. The sky projected view of these regions are displayed in Figure~\ref{fig:map_label_region}.

We exclude the disk region from the NGC 4631 region, and define the {\it IGM region}. This region is mainly for highlighting where the excess $\hi$ dominates the $\hi$ detected by FAST (section~\ref{sec:diffuseHI_distr}), and discussing the hydrodynamical effects in the IGM (section~\ref{sec:discuss_hydrodynamic_environment}).   

\subsection{Multi-wavelength measurements from the literature}
We derive the stellar mass for NGC 4631 with Spitzer IRAC1 and 2 (3.6 and 4.5 $\mu$m) fluxes from the Local Volume Legacy (LVL) project \citep{Dale09}. We use the equation from \citet{Querejeta15}, to derive the IRAC1$-$IRAC2 color dependent IRAC1 mass-to-light ratio. The equation was calibrated using fluxes decomposed into stellar and non-stellar components through independent component analysis. The estimated stellar mass $\log M_*/M_{\odot}=10.25\pm0.1$. We use the star formation rates (SFR) derived in \citet{Lee11} based on the far-ultraviolet and the total-infrared luminosities. The total-infrared luminosity accounts for the dust attenuation of the far-ultraviolet luminosity, and was derived through spectral energy distribution fitting of mid- and far-infrared bands taken by Spitzer as part of the LVL project \citep{Dale09}. The SFR$=4.19\pm0.25~M_{\odot} {\rm yr}^{-1}$. We also obtain these two parameters for NGC 4656 from the same datasets, with $\log M_*/M_{\odot}=9.15\pm0.1$, and SFR$=1.17\pm0.07~M_{\odot} {\rm yr}^{-1}$.

We summarize from the literature and estimate more properties about the N4631g in the appendix.  Particularly, in appendix~\ref{sec:appendix_mass}, we show that, based on the the local grouping of satellites, the characteristic radius $r_{200}$ within which the averaged density is 200 times the cosmic critical density is around 249 kpc. Accordingly the virial temperature of the IGM should be around $8\times10^5$ K, though the near-disk outflowing hot gas reaches a temperature of nearly 2$\times10^6$ K \citep{Wang95}. A single-$\beta$ model of the density profile of the IGM hot gas is presented and discussed in appendix~\ref{sec:appendix_IGMprof}. 

\section{Comparing the FAST and WSRT data }
\label{sec:comparison}

In this section, we provide integral spectra, fluxes and masses of $\hi$ for galaxies in N4631g, and analyze $\hi$ distribution on different angular scales (inverse of spatial frequency) in the FAST and WSRT data.
The difference of the integral measurements for galaxies between the two datasets provides a first-order measure of the excess $\hi$ detected by FAST. 
Comparing integral fluxes of compact sources, and comparing amplitudes in an angular-scale range corresponding to overlapping region in the uv space
help verify the consistency of flux calibrations between the two datasets, which is the basis for characterizing any excess $\hi$ detected by FAST. 
Through comparing the amplitudes of the two datasets on large angular scales, we can further derive the critical angular scale for WSRT to miss extended $\hi$.

\subsection{The integral spectra and integrated fluxes} 
\label{sec:result_spectrum}
\begin{figure} 
\centering
\includegraphics[width=8cm]{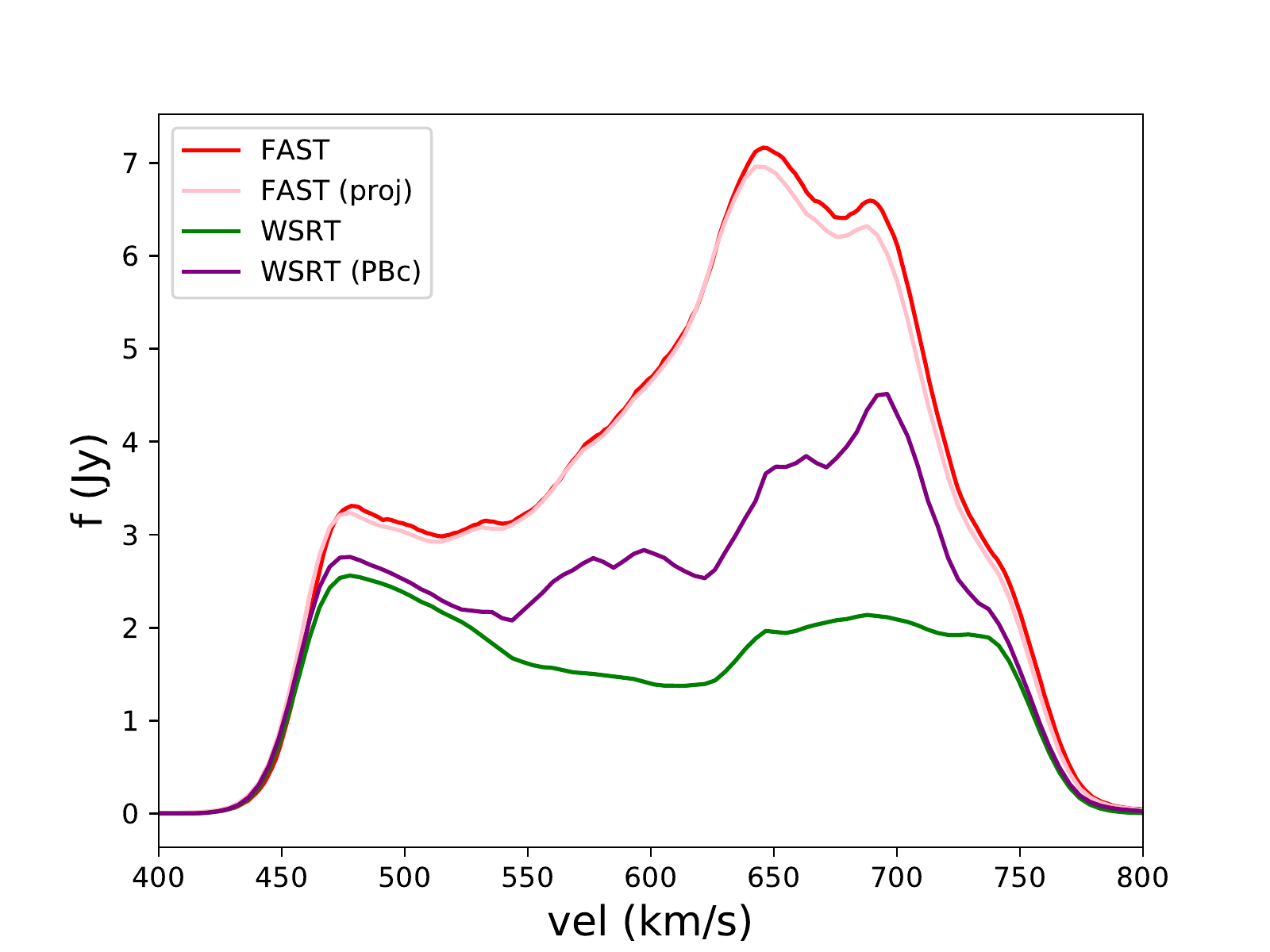}
\caption{Integral $\hi$ spectra of N4631g from different cubes.  The spectra from the FAST cube, the projected FAST cube, the WSRT cube, and the PB-corrected WSRT cube are plotted in red, pink, green and purple, respectively.}
\label{fig:spectrum}
\end{figure}

In Figure~\ref{fig:spectrum}, we show the integral $\hi$ spectra of the N4631g from the FAST data, and compare it with those from the WSRT data. 

The conventional FAST spectrum is slightly higher than the projected FAST spectrum at the high velocity end, consistent with the truncation of the galaxy NGC 4656 at the edge of the field of view of the WSRT observation. The projected FAST $\hi$ flux is in excess of the PB-corrected WSRT one throughout the velocity range. The integral fluxes from the FAST data,  the projected FAST data, the WSRT cube, and the PB-corrected WSRT data are 1345.9$\pm$134.6, 1314.6$\pm$131.5, 593.8$\pm$59.4, and 852.1$\pm$85.2 $\jykms$ respectively. The error bars are dominated by an assumed flux calibration uncertainty of 10\%. 

From the FAST cube, the $\hi$ masses of NGC 4631, its major satellite NGC 4656, and dwarf companion Dwarf A are $10^{10.08\pm0.04}$, $10^{9.77\pm0.04}$, and $10^{7.77\pm0.04}$ $M_{\odot}$  (all assuming the distance of NGC 4656) respectively. In comparison, the corresponding values from the PB-corrected WSRT cube are $10^{9.90\pm0.04}$, $10^{9.38\pm0.04}$, and $10^{7.82\pm0.04}$ $M_{\odot}$. To show the very little influence of resolution in this comparison, we also derive the corresponding values from the PB-corrected smoothed WSRT cube, which are are $10^{9.90\pm0.04}$, $10^{9.39\pm0.04}$, and $10^{7.76\pm0.04}$ $M_{\odot}$. 

There is clear excess $\hi$ detected by FAST for NGC 4631 and NGC 4656. The excess $\hi$ may be caused by the existence of diffuse $\hi$, which has large angular size or low surface densities \footnote{We note that, when the definition of low-surface density is based on the rms level of the WSRT cube, it is influenced by the effect of PB attenuation. We will discuss more on this point in section~\ref{sec:diffuseHI_distr} }. There is no excess $\hi$ detected by FAST for Dwarf A, which is relatively small in angular size.

\subsection{Comparing the amplitude spectra} 
\label{sec:amplitude_spectra}
One concern that arises when comparing the FAST and WSRT data is whether the flux calibrations are consistent. The consistent integral fluxes of Dwarf A support it, but we further justify it by comparing the amplitude spectra between the PB-attenuated FAST cube and the smoothed WSRT cube. 

\begin{figure*} 
\centering
\includegraphics[width=16cm]{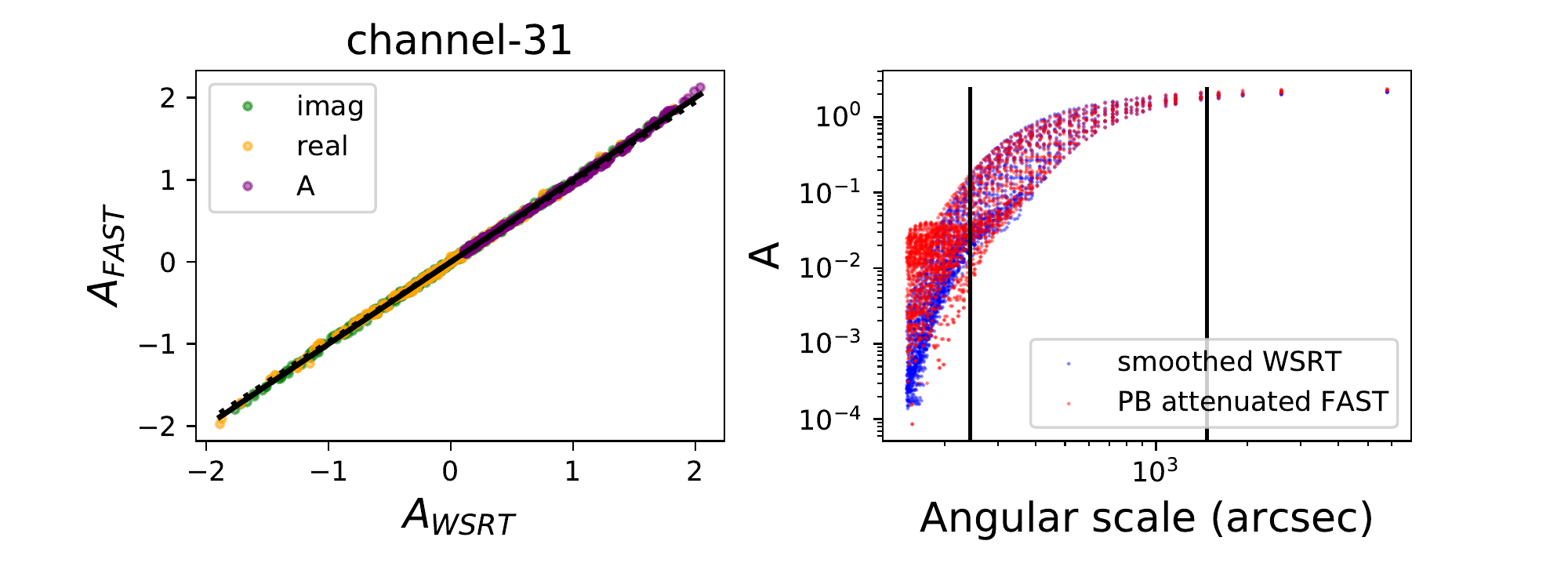}

\caption{ An example of amplitude spectral analysis of channel maps. The two channel maps analyzed have the same channel number 31 (corresponding to a velocity of 477.7 $\kms$), but are from the PB-attenuated FAST cube and the smoothed WSRT cube.  {\bf Left:} a one-to-one comparison in amplitudes (purple), the real parts (yellow), and the imaginary parts (green) between the two types of data, selected to have angular scales between the limiting values marked in the right panel. The dashed line is the $y=x$ line. The solid line shows the best-fit linear relation between the two types of amplitudes. {\bf Right:} the amplitude spectra of the amplitude (A) as a function of the angular scale, for the FAST (red dots) and WSRT data (blue dots). The two black, thick, vertical lines mark the limiting angular scales of 4$'$ and 24.5$'$, between which both FAST and WSRT data should have relatively good sensitivity. Here this channel 31 is arbitrarily chosen, and the emission of that channel map is relatively compact in morphology. In Figure~\ref{fig:power_compare} of appendix~\ref{sec:appendix_powerspectra}, we show a similar comparison between the FAST and WSRT datasets with data from all channels which have significant flux (FAST amplitudes$>$0.15 Jy) putting together. 
 }
\label{fig:power_spectrum_ch31}
\end{figure*}

The analysis exploits modified scripts from the package {\it uvcombine} \footnote{https://github.com/radio-astro-tools/uvcombine/}. Each channel image is Fourier transformed, and becomes a complex image of amplitudes and phases where the position of a pixel reflects the spatial frequency (inverse of the angular scale). The relation between the amplitude $A$ and the angular scale is called the amplitude spectrum.  
The right panel of Figure~\ref{fig:power_spectrum_ch31} shows the amplitude spectra for both datasets at a selected channel. The two spectra converge at intermediate angular scales, largely between a lower and upper limit angular scales of 4$'$ and 24.5$'$. The lower limit is just slightly (1.25 times) higher than the FWHM of the FAST beam,  while the upper one corresponds to the shortest baseline (36 m) of the WSRT array. 
We select the data points of the two datasets between the limiting angular scales, and compare their spectral amplitudes as well as the related real and imaginary parts in the left panel of Figure~\ref{fig:power_spectrum_ch31}. The data points all lie close to the one-to-one line. We select the channels (in total 50) where the maximum FAST amplitudes are higher than 0.15 Jy, and derive the average linear scaling factor of FAST amplitudes over the WSRT amplitudes for each of these channels (more details in appendix~\ref{sec:appendix_powerspectra} and left panel of Figure~\ref{fig:power_compare}). The average scaling factors have a median value and standard deviation of 0.98 and 0.02 respectively. They strongly support the consistency of fluxes from FAST and WSRT observations on the selected overlapping angular scales. We do not correct for this 1.02 scaling difference, but if we do so the amount of excess $\hi$ derived in this work should be systematically enlarged by 2\%.

In the left panel of Figure~\ref{fig:power_spectrum_ch31}, we see a hint of the FAST amplitudes exceeding the WSRT amplitudes on the high amplitude end. It indicates the start of the regime where the WSRT tends to miss large-scale diffuse flux. In order to investigate whether this hint is really, we select channels (50 in total) where the PB-attenuated FAST intensity is higher than 0.15 Jy and higher than the WSRT intensity by more than 10\%. For each channel, we derive the critical angular scale above which the FAST A are higher than the WSRT A by more than 1\%. The critical angular scales have a relative narrow range, and a mean value of 13.6$'\pm1.9'$  (more details in appendix~\ref{sec:appendix_powerspectra} and right panel of Figure~\ref{fig:power_compare}), corresponding to a baseline of 65 m and a physical scale of 29.7 kpc.  This critical angular scale is roughly half the theoretical value derived from the shortest baseline of the WSRT array, possibly due to a combined effect of the PB attenuation, and the limited WSRT sampling density of the shortest baseline which can be exacerbated by RFI flagging. 

\section{The diffuse HI detected by FAST} 
\label{sec:excessHI}
This section characterizes the diffuse $\hi$ detected by FAST, including how much there is, how it is distributed, what its kinematics are, and how it is connected to the higher density $\hi$ that the HALOGAS project detected.

\subsection{Combining the HI data} 
\label{sec:immerge}
Because the FAST beam has a relatively low side-lobe level of around 1\% beyond a radius of $\sim3.5'$ (appendix~\ref{sec:appendix_beam}), we use the MIRIAD procedure {\it immerge} to combine the projected FAST and WSRT cubes, which uses the Gaussian functions to approximate beams. The procedure immerge combines the two types of data in the Fourier domain, with a unit weight for the projected FAST data and a tapering for the WSRT data. The effect of the tappering is to make a Gaussian beam equivalent to the WSRT synthesis beam, after adding the tapered WSRT synthesis beam to the FAST beam in the Fourier domain. 
The output is an image combining the spatial information of both data, but with the same PB attenuation effect as the WSRT data. The procedure also derives a calibration factor of WSRT flux over FAST flux to be 0.98, consistent with the result from our amplitude spectral analysis. 

The combined data give us a visual impression where and how significant FAST detects the diffuse $\hi$ that is lacking in the WSRT observations. The combined moment-0 image is displayed in the bottom-right panel of Figure~\ref{fig:map_NHI}. FAST detects an excess of $\hi$ widely surrounding the denser tidal structures previously detected by WSRT, typically on a scale larger than the critical angular scale for WSRT to miss extended fluxes (see section~\ref{sec:amplitude_spectra}). The {\it immerge} process adds not only a lot of new, diffuse $\hi$ near the WSRT detection limit of $10^{18.86}~\cmsq$, but also thickens the structures at a relatively higher column density of $10^{20}~\cmsq$ (i.e. FAST also detects more relatively high-density gas). 

Because the side-lobe level of the FAST beam cannot be ignored (appendix~\ref{sec:appendix_beam}), the combined data cube and image are mainly for visual inspection here and later in section~\ref{sec:discuss_PSD}. In the following, we analyze the FAST-detected excess, diffuse $\hi$ combining the two datasets, but not directly based on the {\it immerge} combined data. 

\subsection{Relating the diffuse HI to large-angular scale gas  }
\label{sec:diffuseHI_distr}
We classify and quantify the distribution of excess $\hi$ detected in the FAST data with respect to the WSRT data. Unless otherwise specified, we focus on the NGC 4631 region in this section, as NGC 4656 is heavily attenuated by the PB effects.  In the NGC 4631 region, 26.3\% of the flux from the PB-attenuated FAST cube are missed by the WSRT cube. The missed part may be related to the existence of low-surface density or large-angular scale $\hi$, which we refer to together as the diffuse $\hi$. 

We use the rms level of the WSRT data, to separate the PB-attenuated excess $\hi$ into the low-surface density and the large-angular scale types. We remind that, the noise level of the PB-attenuated FAST cube decreases as a function of radius from the image center while that of the WSRT cube remains roughly constant, so the relative level of low-surface density $\hi$ that is missed by WSRT for being below the rms-based threshold should increase toward large radius. This effect biases our analysis toward attributing excess $\hi$ to the low-surface density type at large radius, and undermines the detection of large-angular scale type. 
 
\begin{figure*} 
\centering
\includegraphics[width=16cm]{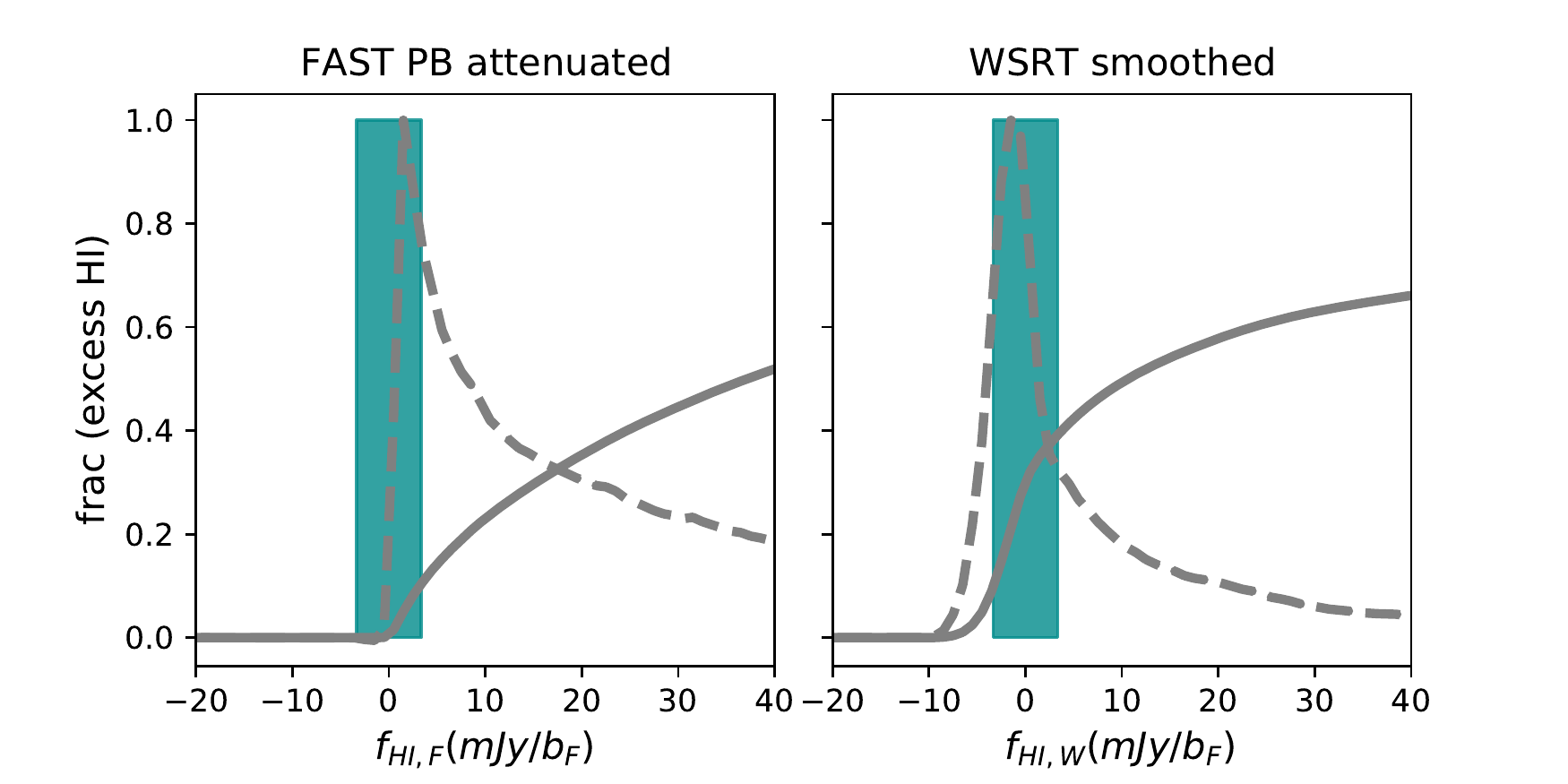}
\vspace{0.5cm}
\caption{ The distribution of attenuated excess $\hi$ as a function of voxel values in data cubes. The left panel is for the attenuated FAST cube, and the right the smoothed WSRT cube. In each panel, the solid grey curve is the cumulative distribution starting from the low-value side, and the dashed grey curve is for the peak-value normalized differential distribution. The cyan band is the  $\pm$3-$\sigma$ range of the smoothed WSRT cube.  }
\label{fig:excessHI_cubedistr}
\end{figure*}

In Figure~\ref{fig:excessHI_cubedistr}, we study the cumulative distribution of the PB-attenuated excess $\hi$ as a function of the associated flux in the PB-attenuated FAST cube. The 3-$\sigma$ detection threshold line of the smoothed WSRT cube is marked in the figure. The distribution to the left of the positive threshold line (i.e. the right-side edge of the cyan band) reflects the part of PB-attenuated excess $\hi$ missed by WSRT due to its low-surface density. There is only 10.3\% of PB-attenuated excess $\hi$ in this part. 
The remaining part (89.7\%) of PB-attenuated excess $\hi$ is likely missed by WSRT due to its large-angular scale distribution. Because the periphery of large-angular scale $\hi$ distribution naturally has low densities, and because of the PB attenuation effects described above, the actual fraction of large-angular scale $\hi$ missed by WSRT should be higher than this value of 89.7\%. 
Thus, the majority of the diffuse $\hi$ are invisible to WSRT not because of the limited sensitivity, but because of the limited shortest baseline.

\begin{figure*} 
\centering
\includegraphics[width=8cm]{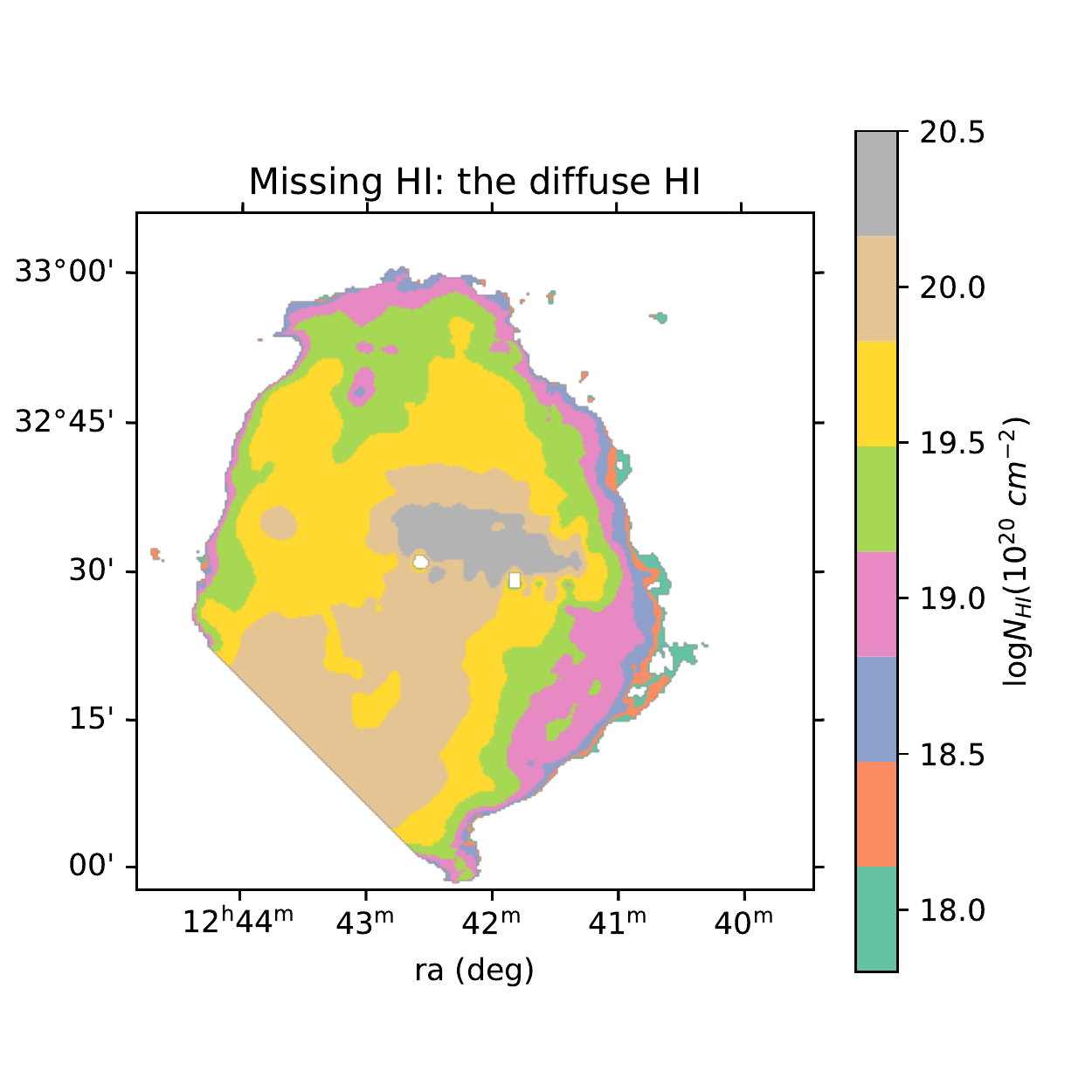}

\includegraphics[width=8cm]{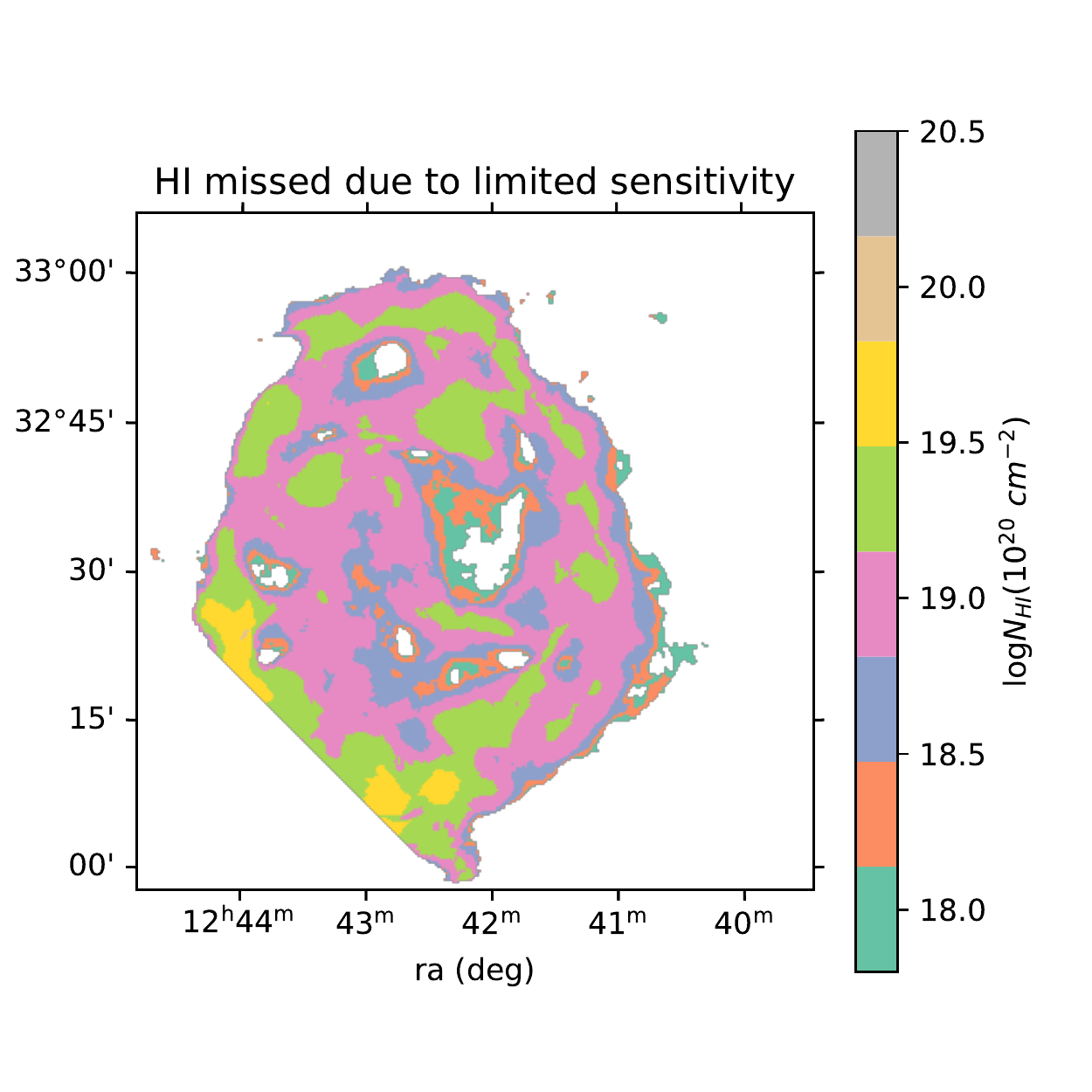}
\includegraphics[width=8cm]{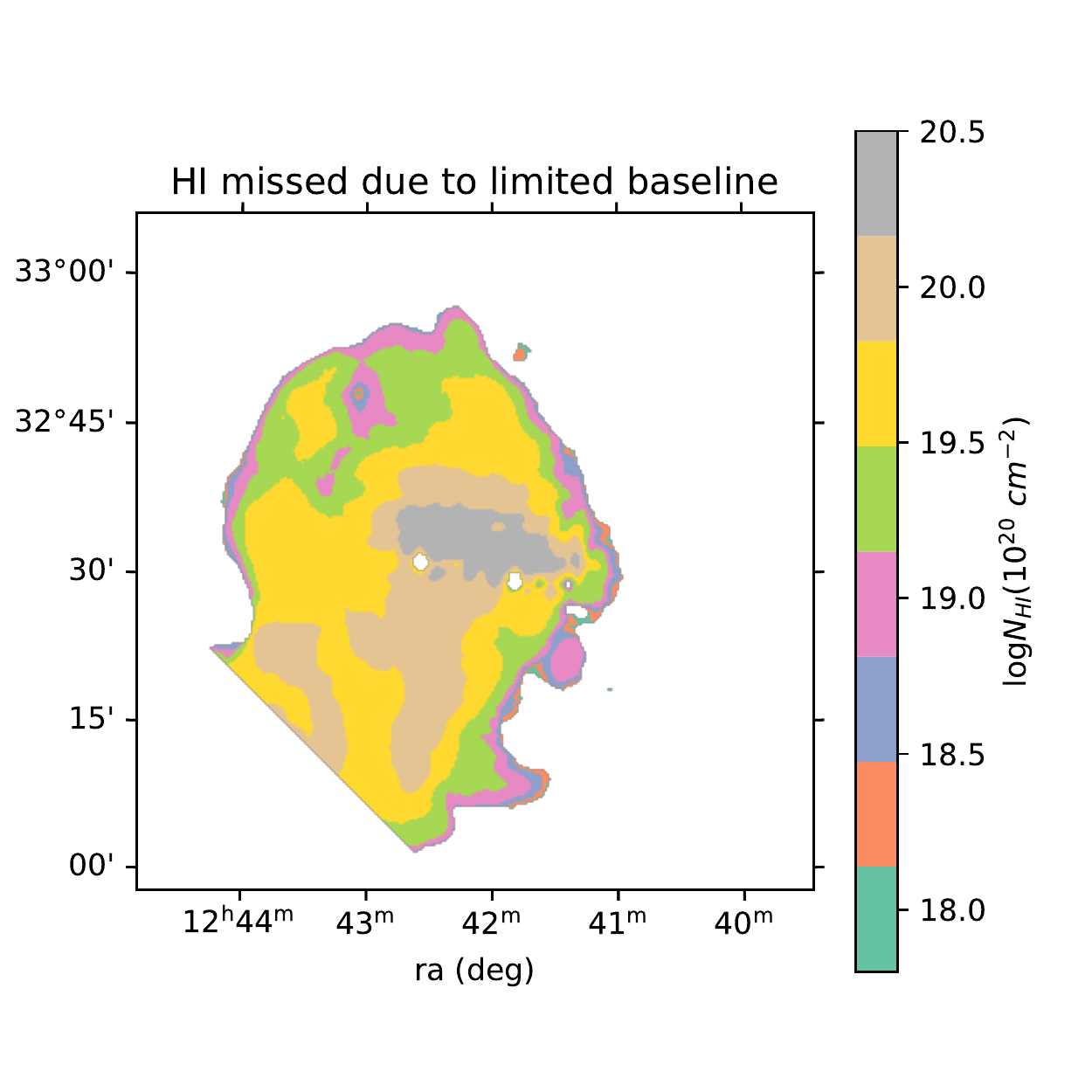}
\caption{ Column density maps of the excess $\hi$ in the NGC 4631 region. The top panel shows the PB-free excess $\hi$ missed by WSRT. The bottom-left and bottom-right panels divide the PB-free excess $\hi$ into two parts: the one missed by WSRT due to the limited sensitivity (the low-surface density $\hi$) and the one missed missed by WSRT due to the limited shortest baseline (the large-angular scale $\hi$), respectively. Some of the low-surface density $\hi$ shown in the bottom-left panel may actually belong to the large-angular scale $\hi$ in the bottom-right panel; please refer to the main text for more details. }
\label{fig:excessHI_maps}
\end{figure*}

We display the column density maps of the diffuse $\hi$ (PB-free excess $\hi$), its low-surface density part (the part below the WSRT detection threshold before applying the correction for PB attenuation), and the large-angular scale part (the diffuse $\hi$ minus the low-surface density part) for the NGC 4631 region in Figure~\ref{fig:excessHI_maps}. As discussed before, the low-surface density and large-angular scale parts displayed here are upper and lower limits of the actual parts. The displayed large-angular scale $\hi$ is almost always higher in level than the displayed low-surface density $\hi$, except for the periphery of the whole region and a small region on the south-west. It confirms that, a considerable fraction of the low-surface density $\hi$ is attached to the large-angular scale $\hi$ in the outskirts. The majority of the excess $\hi$ is by nature the large-angular scale $\hi$. 

It is still questionable whether the diffuse $\hi$ primarily overlaps with or is beyond the region of dense $\hi$ detected in the WSRT cube. 
In Figure~\ref{fig:excessHI_cubedistr}, the right panel is similar to the left panel, but the $x$-axis is replaced by the associated flux of the smoothed WSRT cube. The distribution to the left of the positive threshold line now reflects the part of PB-attenuated excess $\hi$ residing in regions where the WSRT data detects no $\hi$. 
Only 40.3\% of the PB-attenuated excess $\hi$ are found in blank regions of the WSRT data. More than half of the PB-attenuated excess $\hi$ overlaps in regions with where the WSRT detects the dense $\hi$ (i.e. disk region$+$tail region). Another noticeable feature in the right panel of Figure~\ref{fig:excessHI_cubedistr} is that, the WSRT flux distribution is peaked at a value below zero ($\sim-1.5~\mjybf$), likely related to the ``negative bowl'' artifacts discussed in section~\ref{sec:wsrt_data}. Multiplying this absolute peak value with the number of voxels which have smoothed WSRT flux below 3-$\sigma$ but non-zero excess $\hi$ provides a rough estimate of the related uncertainty for the fraction 40.3\% derived above, which is 12.5\%. 

If we further limit the analysis to the IGM region by excluding the disk region of NGC 4631, the fraction of PB-attenuated FAST flux missed by the WSRT data dramatically increases to 71.2\%, the fraction of PB-attenuated excess $\hi$ classified into the low-surface density type slightly increases to 14.5\%, and the fraction found beyond the dense $\hi$ region (equivalent to the tail region) slightly increases to 56.3\%. These fractions also indicate that, in the tail region, the amounts of dense $\hi$ and diffuse $\hi$ are roughly equal. 

\subsection{Relating the diffuse HI to properties of the dense HI}
\label{sec:diffuseHI_denseHI}
In the following, we take advantage of the high resolution of the WSRT data, quantify the localized kinematic properties of dense $\hi$, and search for the preferred kinematic condition traced by the dense $\hi$ to form the diffuse $\hi$. The analysis of this section is limited to the tail region. 

\begin{figure} 
\centering
\includegraphics[width=9.5cm]{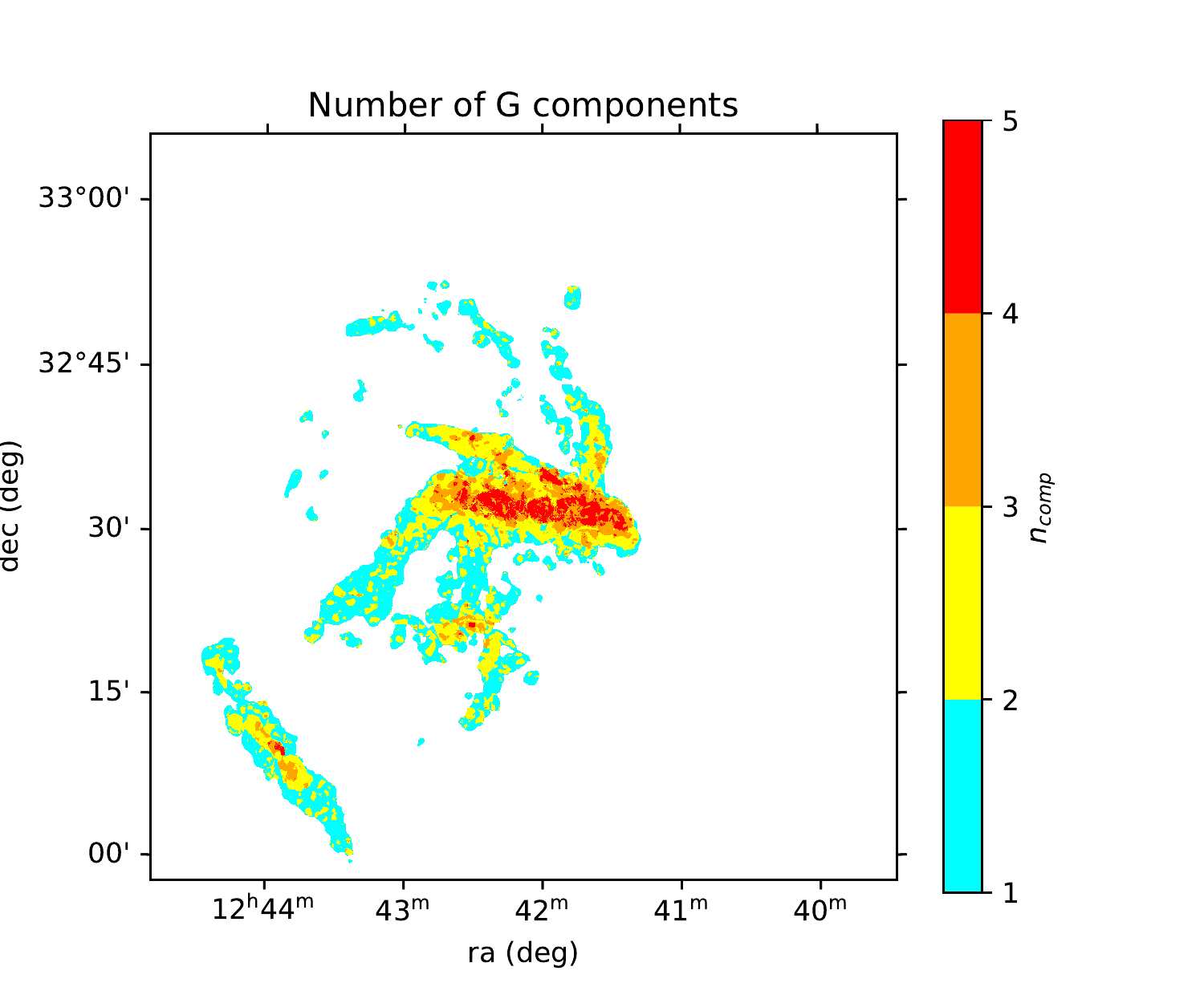}
\caption{ Map of number of Gaussian components from {\it BAYGAUD} fit for the WSRT cube. The numbers 1 to 4 is denoted by colors from cyan to red. }
\label{fig:map_nvcomp}
\end{figure}

We use {\scriptsize BAYGAUD} \citep{Oh22} to fit multi-Gaussian models to the line-of-sight spectra of the WSRT cube. {\it BAYGAUD} uses Bayesian analysis techniques to decide the optimal number of Gaussian components. Figure~\ref{fig:map_nvcomp} shows the map of the number of components.  The maximum number reaches 4, but those line-of-sights with 4 Gaussian components are mostly within the galactic disks. The NGC 4631 disk region has many Gaussian components possibly because of the edge-on geometry, the tidal perturbation, and the energy input from massive young stars. 
These complexities support our decision to leave aside the disk region and focus on the tail region.  

\begin{figure} 
\centering
\includegraphics[width=9cm]{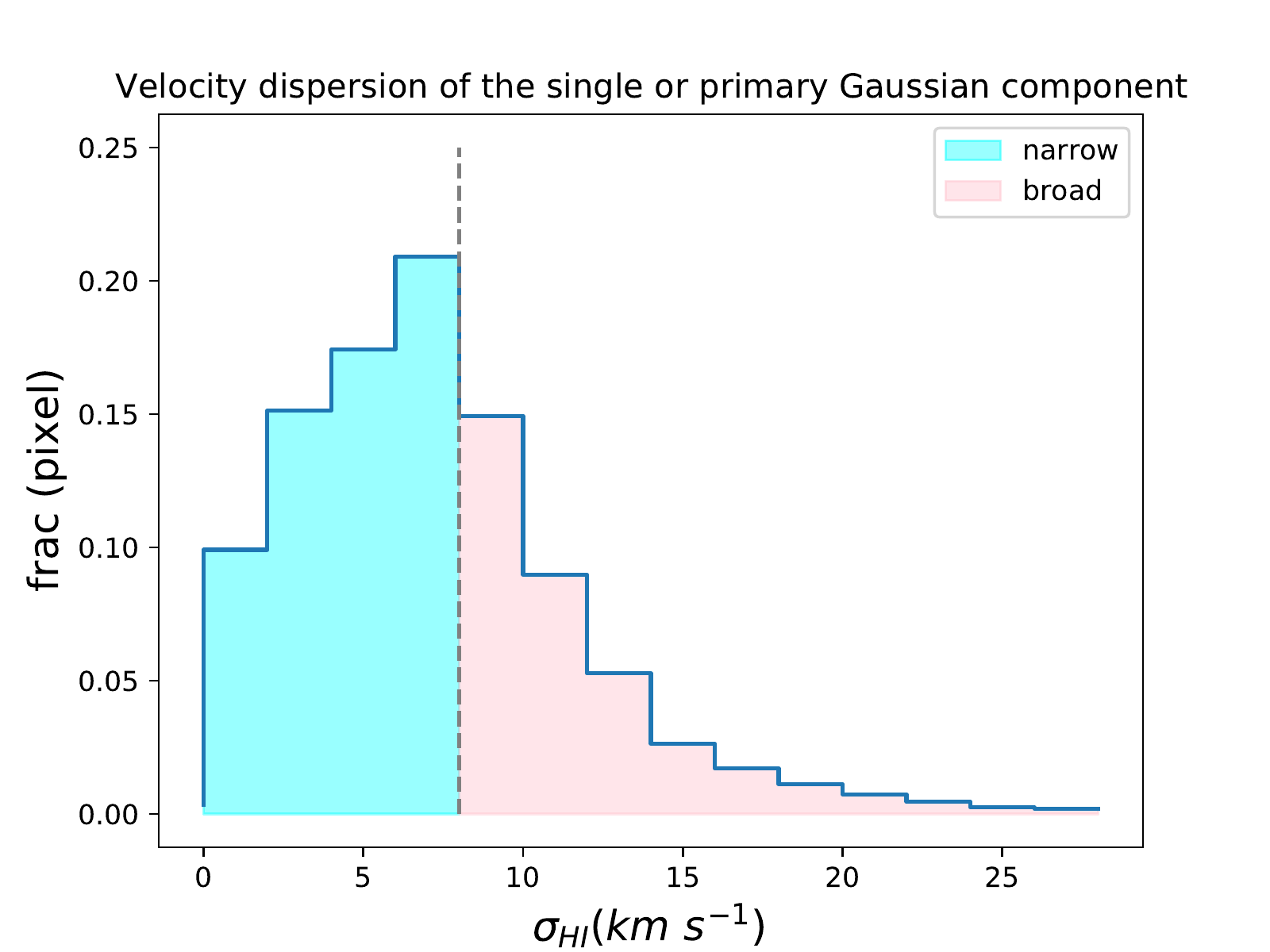}

\caption{ The distribution of the velocity dispersion of the single or primary Gaussian components in the WSRT cube. The vertical and dashed line at 8 $\kms$ divides the Gaussian components into the narrow (warm) and broad (hot) types.   }
\label{fig:vsig_distr}
\end{figure}

The profiles which are best fit with only one Gaussian component are referred to as the single-Gaussian profiles, otherwise the multi-Gaussian profiles.
For each multi-Gaussian profile, we identify the the Gaussian component with the highest intensity as the primary component. 
Figure~\ref{fig:vsig_distr} shows the distribution of $\sigma$ of all the single or primary Gaussian components of dense $\hi$ in the tail region. We divide the Gaussian components into narrow (warm) and broad (hot) ones by a $\sigma$ of 8 $\kms$, thermally corresponding to a temperature of 3600 K though the $\sigma$ here are not really thermal.

We use not only the number of Gaussian components but also profile broadness to indicate the kinematic hotness of the dense $\hi$. The expectations are: 1) for single-Gaussian profiles, velocity dispersion $\sigma$ is indicative of kinematic hotness, and high column densities of the broad (narrow) components tend to be associated with a high level of kinematic hotness (coolness); 
2) in general, single-Gaussian profiles tend to be kinematically cooler than multi-Gaussian profiles if they are not significantly affected by projection effect; 3) for a multi-Gaussian profile, the narrower Gaussian component with the smaller value of $\sigma$ is relatively cooler than the broader ones; 4) for multi-Gaussian profiles,  profiles with narrow components tend to be kinematically cooler than those without, and those with a high fraction of flux in narrow components tend to be cooler than otherwise.

\begin{figure*} 
\centering
\includegraphics[width=7cm]{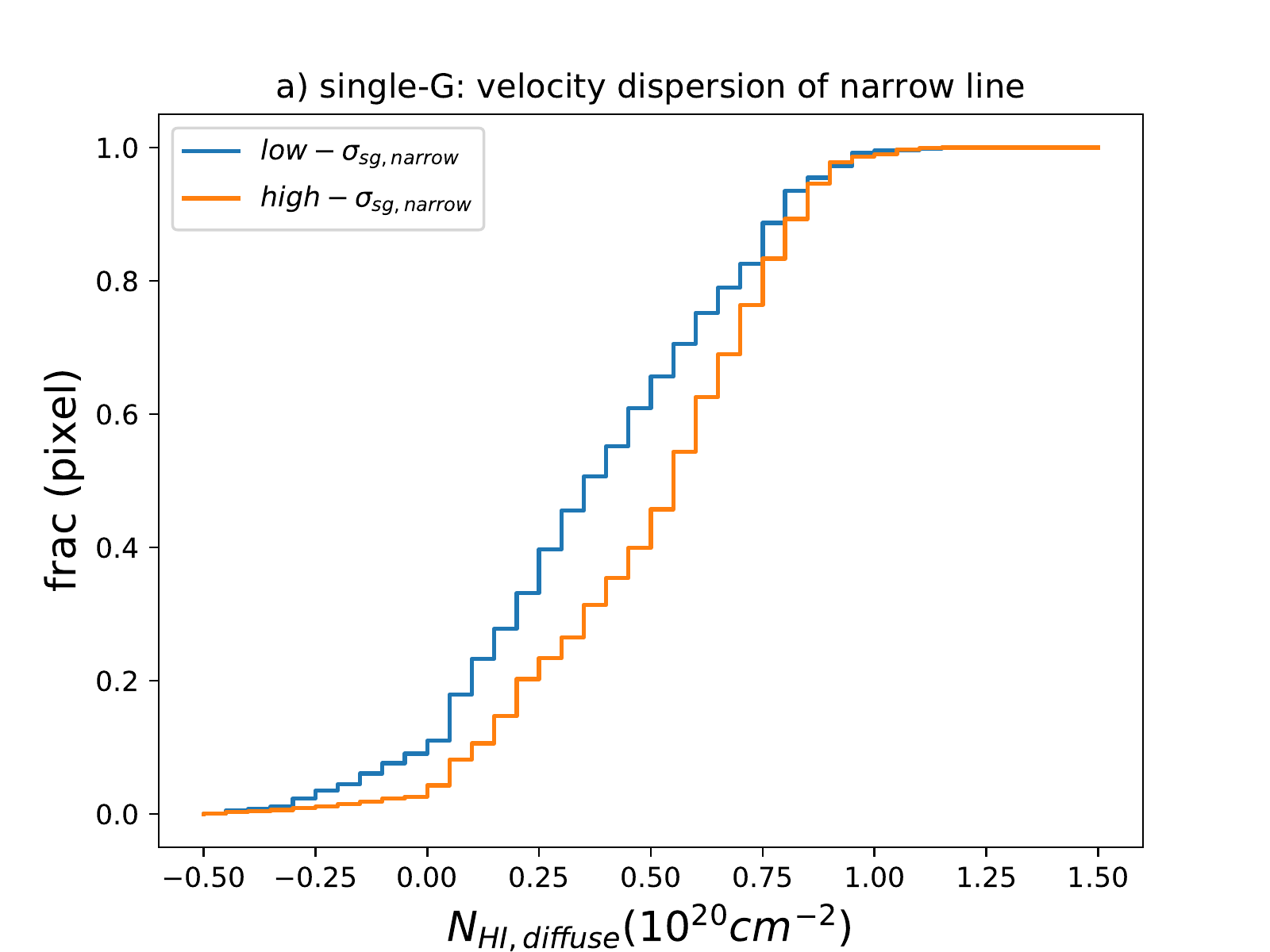}
\includegraphics[width=7cm]{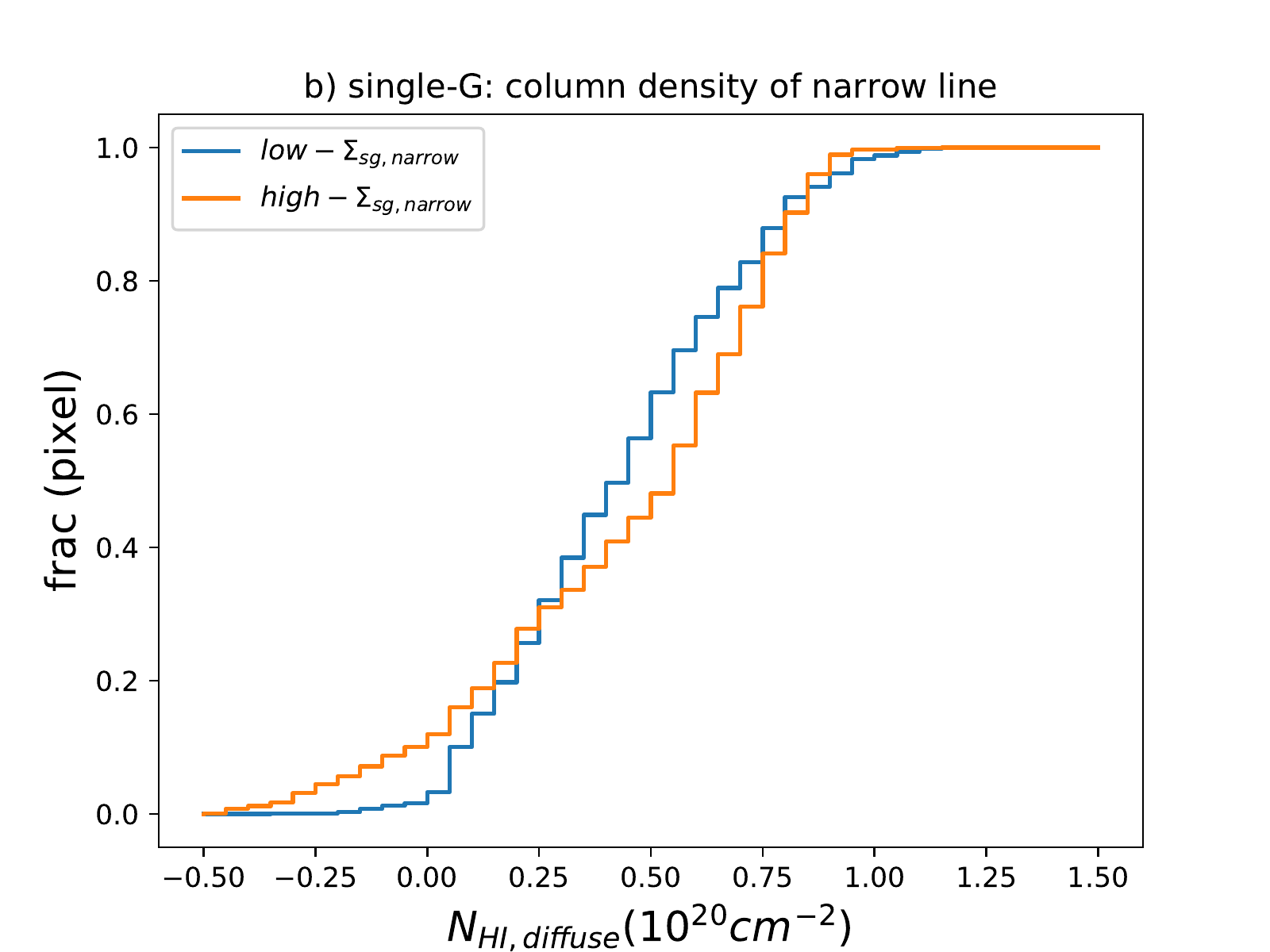}

\includegraphics[width=7cm]{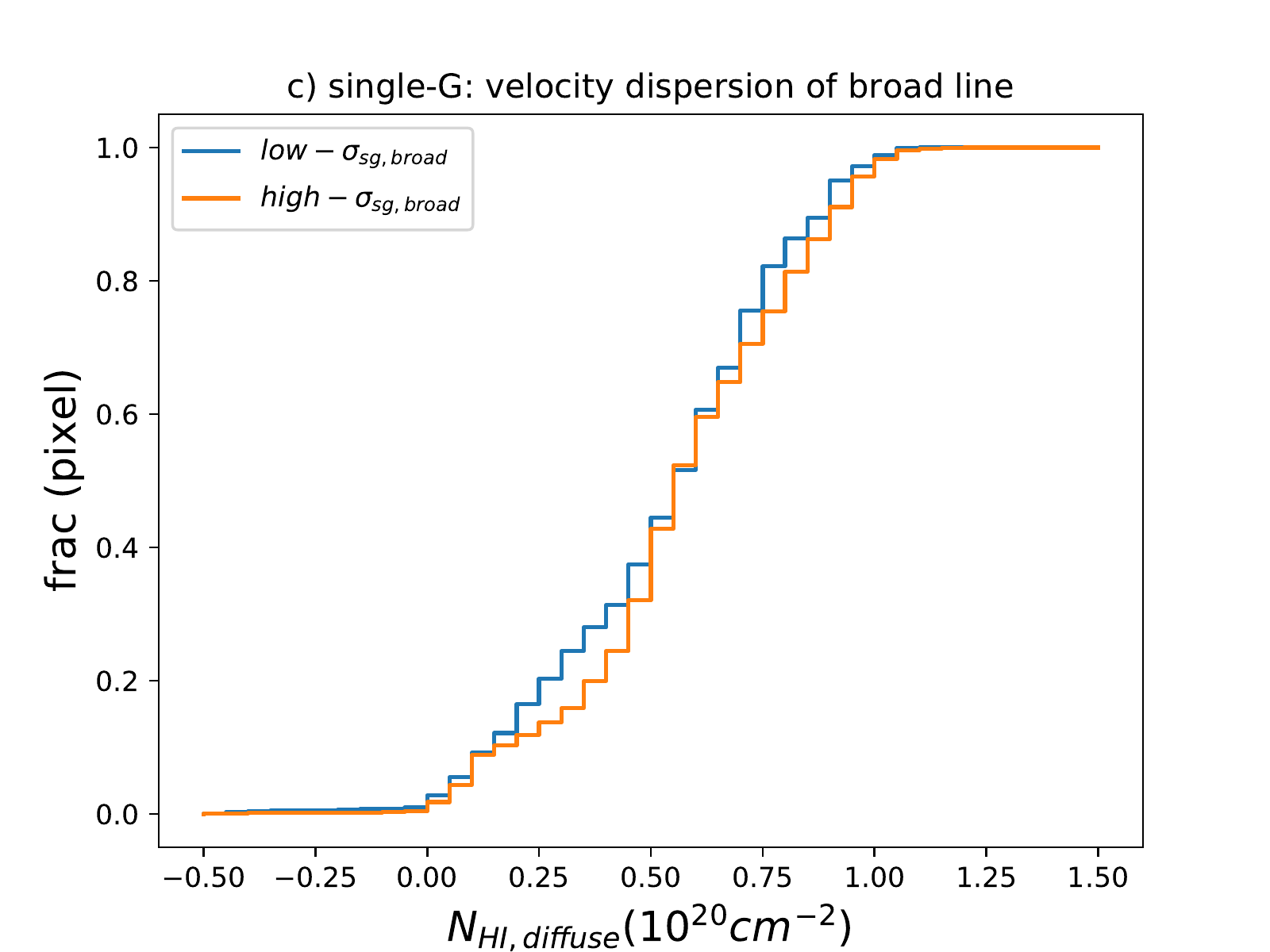}
\includegraphics[width=7cm]{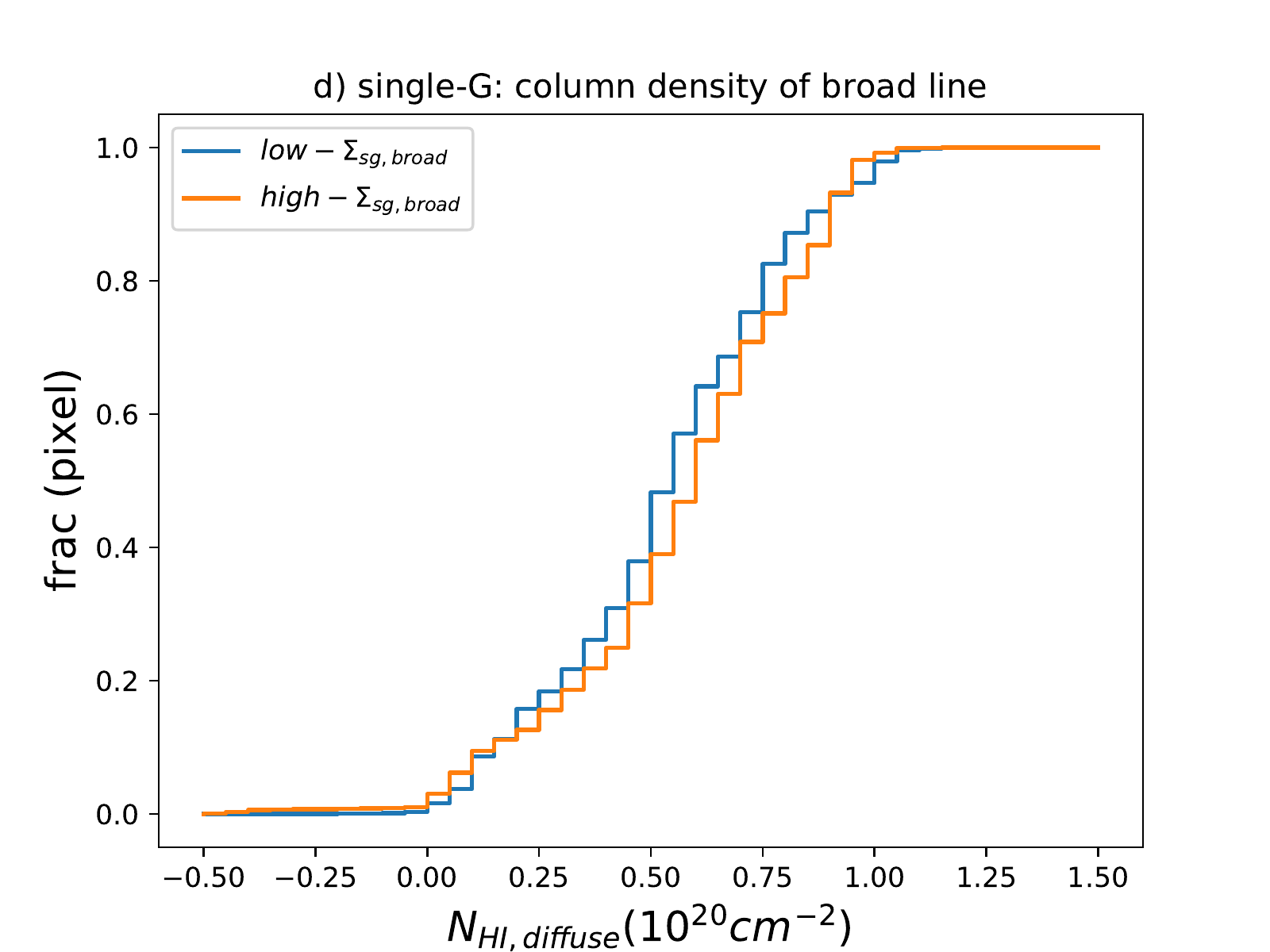}

\includegraphics[width=7cm]{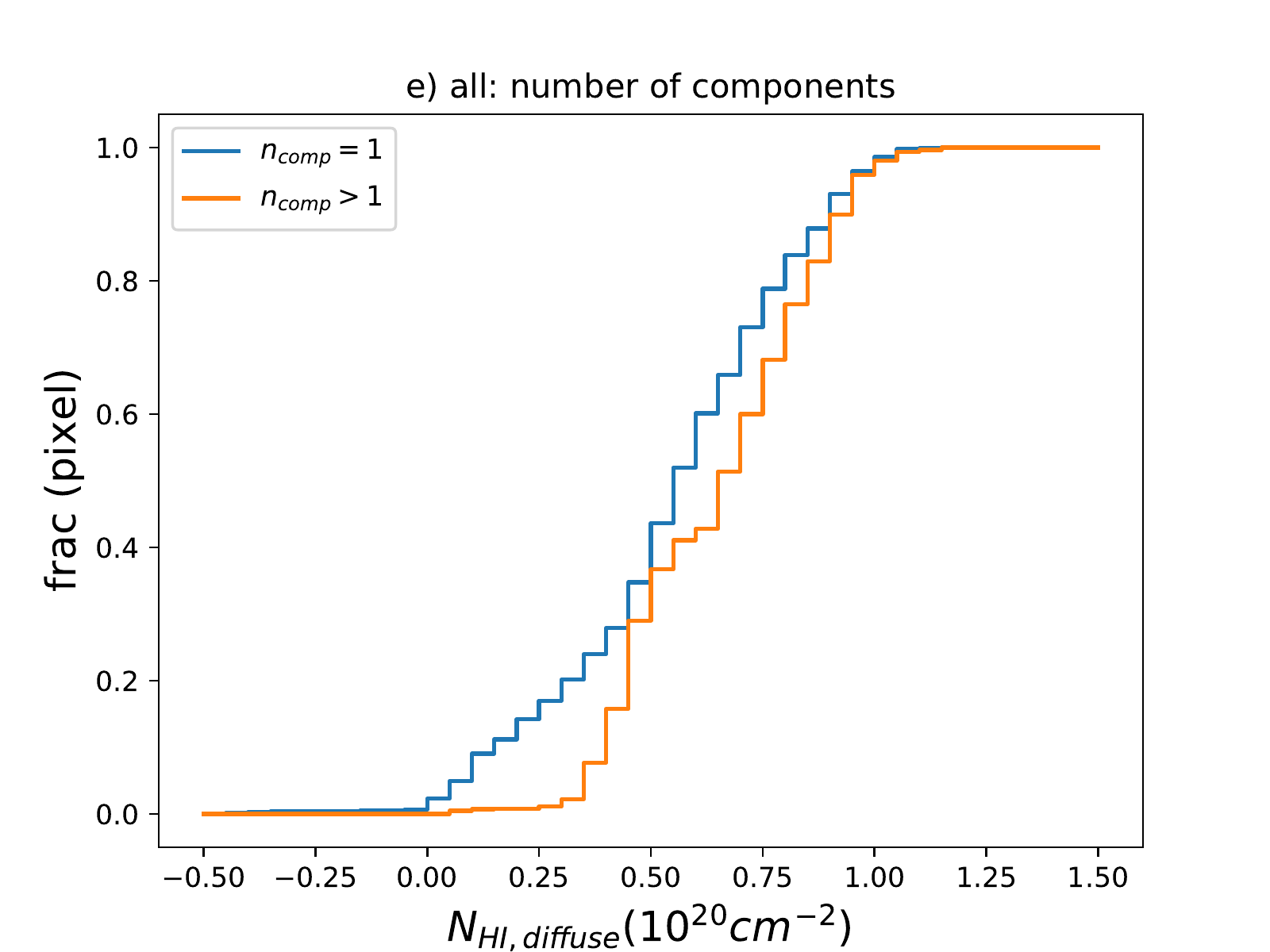}
\includegraphics[width=7cm]{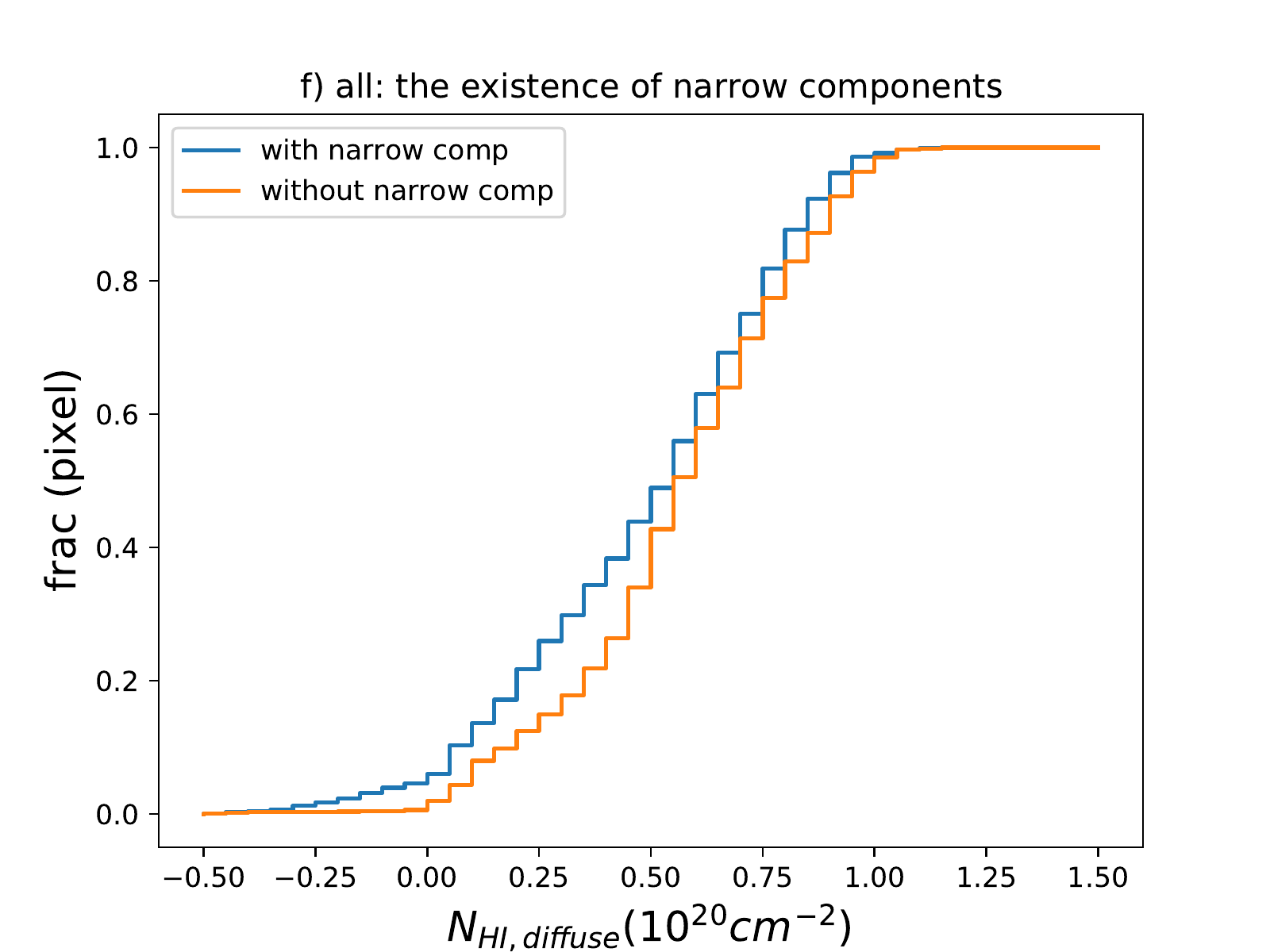}

\includegraphics[width=7cm]{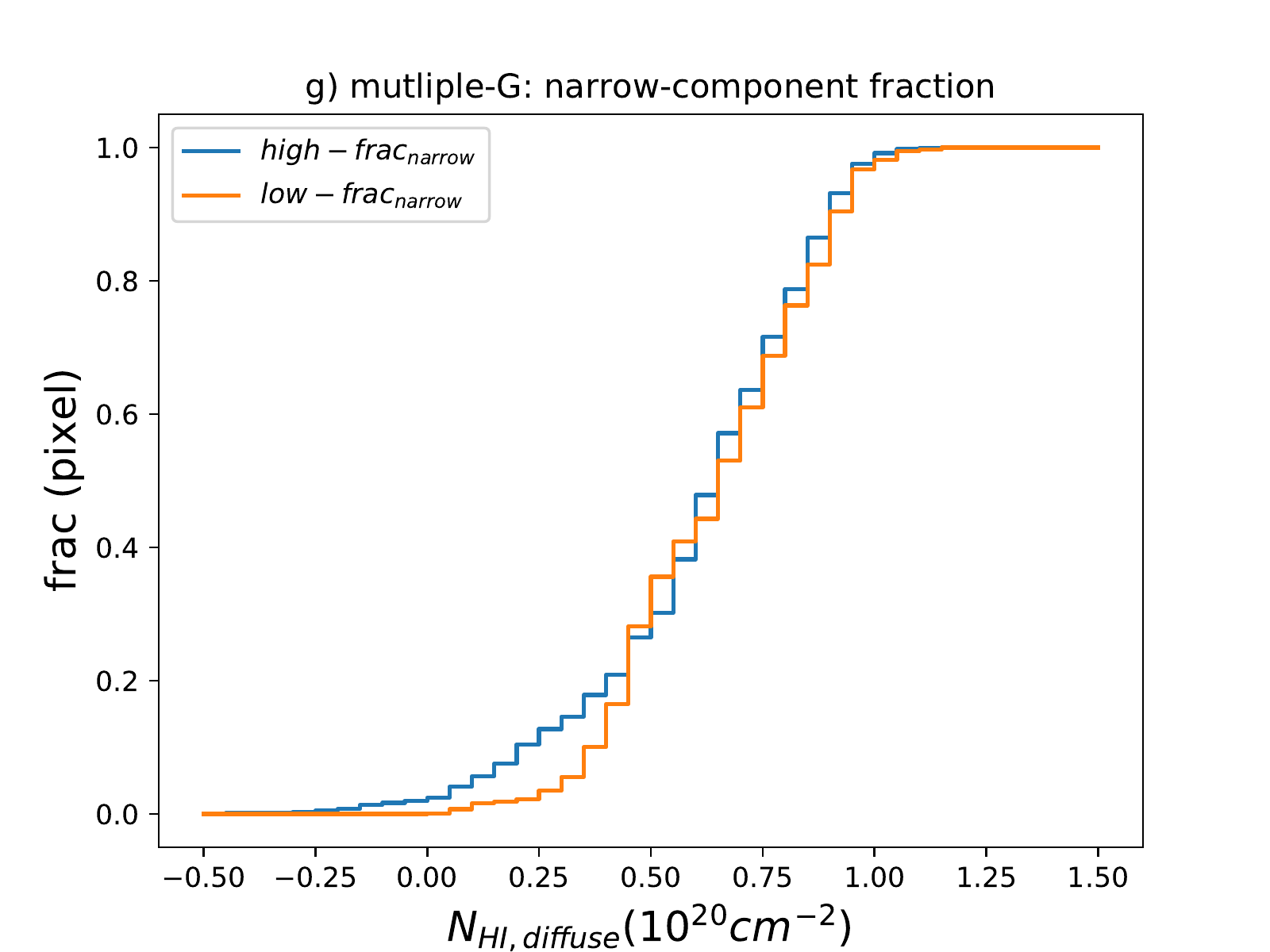}

\caption{ Comparing the cumulative distributions of diffuse $\hi$ column densities between line-of-sights with different localized kinematic properties of the dense $\hi$. The properties considered include 
the velocity dispersion of narrow$\slash$broad single-Gaussian components (panel a$\slash$c), and the column density of narrow$\slash$broad single-Gaussian components (panel b$\slash$d), 
the number of Gaussian components (panel e),
the existence of the narrow component (panel f),
and the flux fraction of narrow components along line-of-sights with multi-Gaussian components (panel g). }
\label{fig:excessHI_baygaud_prop}
\end{figure*}

We study how the column density of diffuse $\hi$ is related to these kinematical properties of the dense $\hi$. 
In each panel of Figure~\ref{fig:excessHI_baygaud_prop}, we select and divide into two subsets the line-of-sights along dense $\hi$ by one type of dense $\hi$ kinematic property described above. We compare the distributions of column density in associated diffuse $\hi$ between the two subsets. 
Systematic trends arise from the comparisons. 
For single-Gaussian narrow profiles, high levels of diffuse $\hi$ prefer those that are broader in widths (panel a), but do not have a clear trend with the column density (panel b).
For single-Gaussian broad profiles, they show a slight tendency toward the broad widths (panel c) and high column densities (panel d). 
For all profiles, they prefer multi-Gaussian profiles over single-Gaussian profiles (panel e), and regions where there is no narrow $\hi$ over otherwise (panel f). 
For multi-Gaussian profiles, they slightly prefer those with low fractions of narrow components (panel g).
Together, these trends indicate that the localized kinematic hotness of the dense $\hi$ and the column density of the diffuse $\hi$ seem to be boosted simultaneously in the tail region. There might be a pipeline of the $\hi$ shifting from the narrow, to the broad, and then to the diffuse status, or in the opposite direction. 

\subsection{The localized kinematics of the diffuse HI}
\label{sec:diffuseHI_vsig}
From the spectra displayed in Figure~\ref{fig:spectrum}, the FAST flux does not extend further in velocity than the WSRT flux. In the literature, such an excess of $\hi$ in the same velocity range is typically attributed to a tidal origin \citep{VerdesMontenegro01}.  Those integral spectra mix the effect of bulk motions and localized kinematics. In the following, we remove the velocity shift due to bulk motions and derive super profiles to reflect localized kinematics. 
In order to minimize the projection effect of multiple velocity components, we select the line-of-sights which have single-Gaussian profiles in the dense $\hi$, which comprise 51\%, 14\%, and 65\% of the line-of-sights in the NGC 4631 region (disk region$+$tail region), the disk region, and the tail region, respectively. 
We keep in mind that in addition to thermal motions, beam smearing, and turbulence should significantly contribute to the broadening of line widths. 

\begin{figure} 
\centering
\includegraphics[width=8cm]{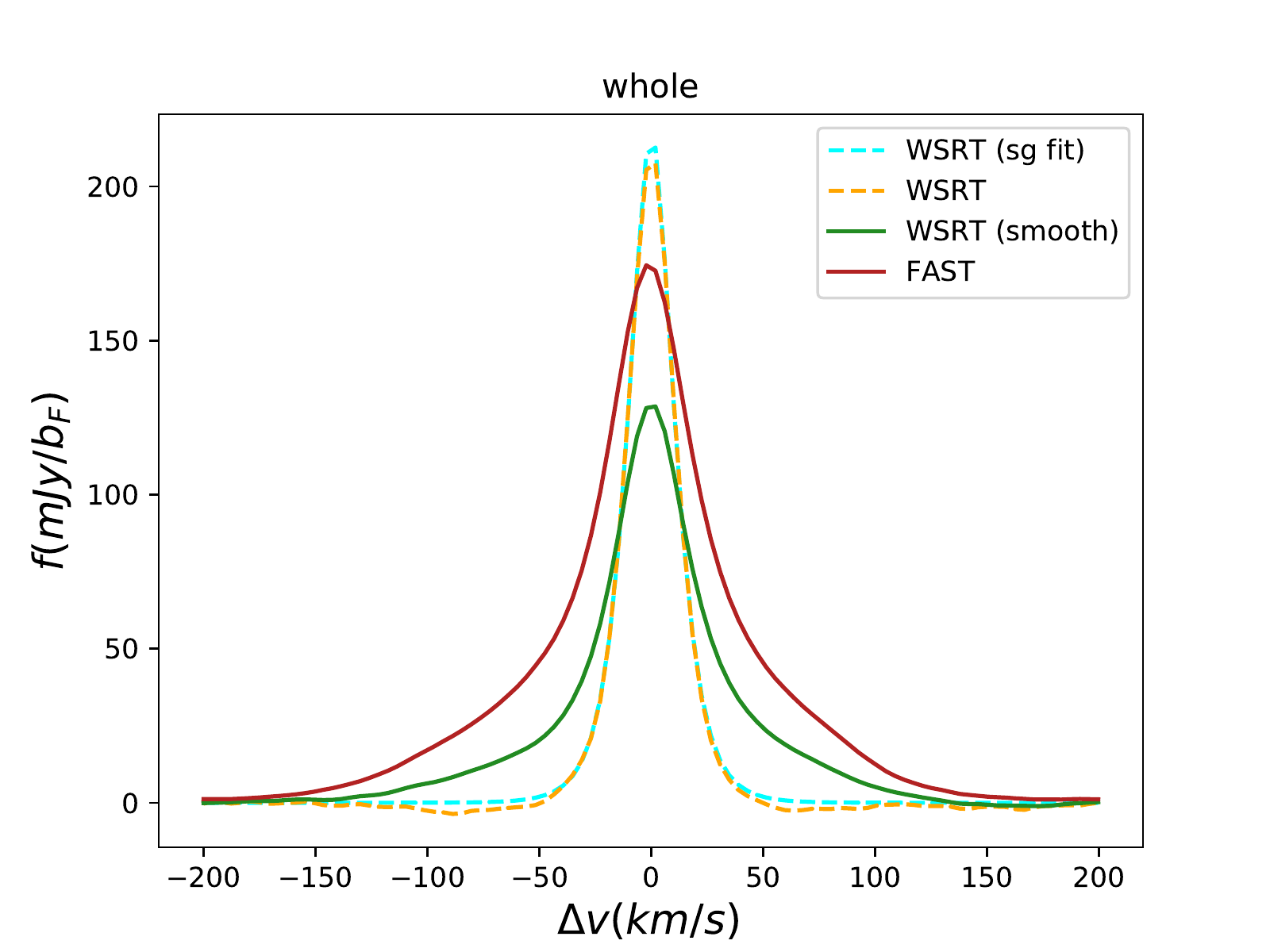}
\includegraphics[width=8cm]{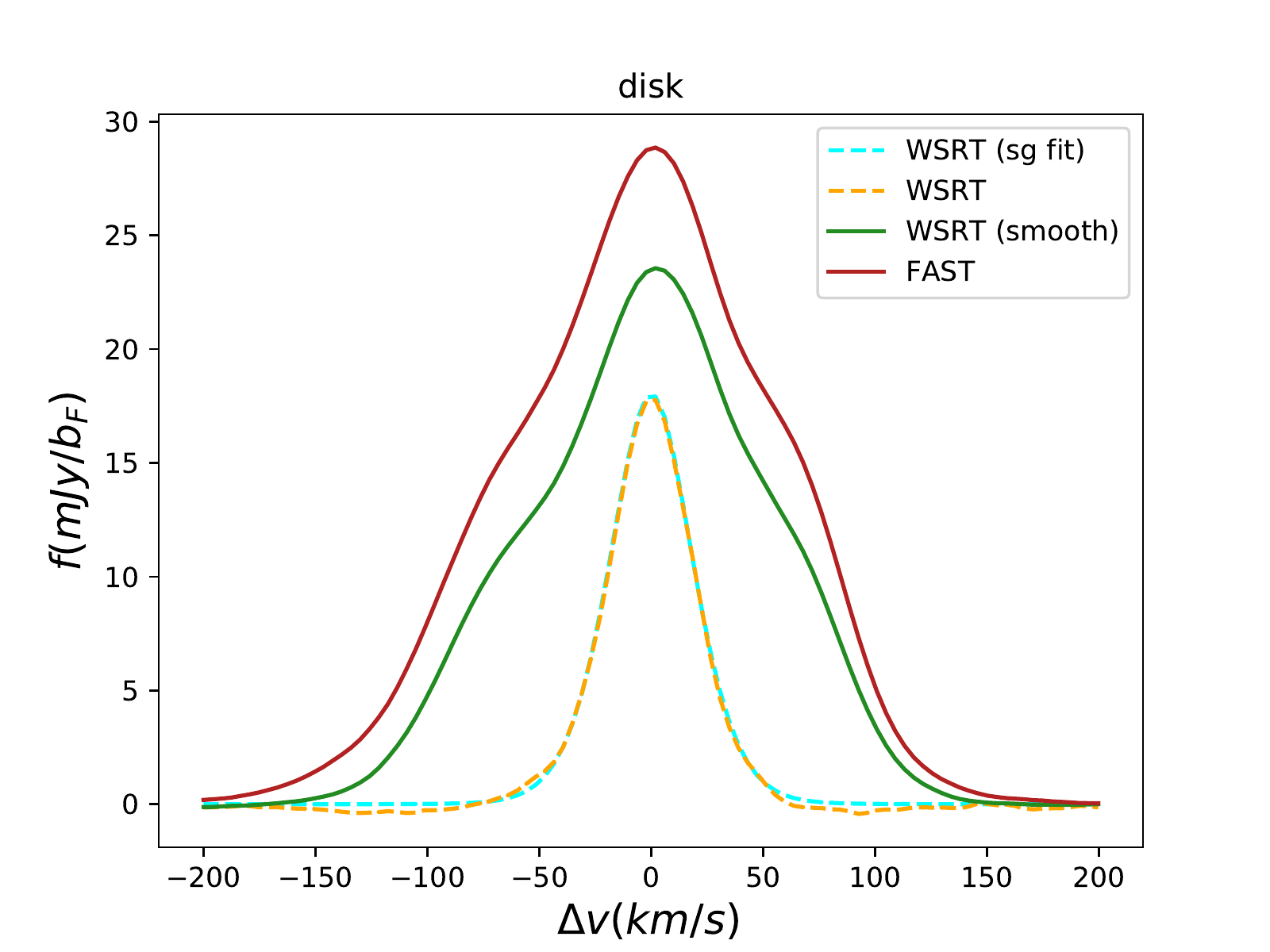}
\includegraphics[width=8cm]{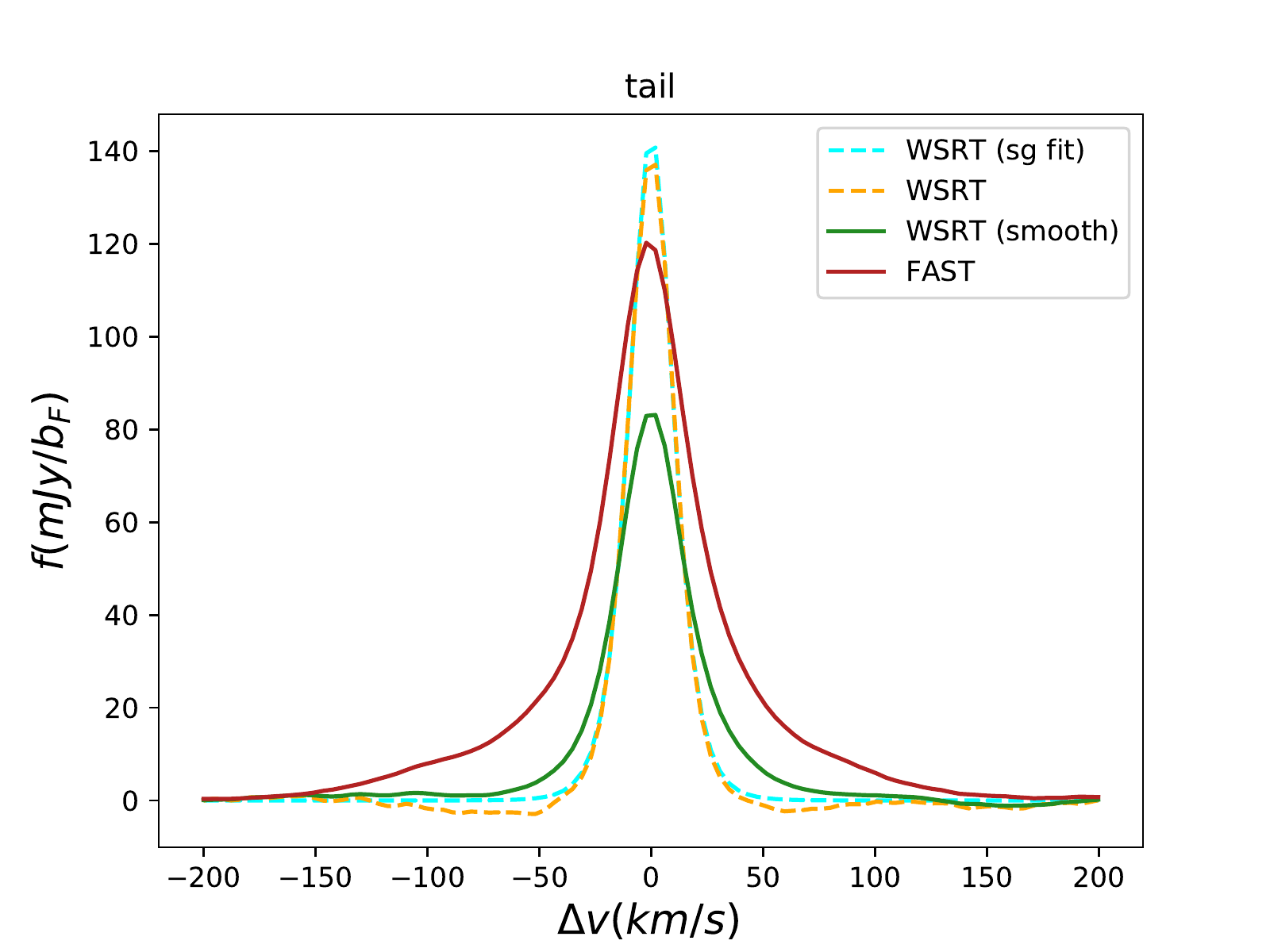}
\caption{ Super profiles of $\hi$ from stacking line-of-sights of data cubes. The line-of-sights are selected to have single-Gaussian profiles in the dense $\hi$. The top, middle, and bottom panels plot the super profiles of the whole WSRT $\hi$ detected region, the disk region, and the tail region, respectively. The cubes include the WSRT cube, the smoothed WSRT cube, and the PB-attenuated FAST cubes, which are plotted in orange, green and red. The stacking centers and stacking regions are determined by single-Gaussian fit to line-of-sights from the WSRT cube; the super profile of the single-Gaussian fits is plotted in cyan.  }
\label{fig:super_prof}
\end{figure}

We use the velocity centers of line-of-sight spectra in the WSRT cube, which have been obtained using {\scriptsize BAYGAUD} \citep{Oh22}.
We stack the line-of-sight spectra of the WSRT cube, after register them to the same velocity center. The stacking is performed for the WSRT-detected NGC 4631 region, the disk region, and the tail region, respectively. We do the same stacking for the smoothed WSRT cube and the PB-attenuated FAST cube, using the same velocity centroid determined from the WSRT cube. We display these super profiles in Figure~\ref{fig:super_prof}. From the top panel, the PB-attenuated excess $\hi$ of FAST is found throughout the localized velocity range, but not preferentially in the wings of the dense $\hi$. The conclusion above holds for both disk and tail regions displayed in the middle and bottom panels, but the super profiles of the former are much broader than the latter, indicating influences from the galactic internal structures, geometry, and stellar feedbacks. 
In the following of this section, we therefore limit the analysis to the tail region to focus on tidal effects. 

We use {\it emcee} \citep{Foreman-Mackey13}, the Python implementation of Goodman \& Weare's Affine Invariant Markov Chain Monte Carlo (MCMC) Ensemble sampler, to fit a double-Gaussian model to each of the super profiles of the tail region. The details and best-fit models are present in appendix~\ref{sec:appendix_dgauss}.

It is interesting to point out that, the narrow-Gaussian component accounts for roughly half the total flux in the FAST super profile, which is close to the ratio of the dense $\hi$ flux over the FAST detected $\hi$ flux in the tail region. Thus, it is possible that the narrow component of the FAST super profile corresponds to the dense $\hi$ detected by WSRT, while the broad component corresponds to a diffuse envelope missed by WSRT. 

We use the square root difference between the velocity dispersions of the WSRT and the smoothed WSRT cubes to correct for beam smearing effects.
After the correction, the narrow and broad Gaussian components of the FAST super profile have $\sigma$ of 13.4 and 51.0 $\kms$ respectively. It is obvious that the width of the broad component is unlikely thermal, but should be possibly dominated by turbulence, and perhaps also some contribution from beam smearing of where no WSRT flux is detected. 

\section{The hydro-dynamic and tidal environments}
\label{sec:environment}
In this section, we investigate the thermal, radiative, and gravitational environments around NGC 4631. We investigate, what is the fate of the $\hi$ and particularly the diffuse $\hi$ in the IGM and tidal region, and what physical mechanisms drive that.

\subsection{The hydro-dynamic effects}
\label{sec:discuss_hydrodynamic_environment}
We come back to the column density map of FAST detected $\hi$ in Figure~\ref{fig:map_NHI}. The high density part, where $\NHI \geq 10^{19} \cmsq$, extends for $\sim$120 kpc across. Near the edge, $\NHI$ drops by nearly 1 dex within a length comparable to the beam size of 3.24$'$, or 7.1 kpc.  

Similar but sharper (due to the use of images with a higher resolution) edges of $\hi$ distribution were noticed before at a similar column density level, particularly by the pioneering work of \citet{Corbelli89} and \citet{vanGorkom93}, in deep $\hi$ imaging of the nearby galaxies M33 and NGC 3198. The truncation of $\hi$ disks was attributed to the ionisation by the cosmic ultraviolet (UV) background \citep{Maloney93}. The prevalence of the truncation and the uniformity of the threshold column density are questioned recently by deep $\hi$ imaging of more galaxies \citep{BlandHawthorn17, Ianjamasimanana18}, as both the local UV background and the clumpiness of $\hi$ affect the ionizing status while both factors are quite uncertain \citep{BlandHawthorn17}.  The condition for $\hi$ to survive and evolve in the hot gas halo of N4631g may also differ from those benchmark galaxies M33 and NGC 3198. Firstly, the tidal $\hi$ reaches far into the IGM while retaining a high column density, which may induce efficient cooling of the hot gas. Secondly, the relatively high SFR of NGC 4631 may enhance the local UV radiation. 

In the following, we discuss the fate of $\hi$ in the IGM region in the context of different hydro-dynamic processes as a function of radius from NGC 4631.
We note that the following discussion is based on first order approximations of the complex interplay between the different phases and dynamics in the IGM. As such they merely provide a first indication of what might be happening to the $\hi$ in the tidal region. 

We point out that, the models discussed are 3-dimensional but the observed $\hi$ is projected. The galactocentric distances can be underestimated,  
and the projection and overlapping of structures can artificially enhance the $\hi$ column density, which may be major sources of uncertainty in the discussion of $\hi$ survival in the IGM. On the other hand, the projected phase-space distribution of flux (Figure\ref{fig:PSD}, discussed in section~\ref{sec:discuss_pPSD}) suggests that the overlapping of structures seems not severe in most parts of tail 1, 3 and 4, which may mitigate the problem. 
To overcome this observational limitation in the future, hydrodynamic modeling specifically conducted to reproduce the $\hi$ distribution in N4631g will greatly help; alternatively, a sample of many interacting systems like N4631g will provide a statistical and representative view for comparison with and constraint on general hydrodynamic simulation of interacting systems.  

Despite the uncertainties, a major advance here is that, the calculations are based on real measurements of the diffuse $\hi$, which were lacking in most previous observations.

\subsubsection{Thermal conduction}
\label{sec:thermal}
Based on the theory of \citet{Cowie77}, we use the following simplified calculation to discuss the status of thermal conduction of the $\hi$ in the IGM region. 

We assume a density distribution for the hot gas in the IGM ($n_{IGM}$)  following the single-$\beta$ models presented in \citet{Eckert11} (See appendix~\ref{sec:appendix_IGMprof} for details). The temperature is assumed to be uniform at the virial temperature of 8$\times10^5$ K (appendix~\ref{sec:appendix_mass}). Based on our measured super profiles, we fix the velocity dispersion of the diffuse $\hi$ to $\sigma_{\rm HI}=51.0 \kms$ (section \ref{sec:diffuseHI_vsig}). We assume the $\hi$ travels in the hot gas halo in an external pressure-confined way, and derive the volume density of the diffuse $\hi$ ($n_{\rm HI}$) accordingly. The value of $\log (n_{\rm HI}/{\rm cm}^{-3})$ drops from $-2.4$ at 10 kpc to $-3.1$ at 60 kpc, consistent with the typical values of tidal $\hi$ discussed in the literature (e.g. \citealt{Borthakur10}). 
 
For an $\hi$ cloud with a radius $r_c$, the dimensionless ``global saturation parameter'' $\sigma_0=\frac{({\rm T_{IGM}/ 1.5\times10^7~K}) ^2}{n_{\rm IGM} ~r_c}$ separates the gas at the interface between $\hi$ and hot gas into regimes of saturated evaporation, classical evaporation, and cooling flows \citep{Cowie77}. 
Using the critical value $\sigma_0=0.027$ \citep{Cowie77}, we derive the critical $r_c$ as a function of distance from NGC 4631. Then we calculate the critical column density $N_{\rm HI, c, evap}=n_{\rm HI}~ r_c=10^{18.5}~ \cmsq$ for classical evaporation in N4631g.  Because N4631g has a relatively low mass and thus low virial temperature, thermal evaporation is only relevant on small scales, corresponding to low column densities. This critical value does not vary significantly with distance to NGC 4631 because the $\hi$ volume density scales with the ICM density in our assumed model. 

\begin{figure} 
\centering
\includegraphics[width=10cm]{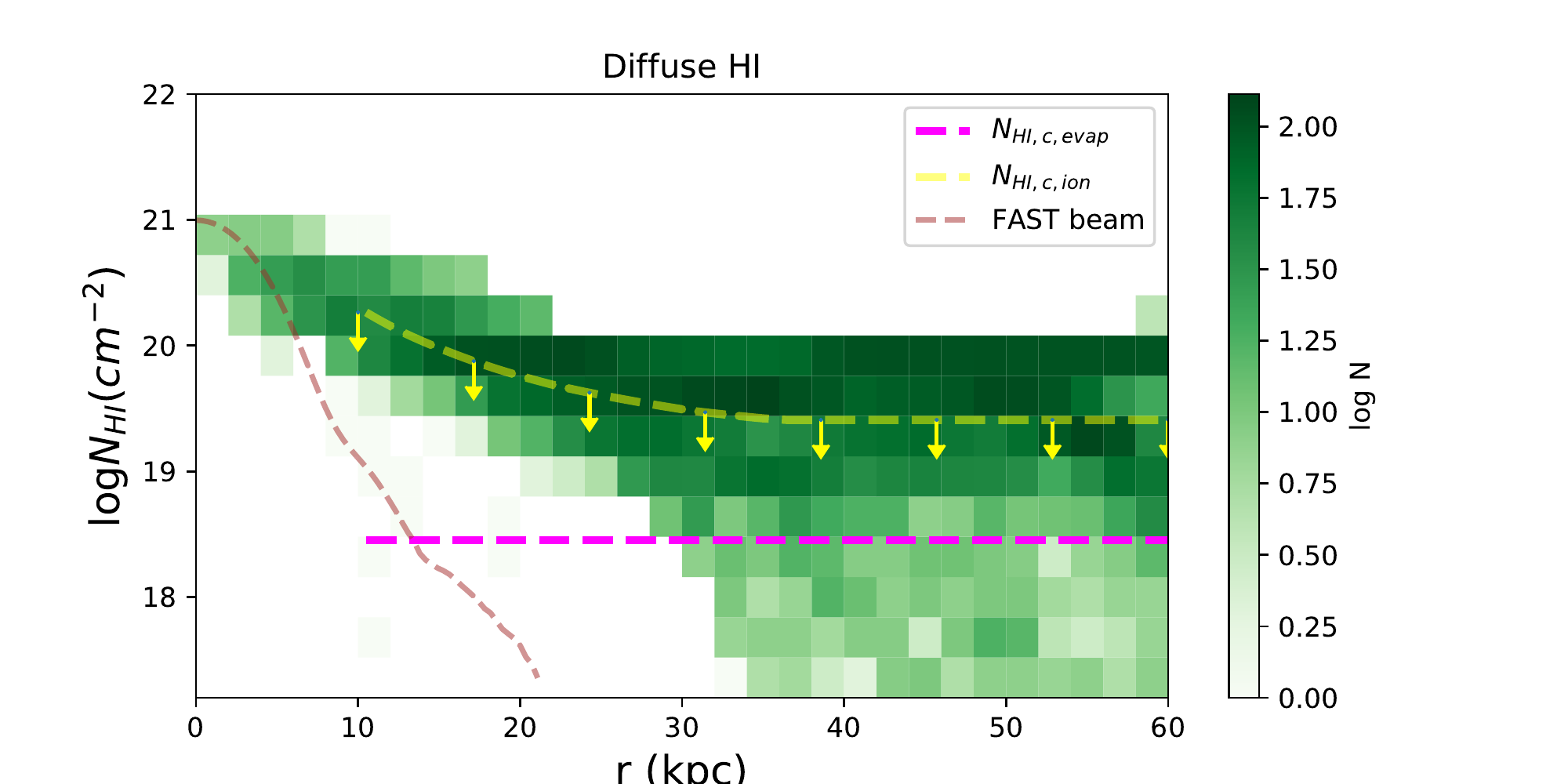}
\includegraphics[width=10cm]{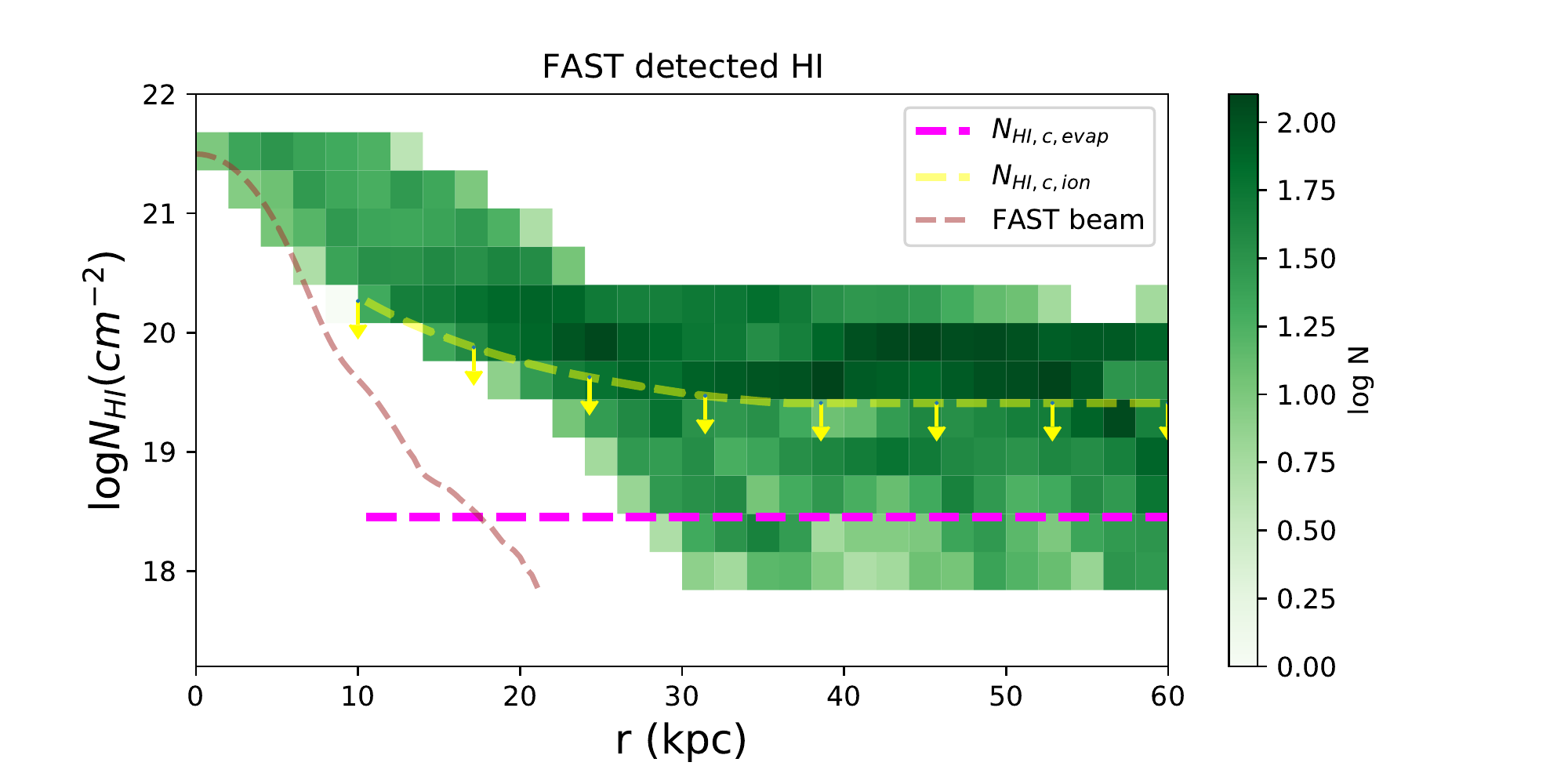}
\caption{ The distribution of $\hi$ column density as a function of projected distance from NGC 4631. The top panel is for the diffuse (excess) $\hi$, while the bottom panel for all $\hi$. Each pixel in the plot is color coded by logarithm of the number of pixels from the relevant column density map. The two dashed curves in each plot show the critical column densities to survive thermal evaporation (magenta), and UV ionisation (yellow) respectively. The brown curves show the shape of the FAST beam.}
\label{fig:Hdens_survive}
\end{figure}

For both the FAST detected $\hi$ and the diffuse $\hi$ (Figure~ \ref{fig:map_NHI} and \ref{fig:excessHI_maps}), this critical column density value is only reached at the very periphery of the IGM region. It is also clear in Figure~\ref{fig:Hdens_survive}, where we plot the number density of pixels as a function of column density of the diffuse $\hi$ and radius. Between a distance of 20 and 60 kpc, the number densities peak where $\log (\NHI/\cmsq)>19.5$, and sharply drop where $\log (N_{\rm HI}/\cmsq)<18.8$. The $N_{\rm HI,c,evap}$ values lie right below where the sharp drop begins. Thus the majority of the $\hi$ in the IGM region is more likely to induce cooling out of the IGM at its surface instead of thermally evaporating itself. 

We make a similar plot replacing the diffuse $\hi$ by all the $\hi$ detected by FAST in the bottom panel of Figure~\ref{fig:Hdens_survive}. The discussion above still applies, but the previous sharp pattern of number density dropping at $\log (N_{\rm HI}/\cmsq)<18.8$ is more blurred. 

In summary, there seems to be efficient cooling instead of evaporation associated with the $\hi$ in the IGM region. We note that the lack of direct measurements on the temperature and density of the IGM, and the magnetic fields \citep{Cowie77a} in the tidal region are major sources of uncertainty in the derivation of the evaporation related parameters.

\subsubsection{Photon ionisation}
The $\hi$ remains neutral in the UV radiation field through self-shelding. The quantity to derive is the critical $\hi$ column density ($N_{\rm HI,c,ion}$) where half of the hydrogen gets ionized due to the photon ionisation by stars plus the cosmic background.
We estimate the dimensionless ionisation parameter of stars ($U_{\rm star}$) from the SFR of NGC 4631 based on the equation in \citet{Tumlinson11}.
 We use {\it Cloudy} \citep{Ferland98} to simulate the ionisation rate of hydrogen at different levels of $U_{star}$ plus the comic background. We derive $N_{\rm HI,c,ion}$ as a function of distance to NGC 4631 based on products of the simulation. More technical details can be found in section~\ref{sec:CLOUDY} of the appendix. We emphasize that in the model $U_{\rm star}$ is only attenuated as a function of radius squared, but the possible absorption of tidal $\hi$ (and possibly also dust) as the photons travel through it is not considered. So the $N_{\rm HI,c,ion}$ derived should be viewed as upper limits.

In Figure~\ref{fig:Hdens_survive}, the $N_{\rm HI,c,ion}$ values lie roughly between where the number densities of pixels concentrate ($\log (\NHI/\cmsq)>19.5$) and sharply drop ($\log (\NHI/\cmsq)<18.8$). The transition in number densities is not sharply defined by $N_{\rm HI,c,ion}$, implying the aforementioned over-estimation of $N_{\rm HI,c,ion}$ and other uncertainties in the modeling, as well as possible counteracting effects of IGM cooling. Despite the likely over-estimation of $N_{\rm HI,c,ion}$, most pixels of diffuse $\hi$ have $\NHI$ above them, indicating that most diffuse $\hi$ are safe against photon ionisation in N4631g.

\subsection{Tidal interactions of the HI} 
\label{sec:discuss_PSD}
\label{sec:discuss_3dPSD}
We investigate the distribution of $\hi$ in N4631g in response to the past and on-going tidal interactions. We visualize the 3-dimensional distribution, and also provide a characterization of the phase-space distribution.

\subsubsection{The 3-dimensional visualization}
We provide snapshots of a 3-dimensional visualization of the $\hi$ distribution in N4631g in Figure~\ref{fig:snapshots}. The visualization is realized using the software SlicerAstro \citep{Punzo17}. We use it to provide a first impression of the complex morphology and kinematics of $\hi$ in the N4631g. Similar discussions were presented in \citet{Rand94} based on channel maps and position-velocity slices of an early WSRT data. 

From the snapshots of FAST data, tail 1 and 2 clearly connect NGC 4631 and NGC 4656. Tail 1 starts from the east and low-velocity side of NGC 4631, and reaches NGC 4656 on its east and high-velocity side (snapshot 1, 2, 3, 5, 6, 8). Tail 2 starts from near the disk center of NGC 4631, and reaches NGC 4656 on its west and low-velocity side (snapshot 1, 2, 5, 6, 8). The connection between the two galaxies by tail 2 was not so clear in the WSRT data of HALOGAS, or the early WSRT data of \citet{Rand94}; probably consequently, \citet{Combes78} tended to attribute the formation of tail 2 primarily to the perturbation of the much smaller but closer companion NGC 4627, and only secondarily to NGC 4656. 

From the snapshots of the FAST data, tail 3 starts from the high-velocity and western side of NGC 4631 (snapshot 1, 3, 4, 5, 8), extends to the intermediate velocity and joins tail 1 in the south (snapshot 2, 5, 8). This link between tail 3 and 1 was tentatively seen but again unclear in the WSRT data. Tail 4 is short in both the FAST and WSRT data. It starts from the west end of the NGC 4631 disk, and extends to the east and low-velocity direction  (snapshot 1, 5, 8). 

\begin{figure*} 
\centering
\includegraphics[width=14cm]{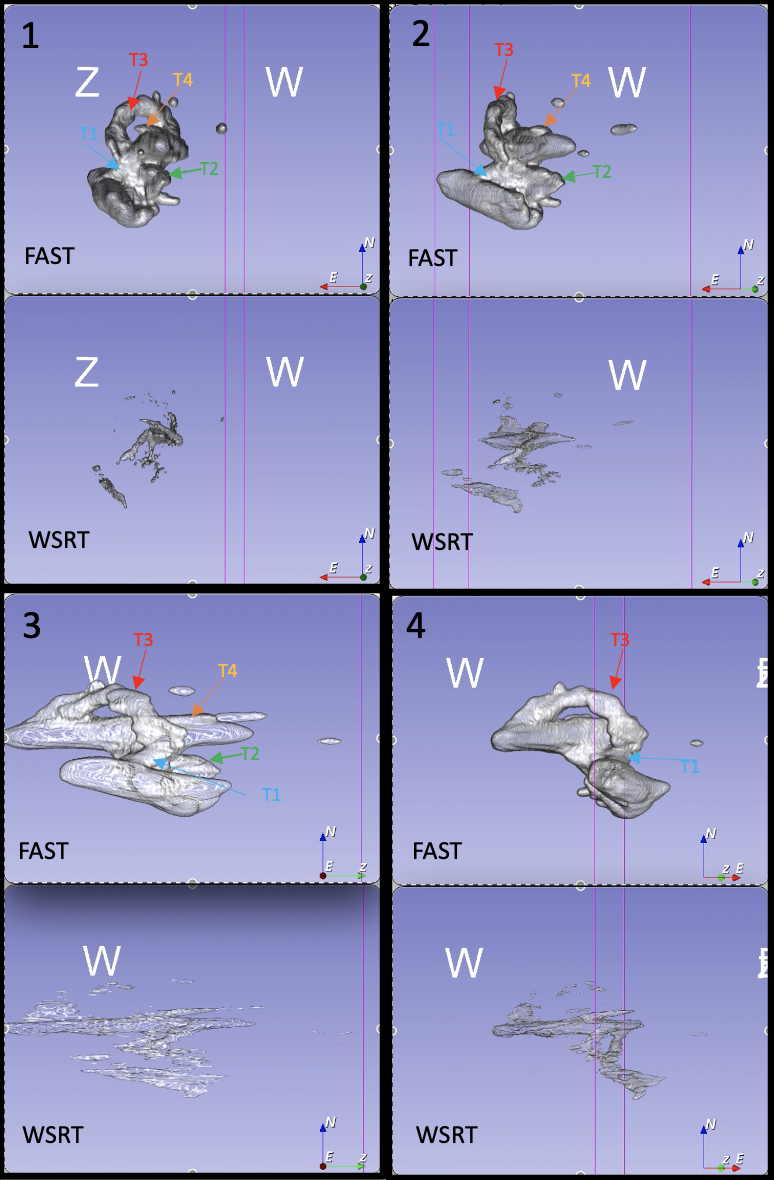}
\caption{ Snapshots of 3-dimensional visualization of $\hi$ distribution in the FAST and WSRT cubes. An animated version of this figure is available as on-line material. 
The duration of the animation is 28 s, and the content is the 3-dimensional visualization of the FAST and WSRT cubes (WCS system registered) continuously rotating by 360$\degree$. The viewing angles are denoted as direction axes, with N, E, W, z and Z pointing toward the north, east, west, low-redshift (velocity), and high-redshift (velocity) direction, respectively. The visualization is realized using the software SlicerAstro \citep{Punzo17}. 
The figure here shows 8 snapshots from that animation, which are evenly distributed in a rotation of 360$\degree$, and are ordered by number denoted in the top-left corner of panels. For visual clarity, the 4 tidal tails are denoted in the figure (but not in the animation), and each pair of FAST and WSRT snapshots are vertically arranged (while horizontally arranged in the animation). To be continued. 
}
\vspace{0.5cm}
\label{fig:snapshots}
\end{figure*}

\addtocounter{figure}{-1}
\begin{figure*} 
\centering
\includegraphics[width=14cm]{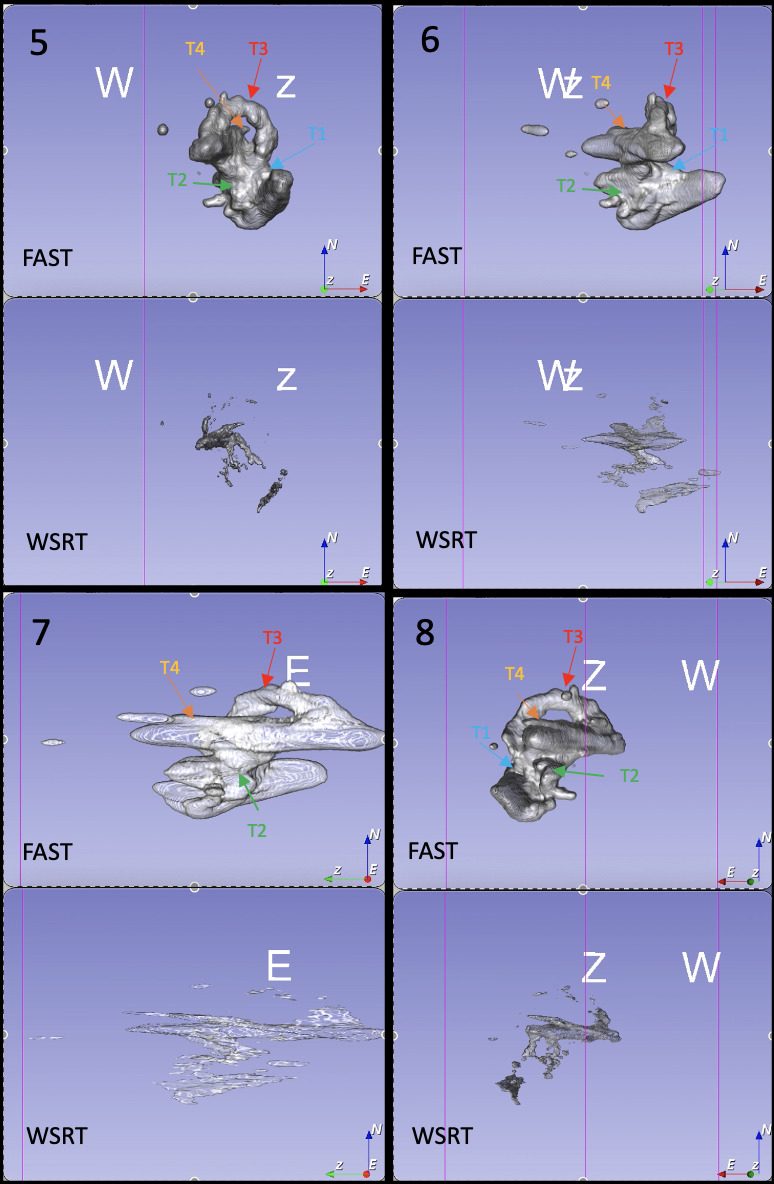}
\caption{ Continued. }
\label{fig:snapshots2}
\end{figure*}

\subsubsection{Analysis of the phase-space distribution}
\label{sec:discuss_pPSD}
We study the projected phase-space distribution of $\hi$ around NGC 4631. The projected phase-space diagram is a diagram of radial velocity offset versus projected distance to the center of NGC 4631. We are limited by observational projections, but a first-order characterization can still be obtained about the bulk motions of $\hi$. 

\begin{figure*} 
\centering
\includegraphics[width=14cm]{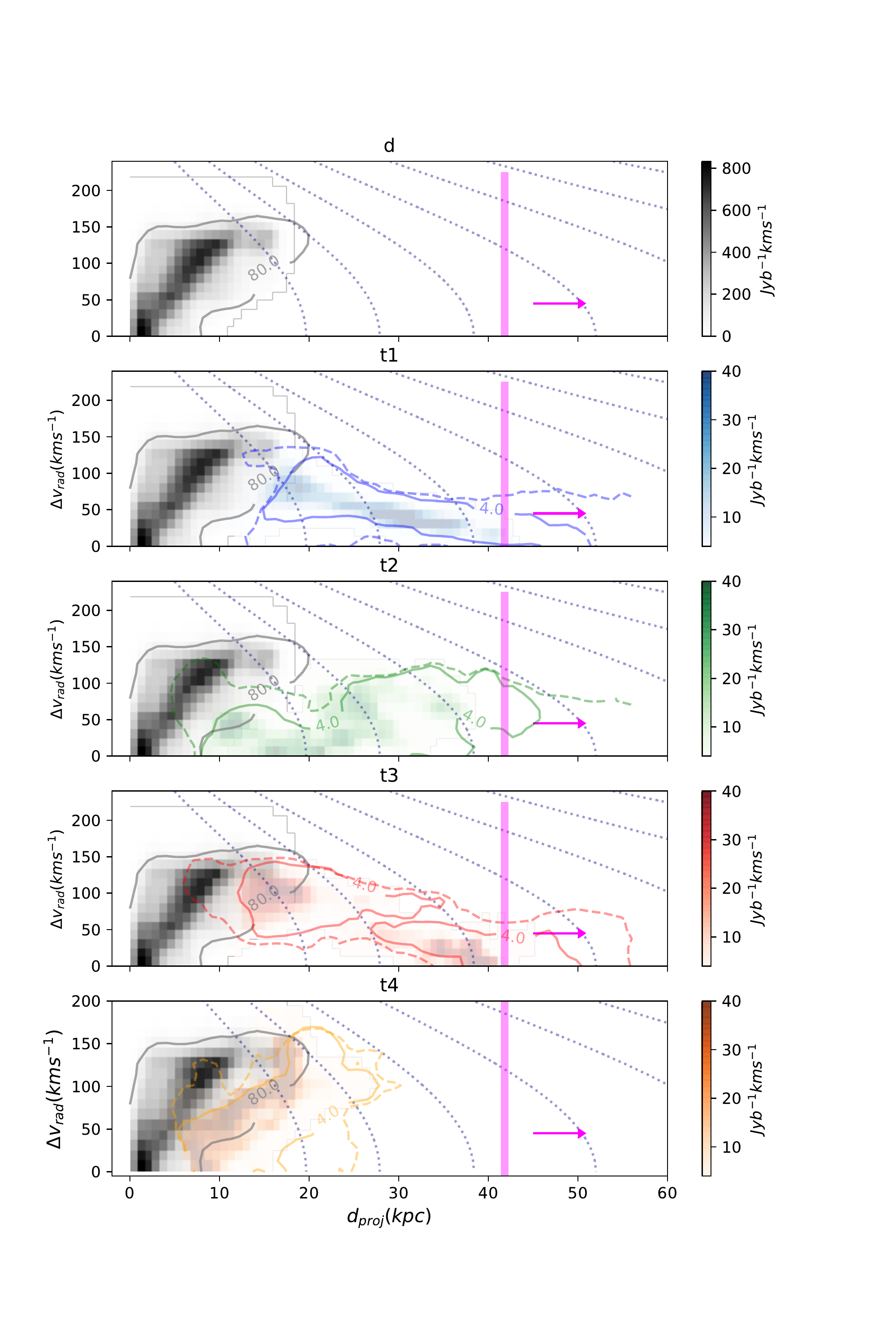}
\caption{ The projected phase-space distribution of $\hi$ flux around NGC 4631. From top to bottom, the distribution of $\hi$ in the disk and the 4 tail regions are plotted respectively. The pixels are color coded by $\hi$ flux from the PB-corrected WSRT cube, and the distributions are outlined by solid contours at the level of 4 (80) $\jybkms$ for the tail (disk) flux. The distribution of $\hi$ flux from the PB-corrected, FAST$+$WSRT combined cube is outlined by dashed contours at the same level.  The disk region is repeated in every panel to guide the eye. Contours of gravitational potential with linear interval steps are plotted as grey dotted curves. The truncation radius for NGC 4656 to strip gas from NGC 4631 is plotted as the thick, pink vertical line. The pink arrow mark the radial velocity deviation of NGC 4631 from NGC 4656. }
\label{fig:PSD}
\end{figure*}

We plot the distribution of $\hi$ in the regions of the NGC 4631 disk and the 4 tails in the projected phase-space diagram in Figure~\ref{fig:PSD}. 
We focus our discussion on the distribution of PB-corrected dense $\hi$, as it is a good tracer of the kinematic skeleton of tidal tails. But we also outline the distribution of diffuse $\hi$ using the PB-corrected, {\it mmerge} combined cube. The distribution of dense $\hi$ in the NGC 4631 disk shows the pattern of a rotating disk with a maximum velocity around 150 $\kms$, which is by construction when defining the disk region. To guide the eye, the distribution of $\hi$ in the disk region is repeated in all panels of the tails. To assist the analysis, we also plot contours of the gravitational potential with linear steps (see details in appendix~\ref{sec:appendix_potential}), and mark the truncation radius imposed by NGC 4656 at 41.9 kpc which we derive using the equation of \citet{Byrd90}. 

We find three distinct patterns of the $\hi$ distribution in the projected phase-space diagram. Tail 1 and 3 both start from the end of the disk, and deaccelerate in relatively radial velocity while extending to large projected distance until reaching around the truncation radius imposed by NGC 4656. Tail 4 is almost a parallel shift of the lower envelope of the disk in the projected phase-space diagram. This linear shape suggests an almost solid-body rotation, which are often found in systems of slow encounters (e.g. M81, \citealt{Sorgho19}). Tail 2 looks much broader in morphology and possibly higher in energy than the other tails. Its furthest end crosses the truncation radius of NGC 4656, while the relative radial velocity is still high. In the projected view, the furthest end reaches a gravitational potential level similar to those of tail 1 and 3, and much higher than that of tail 4. Its high energy and complex morphology suggests it likely to have been perturbed by more than one galaxy (i.e. both NGC 4656 and NGC 4627), which was supported by previous particle simulations to reproduce the dense $\hi$ distribution in N4631g \citep{Combes78}. 

Limited by projection effects, it is difficult to deduce the motion of gas without the aid of hydrodynamic simulations designed to reproduce the morphology. But from the complexity of $\hi$ distribution in the projected phase-space diagram, we can still infer that there are more than one tidal encounters in N4631g, which should explain the widely spreading $\hi$, and may input turbulent energy through shocks to produce the diffuse $\hi$.

\section{Summary and conclusion }
\label{sec:summary}
We present a deep FAST image of $\hi$ in and around NGC 4631. We identify a component of excess $\hi$ detected by FAST but missed by WSRT.  Our major results are summarized below:

1. {\bf The nature of the excess HI is likely large-scale, diffuse HI. }
 This excess $\hi$ has a low spatial frequency, corresponding to a characteristic angular scale $\geq14'$ or 30 kpc, missed by WSRT due to the limited shortest baseline. It is also highly turbulent, with a velocity dispersion around 44 $\kms$.  Around 40\% (70\%) of the excess $\hi$ in the NGC 4631 region (IGM region) is found beyond the regions where dense $\hi$ is detected by the WSRT.

2. {\bf The diffuse HI is more closely related to the dense HI that is kinematically hot than that is warm.} 
When overlapping with the dense $\hi$ in the tail region, the diffuse $\hi$ increases in column density with the dense $\hi$,  and is particularly closely associated with the ``hotter'' part of the dense $\hi$. It is preferentially found where the dense $\hi$ is more dominated by the broad-velocity components, or has multiple velocity components. 
 
 3. {\bf The diffuse HI is likely to induce cooling flows of the hot IGM. } 
 The diffuse $\hi$ in the IGM region typically have a column density $\geq 10^{19.2}~\cmsq$, which is far above the critical column density for thermal evaporation, and likely safe from photon ionization. This relatively high $\hi$ column density is consistent with a condition to induce efficient cooling flows from the hot IGM.

The results above involve gases of four different phases in the IGM region, namely the hot IGM, the diffuse $\hi$ characterized in result 1, the ``hot'' dense $\hi$, and the ``warm'' dense $\hi$, sorted roughly in reverse order of energy.  Result 3 indicates that, except for the periphery of the widely spreading IGM region, the hot IGM is likely cooling into the diffuse $\hi$. An important reason for cooling flows to be induced near $\hi$ gas, is that the thermal temperature of the interface between $\hi$ and the hot IGM produced by turbulent mixing is intermediate between these two gas phases, reaching close to the value of $10^5$ K for the radiative cooling function to peak \citep{Dere09}. Indeed, if we take the radiative cooling function of \citet{Dere09} for the solar metallicity and assume no heating, the isochoric cooling time for hot gas at the virial temperature of N4631g is close to the dynamical time of N4631g ($\sim$400 Myr at a radius of 30 kpc), indicative of difficult cooling. But if the temperature drops to $10^5$ K, the cooling time drops by more than one order of magnitude. 

The diffuse $\hi$ further links to the dense $\hi$ gas. Result 2 indicates a continuous shift in phases between the diffuse $\hi$, the hot dense $\hi$, and the warm dense $\hi$, while the shift could be in either direction. Indeed, tidal interactions can both enhance cooling through mixing of metals, and heating through tidally induced shocks.  If the net effect is the diffuse $\hi$ progressively cooling into the dense $\hi$, then there is a unidirectional accretion pipeline that transfers gas from the hot IGM through the diffuse $\hi$ to the dense $\hi$. If, on the other hand, the dense $\hi$ is being progressively heated into the diffuse $\hi$, the surface area of  $\hi$ in the IGM enlarges, and the cooling rate of the hot IGM increases as a result. In both cases, the diffuse $\hi$ plays an important role in the gas accretion from the hot IGM, either more as a transfer station of the accreted gas, or more a catalyst for cooling from the hot IGM. 

Given such an important role of diffuse $\hi$ in gas accretion, the tidal interaction may have significantly boosted the gas accreting rate (i.e., the integral cooling rate from the hot IGM) of NGC 4631.  N4631g has $M_{200}\sim10^{12.1}~M_{\odot}$ (appendix~\ref{sec:appendix_mass}), putting NGC 4631 in a theoretical regime where heating from cosmic gas accretion and internal feedbacks start to efficiently prevent gas cooling \citep{Keres05}. The gas accretion could be effectively slow in NGC 4631 if it was an unperturbed galaxy. Tidal interaction enhances the gas accreting rate by spreading $\hi$ widely in the IGM. The tidal $\hi$, particularly when in the diffuse phase, greatly increases the area of interface between the $\hi$ and and the hot IGM, thus induce significantly extra cooling. The tidal effects also put the gas in a kinematic status prone to shocks, ram pressure, and tidal compression, causing localized gaseous condensation and thermal instabilities directly relevant for enhanced cooling. The tidal effects further help transport metals throughout the IGM which is important for radiative cooling \citep{Dere09}. If such a scenario of enhanced cooling is true, we speculate the existence of a large amount of warm, ionized gas as an intermediate phase between the hot IGM and the diffuse $\hi$, which was indeed tentatively detected in H$\alpha$ throughout the group \citep{Donahue95}. It might be worth mentioning that, in a recent cosmological zoom-in magnetohydrodynamics simulation by \citet{Sparre22}, one simulated galaxy pair shows a broad $\hi$ bridge, which is qualitatively similar to the structure observed between NGC 4631 and NGC 4656. \citet{Sparre22} found that the gas that had been in the CGM prior to the tidal interaction contributed to nearly half of the mass in the gas bridge, and more than one fourth of the fueling to star formation during the interaction. 

Previous observational studies based on individual or statistical samples of tidally interacting systems found evidence for $\hi$ to be both depleted and replenished in galaxies \citep{VerdesMontenegro01, Hess17, Ellison18}. Theoretically, both effects are physically possible \citep{Boselli06,Hani18,Stevens19}. On the whole, there seems to be a high level of physical complexities in mergers, which may smooth out in statistical analysis of integral $\hi$ measurements, and cannot be fully captured by individual systems. Our study put NGC 4631 in the category where the tidal interaction induces $\hi$ fueling, but the main contribution is characterizing and highlighting the role of the diffuse $\hi$, instead of just adding vote to one side in the question of fueling or depletion.

\begin{figure*} 
\centering
\includegraphics[width=16cm]{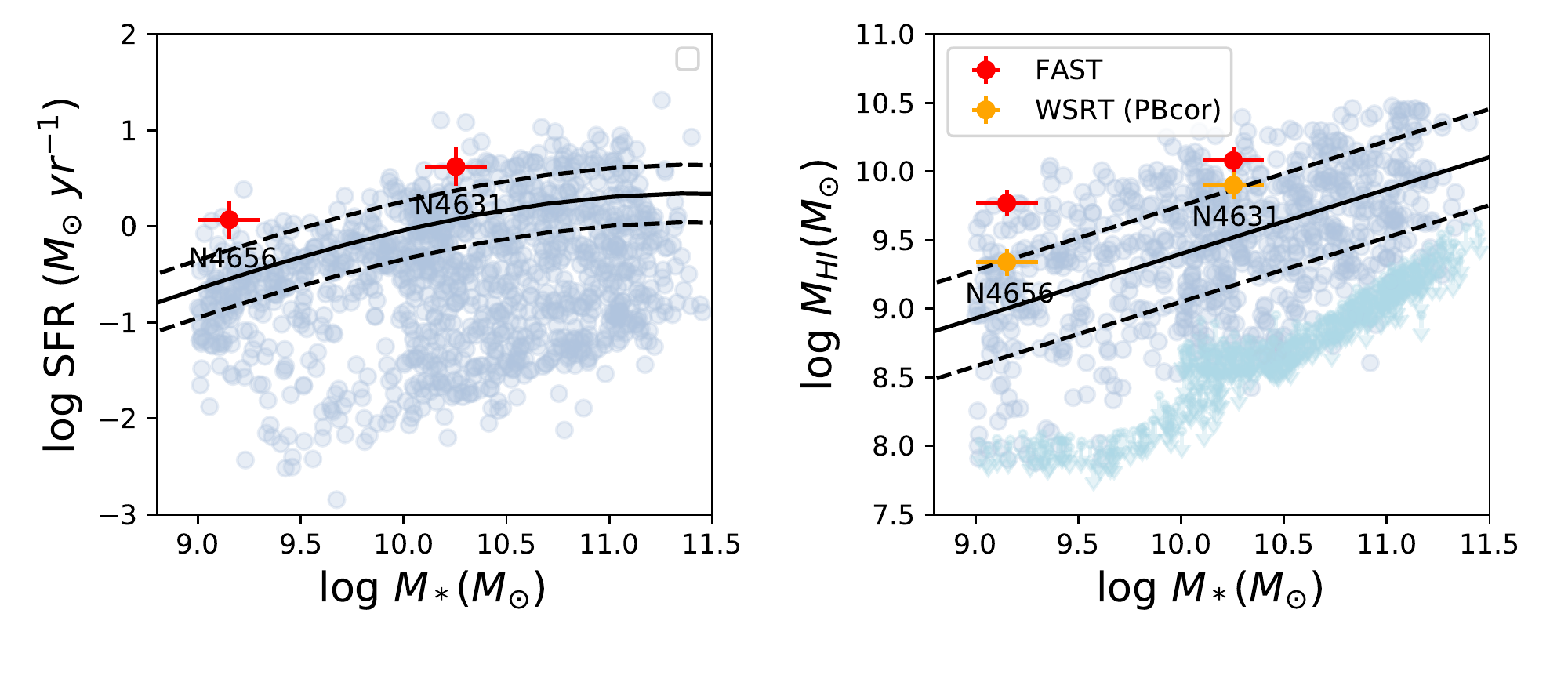}
\caption{The position of NGC 4631 and NGC 4656 in the diagram of SFR versus stellar mass and $\hi$ mass versus stellar mass. The two red or orange dots represent NGC 4636 and NGC 4656, with the former having the higher stellar mass.  The background data points are from the xGASS sample \citep{Catinella18}. Left: the red and orange dots are for $\hi$ masses from the FAST cube and the PB-corrected WSRT cube respectively; the solid and dashed curves are the position and scatter of star forming main sequence from \citep{Saintonge16}. Right: the blue and cyan points in the background represent measurements and upper limits for $\hi$ detected and un-detected galaxies; the solid and dashed curves are the position and scatter of $\hi$ main sequence of star-forming galaxies from  \citet{Janowiecki20}.}
\label{fig:scaling}
\end{figure*}

To put NGC 4631 in the context of general galaxy evolution, in Figure~\ref{fig:scaling}, we compare the SFR and $\hi$ mass of it and NGC 4656 to other galaxies of a stellar mass selected sample, and particularly to the main sequences of the two properties. It is interesting to notice that the FAST measurements of the $\hi$ mass put the two galaxies in the regime of $\hi$-excess galaxies, while the WSRT measurements put them in the $\hi$ main sequence of normal star-forming galaxies. We will need a dataset like the one used in this paper but for a census of interacting galactic systems with different mass ratios, gas richnesses, merging distances, and large-scale environments, as well as for a sample of control galaxies in relative isolations. Such a dataset will be available in the future by combining data from FEASTS and from existing and SKA related interferometry surveys. 

We conclude that, tidal interaction should be an efficient channel to accrete the IGM gas to the galaxy NGC 4631. The excess $\hi$ detected by FAST provides the crucial, new information to reach this conclusion.

\acknowledgments
We thank the anonymous referee for very constructive comments! We thank Xu Kong, Thijs van der Hulst, Zhiyuan Li, Ningyu Tang, Tobias Westmeier, Feng Yuan, Pei Zuo for useful discussions. JW thank support of the research grants from Ministry of Science and Technology of the People's Republic of China (NO. 2022YFA1602902),  the National Science Foundation of China (NO. 12073002, 11721303), and the science research grants from the China Manned Space Project (NO. CMS-CSST-2021-B02).  
S.H.OH. acknowledges support from the National Research Foundation of Korea (NRF) grant funded by the Korea government (Ministry of Science and ICT: MSIT; No. RS-2022-00197685). 
LCH was supported by the National Science Foundation of China (11721303, 11991052, 12011540375) and the China Manned Space Project (CMS-CSST-2021-A04, CMS-CSST-2021-A06). 
KMH acknowledges financial support from the State Agency for Research of the Spanish Ministry of Science, Innovation and Universities through the "Center of Excellence Severo Ochoa" awarded to the Instituto de Astrofísica de Andalucía (SEV-2017-0709), from the coordination of the participation in SKA-SPAIN, funded by the Ministry of Science and Innovation (MCIN), and financial support from grant RTI2018-096228-B-C31 (MCIU/AEI/FEDER,UE).
PK acknowledges financial support by the German Federal Ministry of Education and Research (BMBF) Verbundforschung grant 05A20PC4 (Verbundprojekt D-MeerKAT-II).
Parts of this research were supported by High-performance Computing Platform of Peking University.

This work made use of the data from FAST (Five-hundred-meter Aperture Spherical radio Telescope). FAST is a Chinese national mega-science facility, operated by National Astronomical Observatories, Chinese Academy of Sciences.

\facilities{FAST, GALEX, Spitzer, WSRT }

\software{Astropy \citep{astropy:2013, astropy:2018, astropy:2022}, Astrosclicer \citep{Punzo17}, {\scriptsize BAYGAUD} \citep{Oh22}, 3D-Barolo \citep{DiTeodoro15}, 
{\it Cloudy} \citep{Ferland98}, 
galpy \citep[v1.8.0]{Bovy15}, numpy \citep[v1.21.4]{vanderWalt11}, photutils \citep[v1.2.0]{Bradley19}, Python \citep[v3.9.13]{Perez07}, scipy \citep[1.8.0]{Virtanen20} }

\bibliography{n4631}{}
\bibliographystyle{apj}


\appendix
\section{The average beam image of FAST}
\label{sec:appendix_beam}
We derive a clean average beam image of the 19 beams, gridded in the same way as for the N4631 data. 
We use the calibrated mapping data of point sources from \citet{Jiang20}  to derive the average beam of FAST. The observation was specifically designed to characterize the beam properties of FAST. The data are in total 100 minutes' mapping of point sources in raster scan mode along right ascension or declination directions, with a high sampling rate of per 10$''$.  We refer the readers to \citet{Jiang20} for more details of the data and of the properties of the 19 beams of FAST.  

\begin{figure*} 
\centering
\includegraphics[width=8cm]{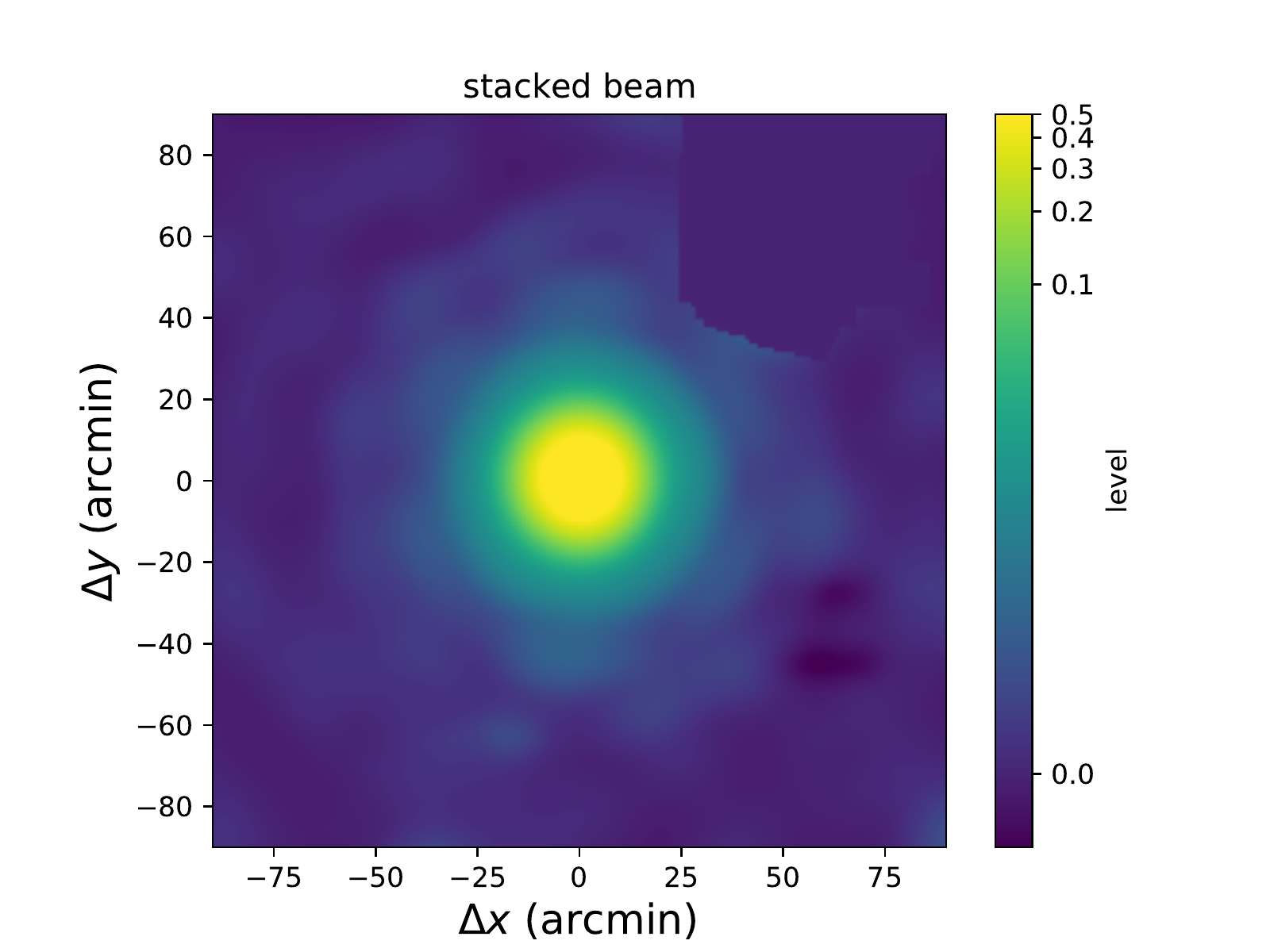}
\includegraphics[width=8cm]{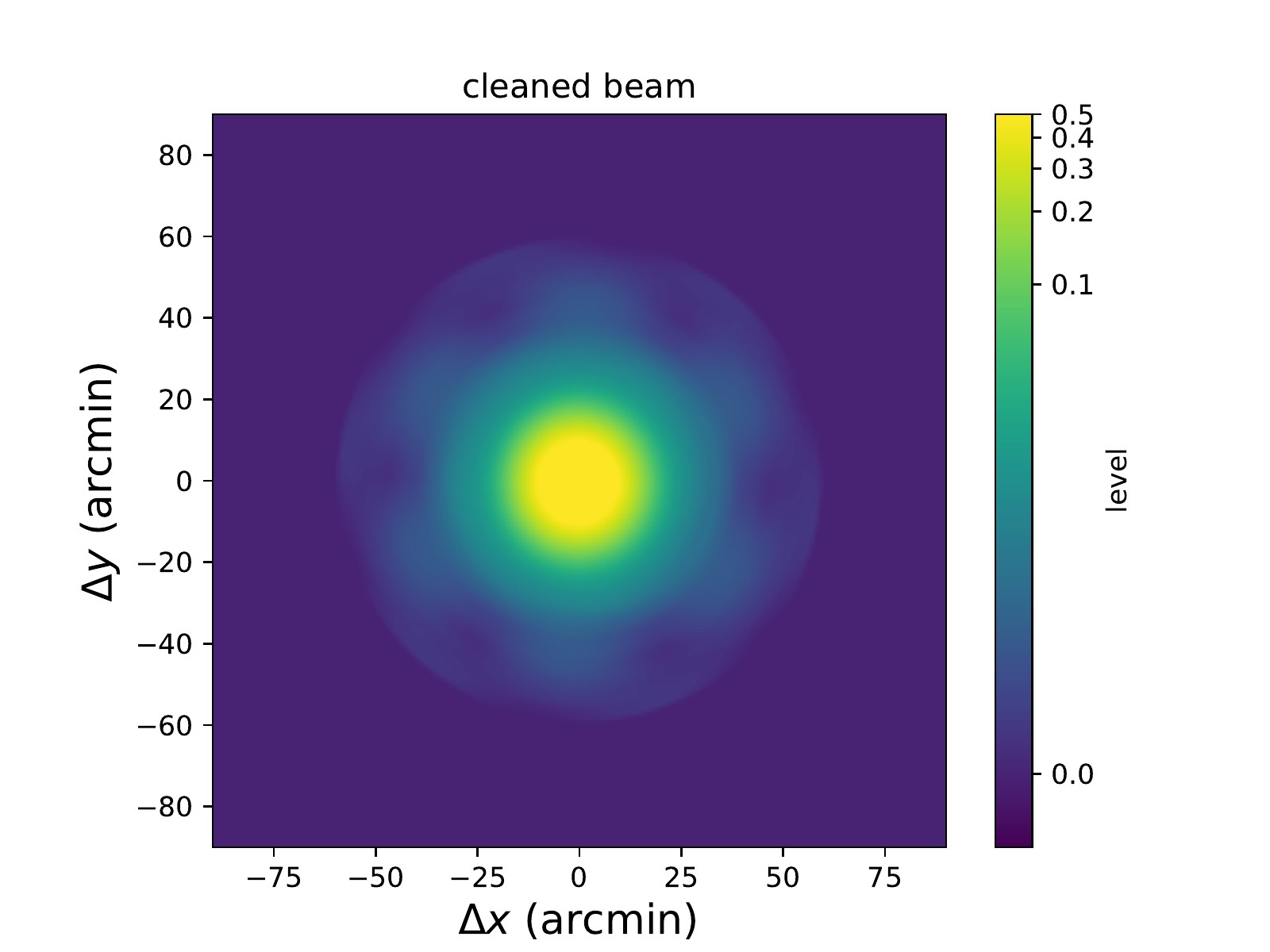}

\includegraphics[width=8cm]{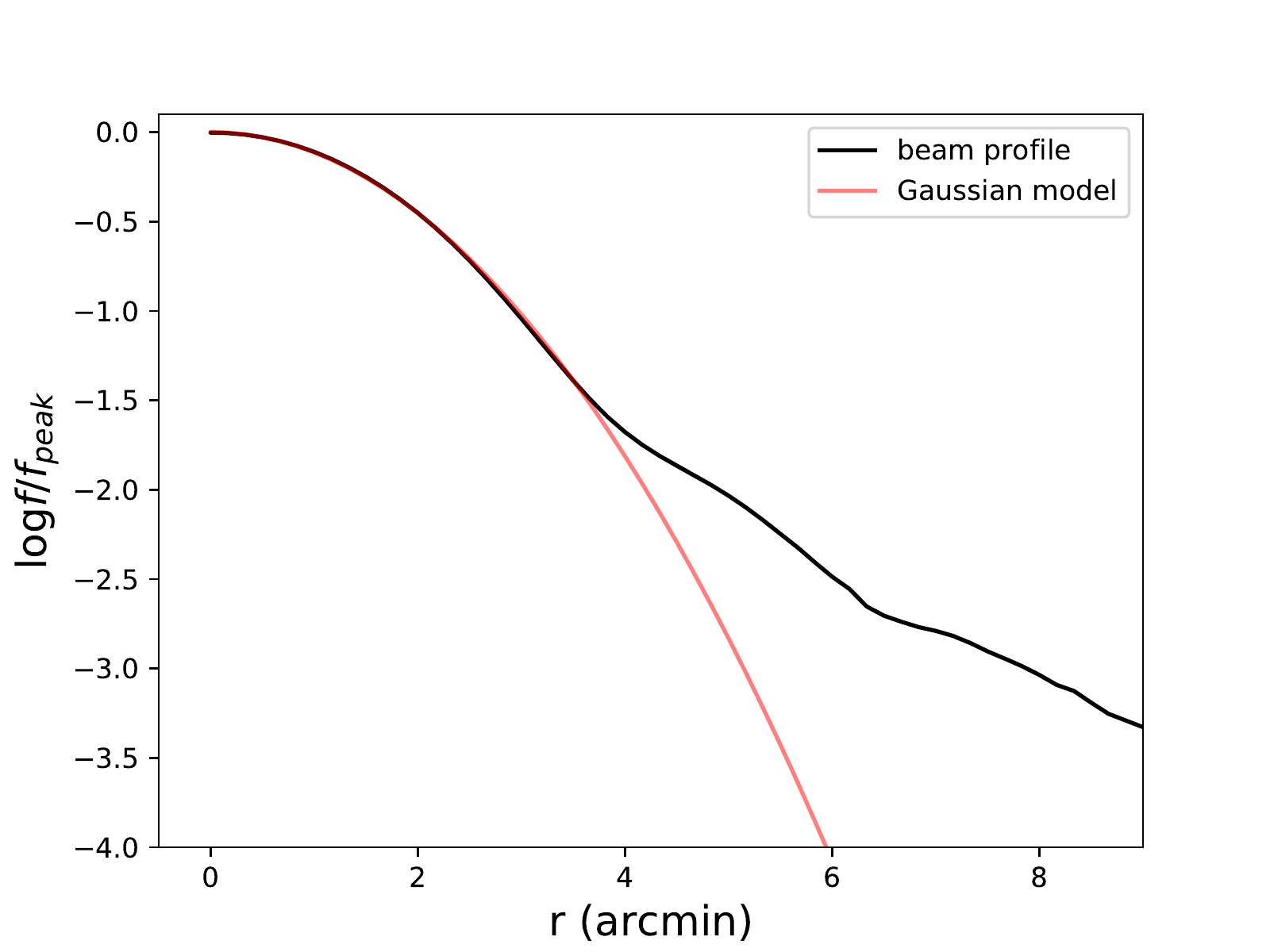}
\caption{The beam of the FAST $\hi$ data. Top-left and top-right panels are the stacked beam and the cleaned beam respectively. The bottom panel is the azimuthally averaged radial profile of the beam, and its best-fit Gaussian model with a FWHM of 3.24$'$. }
\label{fig:beam}
\end{figure*}

We use the same procedure that has produced the NGC 4631 cube in this paper to make the images of the 19 beams separately. We note that, a different 2-dimensional interpolation method was used in \citet{Jiang20} for gridding, which is improper for the NGC 4631 data here whose sampling rate is much lower. After masking contaminating sources in the neighborhood, we follow the steps in \citet{Jiang20}  and fit a skew Gaussian to each of the 19 beam images. We stack the images of the 19 beams after register them to the same Gaussian center. The stacking procedure takes the 3-sigma clipped mean value for each pixel. The directly stacked beam image looks like a smoothed version of beam 1 displayed in \citet{Jiang20} because of additional smoothing in gridding. It has a central core surrounded by a side-lobe ring, and then a axisymmetric periodical pattern with 6 broad peaks.  It still has some noise patterns in the background and imperfectness in the periodical pattern. We clean the beam image by first using the segmentation function of the python package {\it astropy.photutils} to flag and mask the noise patterns in the background. We then perform a Fourier decomposition of the periodical pattern, and find that the pattern can be well represented by only retaining the $m=6$ mode of the Fourier components. After these two steps, we obtain a relatively clean and average beam image for the NGC 4631 FAST image data. We show the directly stacked beam image, the cleaned beam image, and a difference map of the two images in Figure~\ref{fig:beam}. 

In Figure~\ref{fig:beam}, we also present an azimuthally averaged radial profile of the cleaned beam image. Within a radius of $\sim3.5'$, the inner region of the profile is well fitted by a Gaussian function with a FWHM of 3.24$'$. Beyond that, the real beam deviate from the Gaussian approximation, with a level of around 1\%. The level of the first side-lobe is around $1/10$ that of Arecibo \citep{Heiles01}, indicating the power of FAST to map low-surface density, extended $\hi$. However, the level also suggests that the scattered light due to side-lobes cannot be fully ignored when we investigate extended $\hi$ with column densities close to $10^{18}\cmsq$.  Therefore, in section~\ref{sec:ancillary}, we convolve the WSRT cube with the real beam of FAST, before comparing the distribution of $\hi$ fluxes between the WSRT and FAST data. 

We caution that, the beam shape of the NGC 4631 data may differ from the data of in \citet{Jiang20}, as the observing times are quite different. Moreover, the beam shapes, particularly the side-lobes, differ between the 19 beams, as shown in \citet{Jiang20}.  However, this average beam image is the best we can achieve with the resources in hand. And, the variation among beams is an intrinsic systematic uncertainty of the 19-beam mapping, which is a necessary compromise for the mapping efficiency.

\section{The disk and tidal tail regions of NGC 4631}
\label{sec:appendix_label_region}
We manually separate the FAST-detected region of the NGC 4631 and NGC 4656 system by arbitrarily drawing a line roughly along the disk direction of NGC 4656 (the white dashed line in Figure~\ref{fig:map_label_region}). We separate the WSRT-detected NGC 4631 region into regions of the main disk and 4 tidal tails.  We firstly determine the region of the main disk of NGC 4631 in the data cube. We take the parameters of the kinematic model of the main disk of NGC 4631 from \citet{Rand94}, and use 3D-Barolo \citep{DiTeodoro15} to make a 3 dimensional data cube based on the parameters. We arbitrarily take a density threshold equivalent to 0.02\% the peak density of the model to draw a mask of the NGC 4631 disk region in the cube. The region belonging to the two satellite galaxies NGC 4656 and Dwarf A are already labeled by SoFiA. The cube space beyond the disk region of NGC 4631, NGC 4656 and Dwarf A,  but are within the SoFiA mask of the WSRT cube are considered the tail region.

The separation of the tail region into different tails is performed with the 3-dimensional watershed algorithm of the python package {\it skimage.segmentation}. The outputs are 3-dimensional flagging masks of the separate components. A previous study, \citet{Combes78} manually separated the tidal features into four tails, denoting them by number 1 to 4, and used numerical simulation of interaction to reproduce the morphology and kinematics of these 4 tails. The same denoting system has been adopted by studies later to have a coherence context of discussion (e.g. \citealt{Rand94}). Our watershed results directly flag tail 3 and 4, but 1 and 2 are blended. We manually and arbitrarily draw a division line in the sky plane to separate tail 1 from 2 in the blended region to qualitatively match the separation in \citet{Rand94}. 

The sky projected view of these regions are displayed in Figure~\ref{fig:map_label_region}.

\section{The temperature of the hot gas near the disk}
\label{sec:appendix_temperature}
There are abundant studies in the literature on the properties of the X-ray emitting hot gas near the disk within a distance of around 10 kpc. This part of the hot gas halo mostly represents hot gas outflows from the galactic disk, especially from its inner actively star-forming region.  Therefore its temperature should be higher than, and can be used as upper limit of that in a hydrostatic equilibrium in the galaxy's potential.

\citet{Wang95} used ROSAT data to detect the soft X-ray radiation of the hot gas of NGC 4631 out to 8 kpc above the disk plane. They estimated a characteristic thermal temperature of 0.25$\pm0.03$ keV. \citet{Wang01} used Chandra to detect the halo out to a similar distance. They performed a 2-component thermal plasma model fit, obtaining a hot component of 0.61$\pm0.12$ keV close to the disk and a cooler component of 0.18$\pm0.02$ keV dominating the outer corona. As in this study we focus on the tidal $\hi$ far away from the disk, we only take the temperature of the further component. \citet{Tullmann06} used XMM-Newton to derive the temperature out of 3 stripes south of the disk, and 5 stripes north of the disk, reaching out to nearly 11 kpc. They used 5 bands ranging from super-soft to hard, so they managed to derive two characteristic temperatures for both a soft and a hard components. The hard component has a mean temperature of 0.24$\pm$0.03 keV, and a slightly higher but comparable number density of IGM throughout the analysis regions. The soft component has a temperature that is roughly 4 times lower, and we thus take the temperature of the more energetic hard component. Finally, \citet{Yamasaki09} used the Imaging Spectrometer of Suzaku to trance X-ray halo out to about 10 kpc from the disk. They fit 2-component thermal models to the disk and the halo regions separately. In the halo region, the hard component of 0.3$\pm0.016$ keV is dominating over the soft component by 5 times more flux. 

 Taking together these four sets of previous measurements, we calculate a mean value of 0.24$\pm$0.03 keV,  equivalent to $2.8\times10^6$ K, as the temperature of the hot gas halo within 10 kpc around the NGC 4631 disk.

\section{The dark matter halo mass of N4631g}
\label{sec:appendix_mass}
 We use $M_{500}$ as the fiducial measure of the dark matter halo mass, the mass within $r_{500}$ the radius where the average density is 500 times the critical density of the universe. Characteristic masses are also defined at alternative averaged density levels, like $M_{200}$ and $M_{101}$ (the virial mass at redshift z$=0$ in $\Lambda$CDM cosmology). They are convertible with each other assuming a NFW model \citep{Navarro97} of the dark matter halo with a concentration index $c_{\Delta}$ of 8, as is expected for a halo of roughly $10^{12} ~M_{\odot}$ at redshift $z=0$ \citep{Dutton14}. 

We use the stellar mass-halo mass relation in \citet{Behroozi10} to derive a lower limit of $\log M_{500}/M_{\odot}=11.61$. It is viewed as a lower limit because from halo occupation distribution studies, dark matter halos with a mass around $10^{12} ~M_{\odot}$ should have the number of satellites which have stellar masses above $10^{9.28}~M_{\odot}$ far less than unity \citep{Bose19}. 

We use the $M_{500}$-IGM temperature relation from \citet{Reichert11}, in combination with the characteristic temperature of the near-disk hot gas summarized above, to derive an upper limit of $\log M_{500}/M_{\odot} =12.13\pm0.06$. 

The N4631g can be found in the group catalog of \citet{Kourkchi17} with a PGC id of 42637. From that group catalog, it has 10 member galaxies. These member galaxies have a radial velocity dispersion $\sigma_c$ of 217 $\kms$, and a projected gravitational radius $R_{g}$ of 92 kpc. Based on the equation of \citet{Tully15}, we derive$\log M_{500}/M_{\odot}$ of 11.97. This value is between the lower and upper limits derived above, and is taken to be the final estimate.

Accordingly, the $r_{500}$, $r_{200}$, and $M_{200}$ of N4631g are 148 kpc, 224 kpc, and $10^{12.1}~M_{\odot}$ respectively. The corresponding virial mass implies a virial temperature of 8$\times10^5$ K, considerably lower than that of the X-ray emitting and outflowing hot gas near the disk.

\section{The density of the hot gas halo}
\label{sec:appendix_IGMprof}
We base on the $M_{500}$ estimated in appendix~\ref{sec:appendix_mass}  to derive $M_{500,gas}$, the hot gas mass within $R_{500}$. We use the $M_{500,gas}$-$M_{500}$ relations from \citet{Andreon17} and \citet{Ettori15}, which derive $\log M_{500,gas}/M_{\odot}$ of 10.56$\pm0.17$ and 10.51$\pm0.10$ respectively. The two values are consistent within the error bar, and we take the one with smaller error.

\begin{figure} 
\centering
\includegraphics[width=8cm]{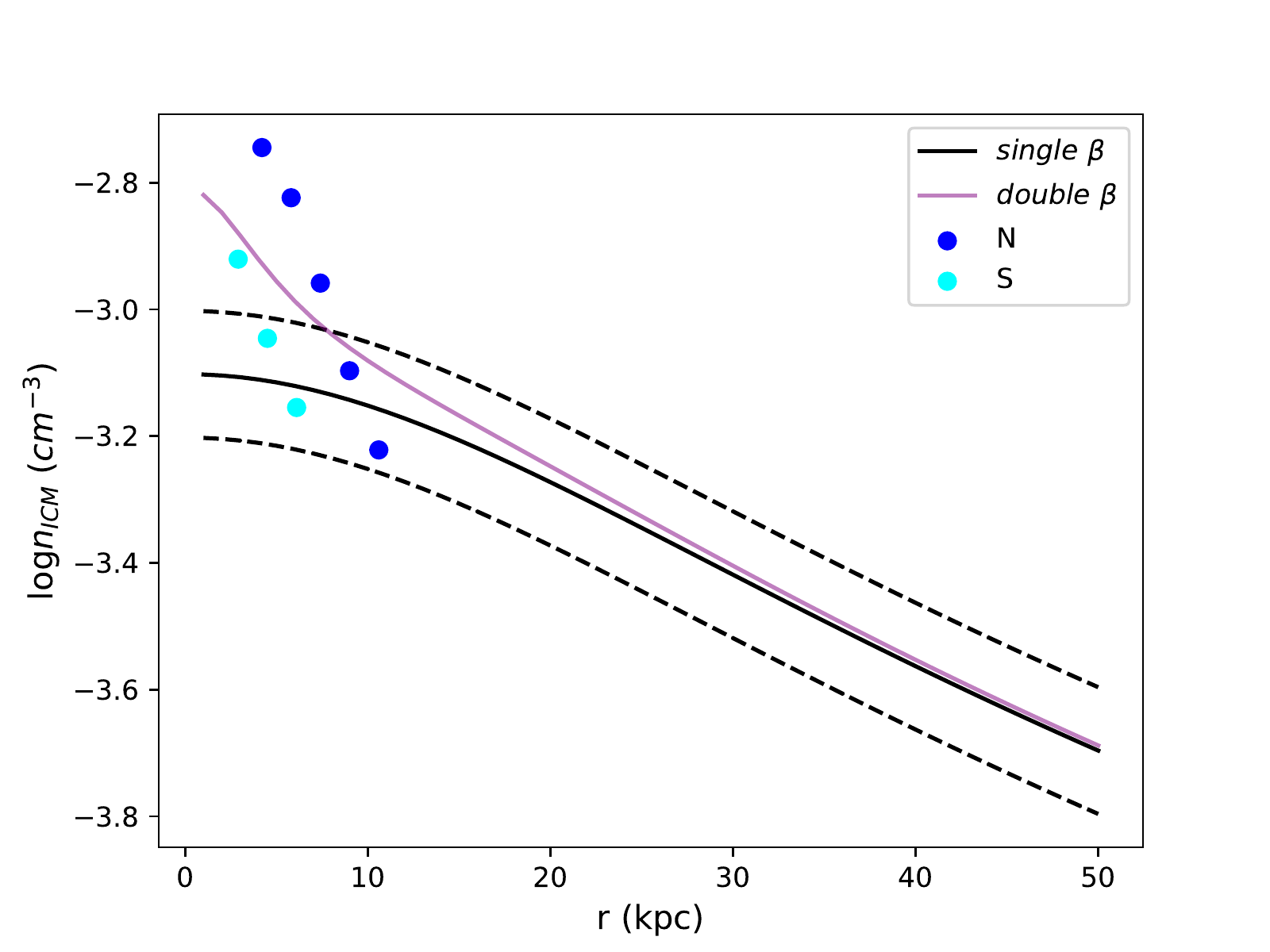}
\caption{ IGM density profiles of N4631g. The black and pink solid lines are the single and double-$\beta$ models. The black dashed lines are the 1-$\sigma$ uncertainty range of the single-$\beta$ model due to uncertainty in the estimate of $M_{500,gas}$. The blue and cyan dots are measurements from Yamasaki et al. (2018) at distances from the north and south sides of the disk.  }
\label{fig:IGM_prof}
\end{figure}

We consider a single-$\beta$ model distribution of the IGM. Following the specifics in \citet{Eckert11}, we assume the $\beta$ value to be 0.64, and the core radius $r_c=0.19r_{500}$. Cumulating the IGM model profile from center to $R_{500}$, we derive a central density of 5.778$\times10^{-4}~ cm^{-3}$. We also consider a double-$\beta$ model, to match the fact that many previous studies found two thermal components in the hot gas halo. Following \citet{Eckert11}, we set the outer component to have core radius $r_c=0.03r_{500}$. We arbitrarily set the central density of the outer and inner components to be equal, partly motivated by the fact that in \citet{Tullmann06} the density of the hot and cold components are roughly equal in the inner corona. We plot both models in Figure~\ref{fig:IGM_prof}. Although the double-$\beta$ model fit the measured densities from \citet{Yamasaki09} better, both models are close beyond a radius of 10 kpc in the IGM region. We thus adopt the single-$\beta$ model for simpler assumptions.

\section{Amplitude spectral analysis throughout applicable channels}
\label{sec:appendix_powerspectra}
\begin{figure*} 
\centering
\includegraphics[width=8cm]{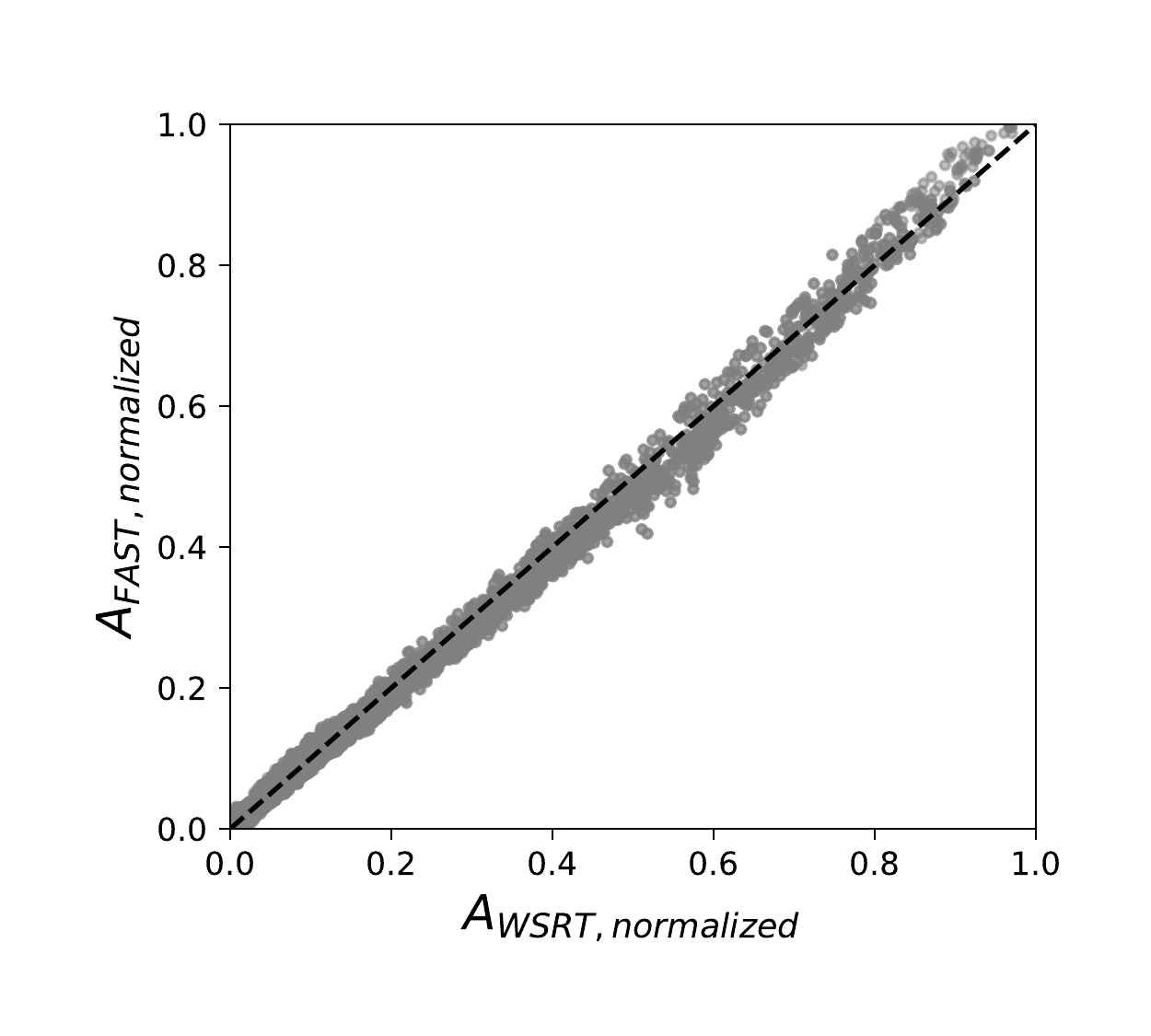}
\includegraphics[width=8cm]{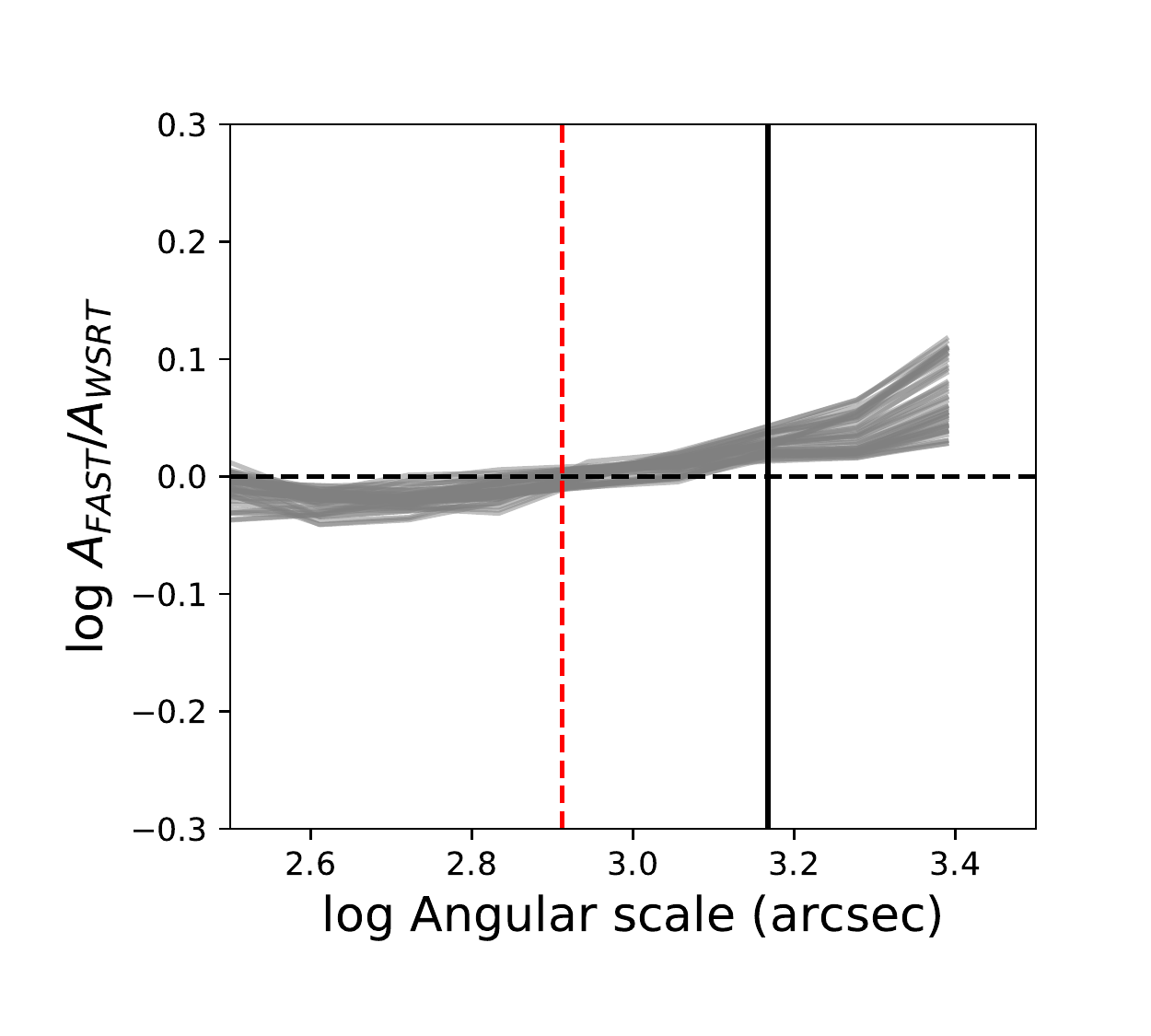}
\caption{ The comparison between amplitudes of amplitude spectra from the WSRT cube and PB-attenuated FAST cube. Left:  relation of normalized amplitudes from the two types of cubes selected from the angular scale range between 4$'$ and 24.5$'$ as in Figure~\ref{fig:power_spectrum_ch31}. The dashed line mark the $y=x$ line. Right: the median curves of the ratio of amplitudes from the two cubes as a function of the angular scale. Each curve corresponds to one channel. The horizontal dashed line mark the y position of zero. The red vertical line marks the mean critical angular scale of 13.6$'$ for FAST amplitudes to exceed the WSRT amplitudes by 1\%. The black vertical line marks the angular scale corresponding to the shortest baseline of the WSRT array configuration. }
\label{fig:power_compare}
\end{figure*}

To demonstrate the consistency of amplitudes cross the applicable channels (which have flux intensity greater than 0.15 Jy), we plot the relation between the normalized amplitudes of all these channels in the left panel of Figure~\ref{fig:power_compare}. The normalization factor of each channel is taken to be the maximum amplitude from the  PB-attenuated FAST cube in the selected angular scale range (4 to 24.5$'$). The data points distribute close to the $y=x$ line.  To demonstrate the deviation of critical angular scales, in the right panel of Figure~\ref{fig:power_compare}, we show the relation between amplitude ratios and angular scales from the selected channels. Each curve represents the median relation from a channel map, and starts from 4$'$. The curves start to exceed unity near the mean critical angular scale of 13.6$'$.

\section{MCMC result of double-Gaussian fit to the super profiles}
\label{sec:appendix_dgauss}
\begin{figure*} 
\centering
\includegraphics[width=14cm]{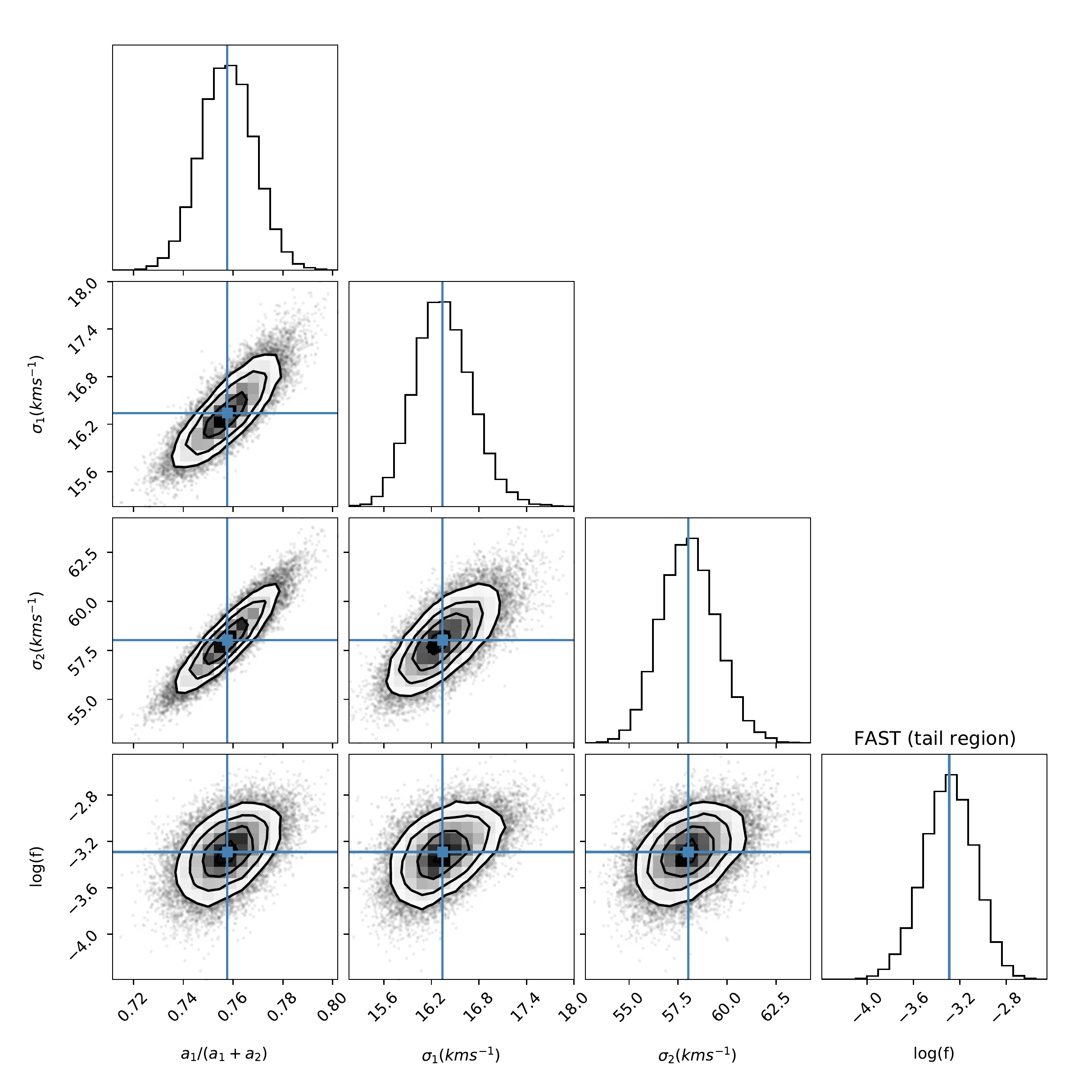}
\caption{Corner figure of parameter probabilities of the double-Gaussian model for the $\hi$ super profile of FAST data cube in the tail region.  }
\label{fig:mcmc_fast}
\end{figure*}

We use emcee \citep{Foreman-Mackey13} to conduct a double-Gaussian fit to the super profile stacked from the line-of-sights with single-Gaussian spectra in dense $\hi$ in the tail region of the projected FAST cube. The amplitude $a$ and $\sigma$ of the narrow and broad Gaussian components are denoted by 1 and 2 respectively. We also include a fraction uncertainty of the model $f$ in the fitting. The corner figure of probability distribution of parameters are displayed in Figure~\ref{fig:mcmc_fast}.  We do not find strong degeneracy between model parameters from the corner figures, where the probability distribution are projected onto 2-dimensional diagrams of the parameters. The fractional uncertainty of the model is low. So the double-Gaussian model seems a good description of the super profiles. 

\begin{figure*} 
\centering
\includegraphics[width=14cm]{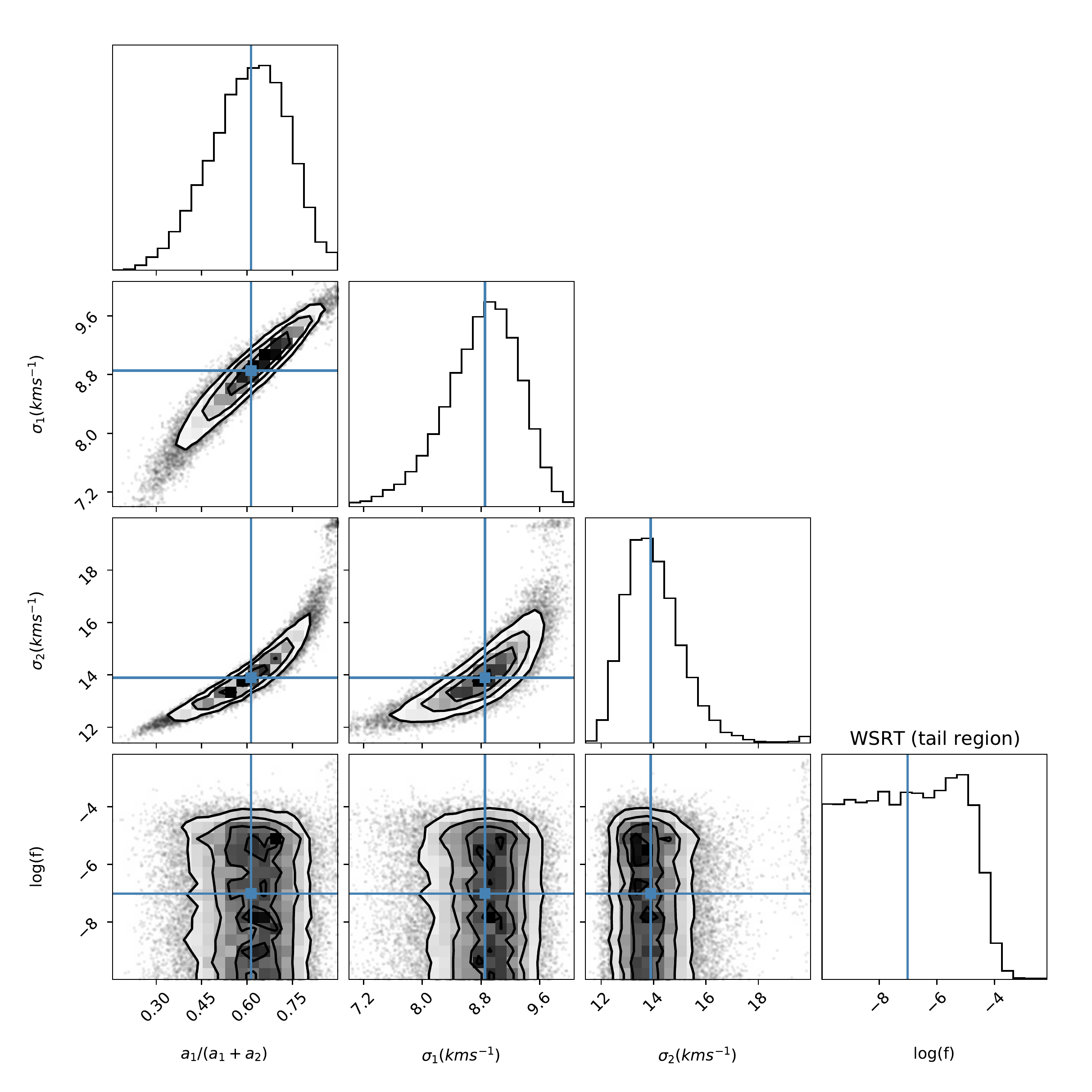}
\caption{Corner figure of parameter probabilities of the double-Gaussian model for the $\hi$ super profile of WSRT data cube in the tail region. }
\label{fig:mcmc_wsrt}
\end{figure*}

\begin{figure*} 
\centering
\includegraphics[width=14cm]{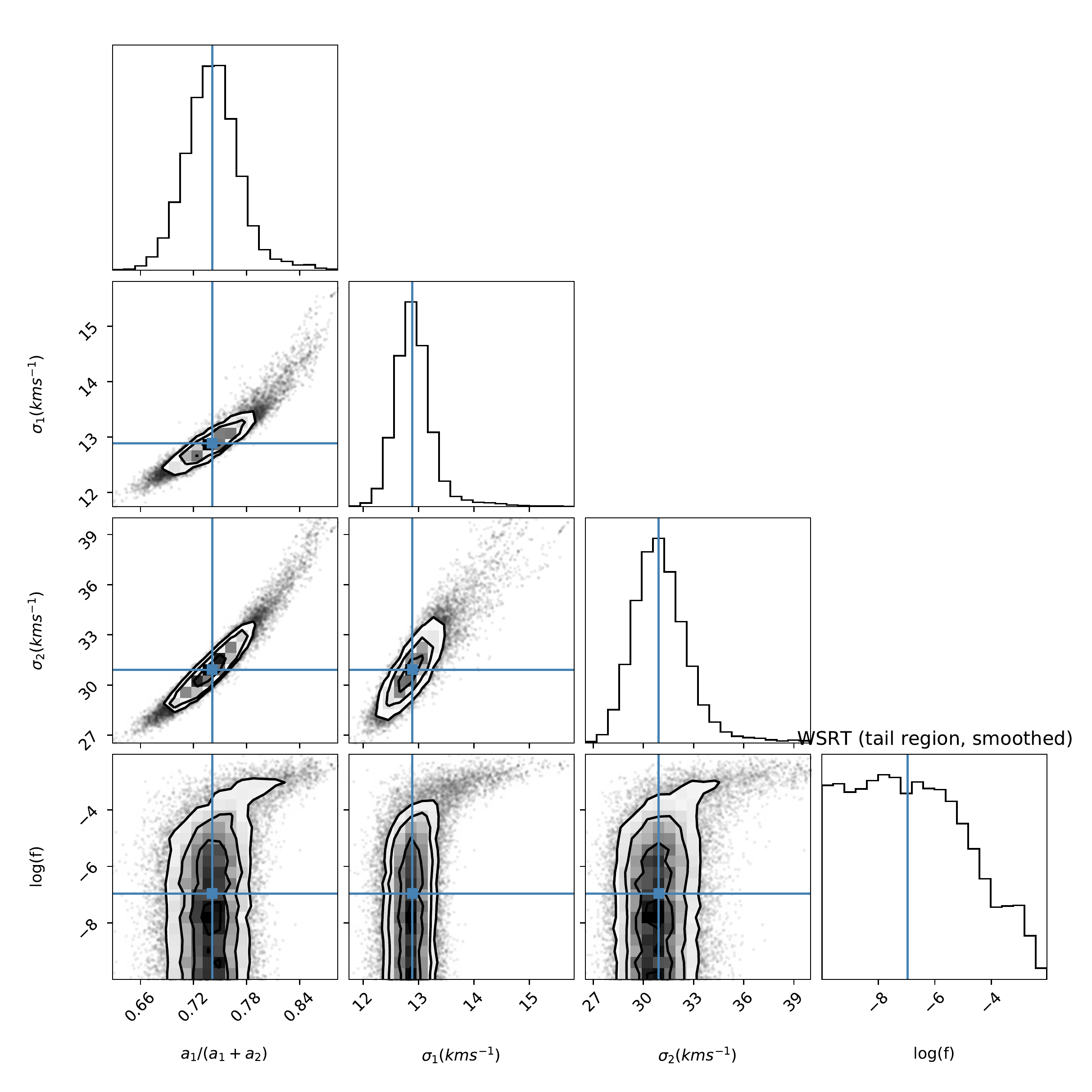}
\caption{Corner figure of parameter probabilities of the double-Gaussian model for the $\hi$ super profile of smoothed WSRT data cube in the tail region. }
\label{fig:mcmc_wsrt_smoothed}
\end{figure*}

We do the same for the smoothed WSRT cube, and the original WSRT cube. The results are displayed in Figure ~\ref{fig:mcmc_wsrt}, and ~\ref{fig:mcmc_wsrt_smoothed} respectively. The degeneracies of $\sigma_1$ and $\sigma_2$ with $a_1/(a_1+a_2)$ become stronger, but the probability distributions are still relatively narrow. So the double-Gaussian model seems still a reasonable description of the super profiles. 

The best-fit $\sigma$ of the narrow and broad Gaussian components, and the ratio of peak intensities between them are $16.3_{-0.33}^{+0.37}~\kms$, $58.0_{-1.39}^{+1.43} ~\kms$, and 0.76$_{0.010}^{+0.011}$ for the super profile of FAST data. The values are $12.9_{-0.29}^{+0.30} ~\kms$, $30.9_{-1.38}^{+1.63} ~\kms$, and $0.74_{-0.028}^{+0.028}$ for the super profile of the smoothed WSRT cube, and $8.9_{-0.55}^{+0.45} ~\kms$, $13.9_{-0.91}^{+1.17} ~\kms$ and $0.61_{-0.136}^{+0.118}$ for the super profile of the WSRT cube.

\section{Using Cloudy to predict the critical column density of HI for photon ionisation}
\label{sec:CLOUDY}
The procedure below is a modified version of the one described in \citet{Borthakur15}.  

We consider two sources contributing to the UV photons, the cosmic background UV radiation, and the photons from the young stars in NGC 4631. For the UV photons associated with young stars, we use the equation from \citet{Tumlinson11} to estimate the dimensionless ionisation parameter $U_{star}$ which depends on the SFR of NGC 4631 and the distance squared. It also assumes a uniform fraction of 0.1 for the UV photons to escape from the interstellar medium.  We have ignored the UV photons from NGC 4656, as its SFR is around one fourth, thus the distance to have the same level of $U_{star}$ is half that of NGC 4631. The UV photons from NGC 4656 only start to be important when the distance from NGC 4631 is larger than 52.5 kpc along the direction connecting these two galaxies. Diffuse stellar features have been found around NGC 4631, but mostly consisting of old stars \citep{MartnezDelgado15}, unlikely to provide additional UV photons.

\begin{figure} 
\centering
\includegraphics[width=8cm]{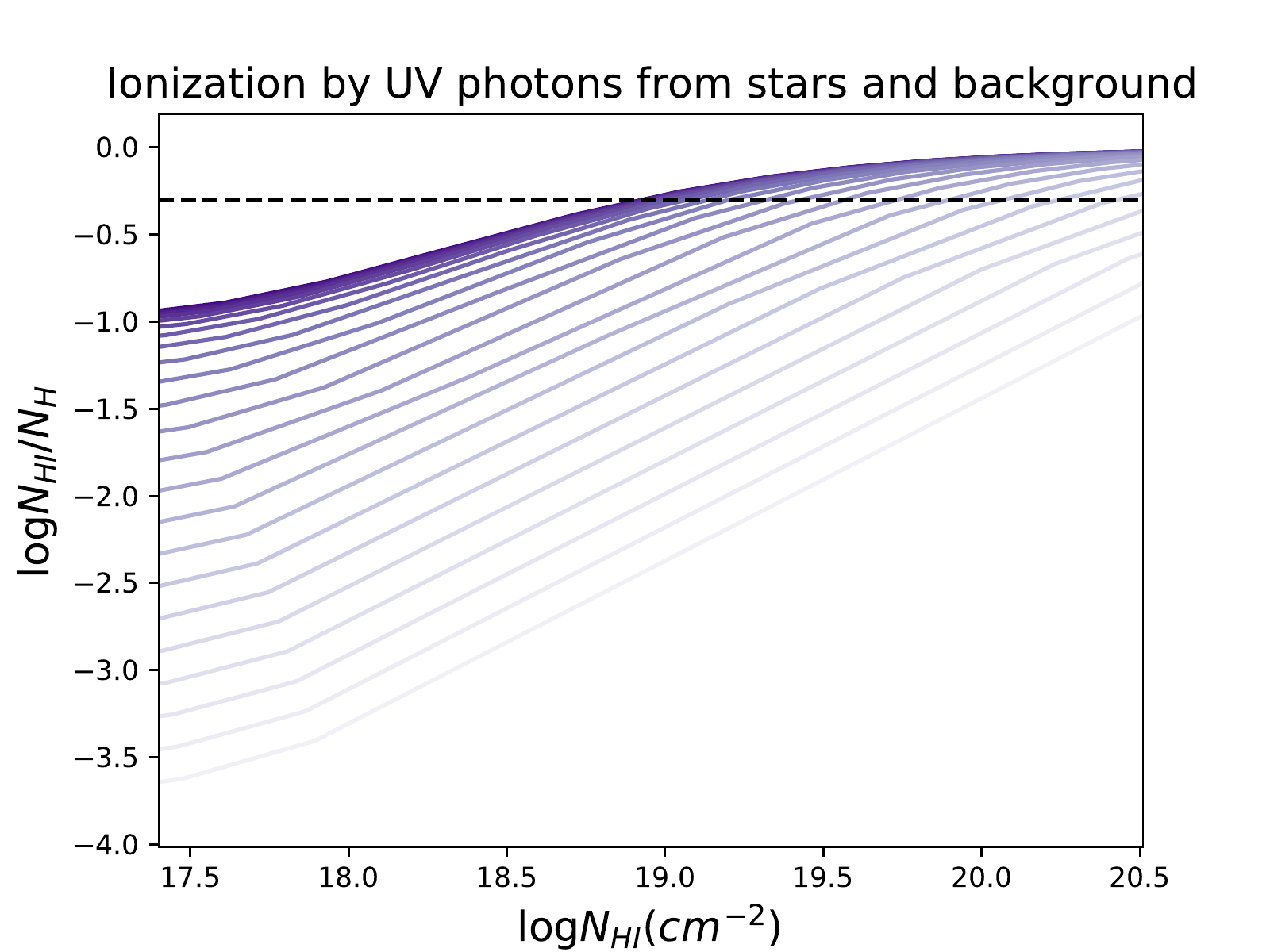}
\includegraphics[width=8cm]{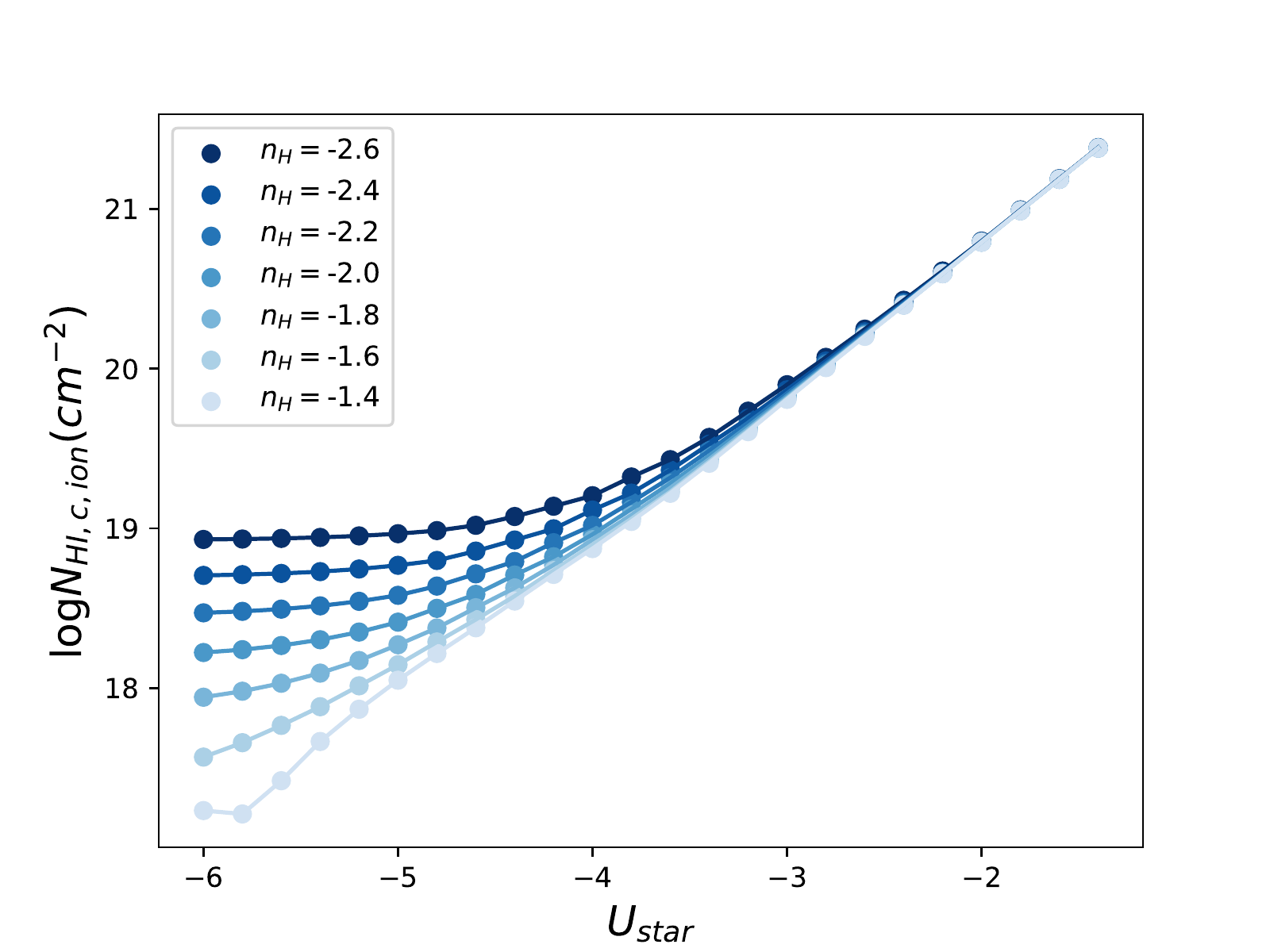}
\caption{Photon ionisation of hydrogen in grids of stellar ionisation parameter, hydrogen density and hydrogen column density. Left: the neutral ratio of hydrogen, $N_{\rm HI}/N_{\rm H}$, is plotted as a function of the $\hi$ column density in bins of stellar ionisation parameter $U_{star}$. The values of $U_{star}$ range from -6 to -1.4 with a linear step of 0.2, and the curves in lighter purple colors correspond to higher values of $U_{star}$. 
 The dashed, horizontal line mark where the neutral ratio is 50\%. Right: the critical $\hi$ column density, where the ionisation or neutral ratio is 50\%, as a function of $U_{star}$ in different bins of $n_{\rm H}$. The darker blue colors correspond to  lower values of $n_{\rm H}$. }
\label{fig:grid_ionisation}
\end{figure}

We use {\it starburst99} \citep{Leitherer10} to generate a young stellar population with a solar metallicity and an age of 4 Myr.  We use the spectral energy distribution of this stellar population as input for {\it Cloudy}. We generate a three dimensional grid of hydrogen density $n_{\rm H}$, the hydrogen column density $N_{\rm H}$, and the stellar ionisation parameter $U_{star}$. $\log n_{\rm H}$ ranges from -2.6 to -1.8 with a step of 0.2,  $\log N_{\rm H}$ range 17.5 to 22.5 with a step of 0.25, and $U_{star}$ from -6 to -1.4 with a step of 0.2. We use the default ``background'' and ``Background cosmic ray'', to add the ionizing effects of the cosmic UV background and the cosmic ray background. In the top panel of Figure~\ref{fig:grid_ionisation}, we plot the resulting neutral fraction of hydrogen ($N_{\rm HI}/N_{\rm H}$), as a function of $\log N_{\rm HI}$ in different bins of $U$, fixing $\log n_{\rm HI}$ at -2.6. We can see that toward the low values of $U_{star}$ (dark purple), $N_{\rm HI}/N_{\rm H}$ converges to the highest possible value at a given $N_{\rm HI}$,  because the comic UV background starts to dominate there. Setting $N_{\rm HI}/N_{\rm H}=0.5$, we derive the critical $\hi$ column density ($N_{\rm HI,c,ion}$) from each curve. We plot $N_{\rm HI,c,ion}$ as a function of $U_{star}$ for different values of $n_{\rm H}$ in the righ panel of Figure~\ref{fig:grid_ionisation}. We assume $n_{\rm H}=2n_{\rm HI}$, and interpolate in this parameter space to derive the critical column density of $\hi$ as a function of distance.

We plot $N_{\rm HI,c,ion}$ as a function of radius to NGC 4631 in the right panel of Figure~\ref{fig:Hdens_survive}. The values of $N_{\rm HI,c,ion}$ flatten around $10^{19.2}~\cmsq$ beyond a distance of 30 kpc. The value of $10^{19.2}~\cmsq$ is close to many previously derived when only accounting for the cosmic background of UV radiation \citep{Maloney93}, but if we remove the effect of young stars in NGC 4631, $N_{\rm HI,c,ion}$ would drop to one third of its current value at a radius of 30 kpc.

 We note that the leakage fraction of UV photons, the extent of shielding by $\hi$ in tails at smaller distances, the lack of information on the filling factor and clumpiness of $\hi$, and the contribution of ionizing energy from shocks, are major sources of uncertainties in the deviation of the ionisation related parameters. 

\section{The gravitational potential around NGC 4631}
\label{sec:appendix_potential}
We use {\it galpy} \citep{Bovy15} to model the mass distribution and gravitational potential around NGC 4631. We use the Miyamoto-Nagai model with the Milky Way specifics (scale length 3 kpc and scale height 0.28 kpc) to represent the disk, the NFW model with scale radius equal to $r_{200}/c_{\Delta}$ to represent the dark matter, and use the rotational velocity of 145 $\kms$ at a radius of 8 kpc \citep{Rand94} to calibrate the normalization. The resulted mass model has a disk mass of $10^{10.2}~M_{\odot}$ within 8 kpc, and a halo mass of  $10^{12.02}~M_{\odot}$ within $r_{500}$ derived in appendix~\ref{sec:appendix_mass}.  Thus the mass model is close to the observed stellar mass and $M_{500}$ of NGC 4631. Then we use the {\it evaluatePotentials} task of {\it galpy} to evaluate the potential distribution around NGC 4631.

\end{CJK*}
\end{document}